\pdfoutput=1

\newcommand{\nin}{\noindent}

\newcommand{\non}{\nonumber}

\newcommand{\yskip}{\penalty -50 \vskip 3pt plus 3pt minus 4pt}

\newcommand{\qua}{\ {\vbox{\hrule\hbox{\vrule\hbox to 4pt{\vbox to 8pt{}}
\vrule}\hrule}}\ }
\newcommand{\blank}{\rm{\vbox{\hbox{\vrule\hbox to 3pt{\vbox to 2.5pt{}}\vrule}\hrule}}}

\newcommand{\co}[2]{\par\yskip
                    \noindent\hangindent #1cm
                    \hbox to #1cm
                    {#2\hss\hskip 10pt}}
\newcommand{\cond}[1]{\par\yskip
                   \hangindent 0.7cm \noindent
                   \hbox to 0.7cm{#1\hfill}}
\newcommand{\condition}[1]{\par\yskip
                   \hangindent 30pt \noindent
                   \hbox to 30pt{\hss#1\hskip10pt}}
\newcommand{\condvar}[2]{\par\yskip
                   \hangindent #1cm \noindent
                   \hbox to #1cm{\hss#2\hskip10pt}}

\newtheorem{@theorem}{Theorem}
\newcommand{\theorem}[1]{\begin{@theorem}#1\end{@theorem}}

\newtheorem{@satz}{Satz}
\newcommand{\satz}[1]{\begin{@satz}#1\end{@satz}}

\newtheorem{@claim}{Claim}
\newcommand{\claim}[1]{\begin{@claim}#1\end{@claim}}

\newtheorem{@behauptung}{Behauptung}
\newcommand{\behauptung}[1]{\begin{@behauptung}#1\end{@behauptung}}

\newtheorem{@proposition}{Proposition}
\newcommand{\proposition}[1]{\begin{@proposition}#1\end{@proposition}}

\newtheorem{@lemma}{Lemma}
\newcommand{\lemma}[1]{\begin{@lemma}#1\end{@lemma}}

\newtheorem{@corollary}{Corollary}
\newcommand{\corollary}[1]{\begin{@corollary}#1\end{@corollary}}

\newtheorem{@korollar}{Korollar}
\newcommand{\korollar}[1]{\begin{@korollar}#1\end{@korollar}}

\newtheorem{@definition}{Definition}
\newcommand{\definition}[1]{\begin{@definition}#1\end{@definition}}

\newtheorem{@assumption}{Assumption}
\newcommand{\assumption}[1]{\begin{@assumption}#1\end{@assumption}}

\newtheorem{@annahme}{Annahme}
\newcommand{\annahme}[1]{\begin{@annahme}#1\end{@annahme}}

\newtheorem{@remark}{Remark}
\newcommand{\remark}[1]{\begin{@remark}#1\end{@remark}}

\newtheorem{@bemerkung}{Bemerkung}
\newcommand{\bemerkung}[1]{\begin{@bemerkung}#1\end{@bemerkung}}

\newtheorem{@example}{Example}
\newcommand{\example}[1]{\begin{@example}#1\end{@example}}

\newtheorem{@beispiel}{Beispiel}
\newcommand{\beispiel}[1]{\begin{@beispiel}#1\end{@beispiel}}

\def\IIbox#1#2#3 #4 #5{\noindent\hbox{\relax
  \vtop{\hsize #1cm\noindent\strut #4}\hskip #2cm
  \vtop{\hsize #3cm\noindent\strut #5}\hss}\eol}

\newcommand{\gl}{\lower4pt\hbox{${{\displaystyle >}\atop{\displaystyle <}}\atop\/ $}}

\renewcommand{\lg}{\lower4pt\hbox{${{\displaystyle <}\atop{\displaystyle >}}\atop\/ $}}

\newcommand{\sge}{\mathop{\lower 2pt\hbox{$\buildrel >\over =$}}}

\newcommand{\sle}{\mathop{\lower 2pt\hbox{$\buildrel <\over =$}}}

\newcommand{\slg}{\lower1.6truept\hbox{${{{\scriptstyle >}\atop{\scriptstyle =}}\atop {\scriptstyle <}}$}}

\newcommand{\sgl}{\lower1.6truept\hbox{${{{\scriptstyle <}\atop{\scriptstyle =}}\atop {\scriptstyle >}}$}}

\newcommand{\eax}{{{\lower4truept\hbox{$=$}\atop {\textstyle x}}\atop \/}}

\newcommand{\notprecsim}{\hskip 2 pt
                \not\kern -0.3 em{\hbox{$\precsim$}}
                \hskip 2 pt }
\newcommand{\notsuccsim}{\hskip 2 pt
                \not\kern -0.3 em{\hbox{$\succsim$}}
                \hskip 2 pt     }
\newcommand{\precsimi}{\hskip 2 pt
              \hbox{$\precsim_{\kern -0.6 em\raise 0.4 ex\hbox{
              $\scriptstyle i$}}$}
              \hskip 2 pt        }
\newcommand{\succsimi}{\hskip 2 pt
              \hbox{$\succsim_{\kern -0.6 em\raise 0.4 ex\hbox{
              $\scriptstyle i$}}$}
              \hskip 2 pt}
\newcommand{\precsima}{\hskip 2 pt
              \hbox{$\precsim_{\kern -0.6 em\raise 0.1 ex\hbox{
              $\scriptstyle a$}}$}
              \hskip 2 pt}
\newcommand{\precsimA}{\hskip 2 pt
              \hbox{$\precsim_{\kern -0.6 em\raise 0.1 ex\hbox{
              $\scriptstyle A$}}$}
              \hskip 2 pt}
\newcommand{\succsima}{\hskip 2 pt
              \hbox{$\succsim_{\kern -0.6 em\raise 0.1 ex\hbox{
              $\scriptstyle a$}}$}
              \hskip 2 pt}
\newcommand{\succsimA}{\hskip 2 pt
              \hbox{$\succsim_{\kern -0.6 em\raise 0.1 ex\hbox{
              $\scriptstyle A$}}$}
              \hskip 2 pt}
\newcommand{\precsimb}{\hskip 2 pt
              \hbox{$\precsim_{\kern -0.6 em\raise 0.1 ex\hbox{
              $\scriptstyle b$}}$}
              \hskip 2 pt}
\newcommand{\precsimB}{\hskip 2 pt
              \hbox{$\precsim_{\kern -0.6 em\raise 0.1 ex\hbox{
              $\scriptstyle B$}}$}
              \hskip 2 pt}
\newcommand{\succsimb}{\hskip 2 pt
              \hbox{$\succsim_{\kern -0.6 em\raise 0.4 ex\hbox{
              $\scriptstyle b$}}$}
              \hskip 2 pt}
\newcommand{\succsimB}{\hskip 2 pt
              \hbox{$\succsim_{\kern -0.6 em\raise 0.4 ex\hbox{
              $\scriptstyle B$}}$}
              \hskip 2 pt}
\newcommand{\precsimn}{\hskip 2 pt
              \hbox{$\precsim_{\kern -0.6 em\raise 0.1 ex\hbox{
              $\scriptstyle n$}}$}
              \hskip 2 pt }
\newcommand{\succsimn}{\hskip 2 pt
              \hbox{$\succsim_{\kern -0.6 em\raise 0.1 ex\hbox{
              $\scriptstyle n$}}$}
              \hskip 2 pt }
\newcommand{\succsimai}{\hskip 2 pt
               \hbox{$\succsim_{\kern -0.6 em\raise 0.4 ex\hbox{
               $\scriptstyle(a,i)$}}$}
               \hskip 2 pt }

\newcommand{\C}{\hbox{C\hskip-0.5em\lower-0.1ex\hbox{\vrule
                      height1.34ex width0.07em }}\hskip0.50em}

\newcommand{\G}{\hbox{G\hskip-0.525em\lower-0.081ex\hbox{\vrule
                      height1.4ex width0.07em }}\hskip0.50em}

\renewcommand{\O}{\hbox{O\hskip-0.525em\lower-0.095ex\hbox{\vrule
                      height1.45ex width0.07em}}\hskip0.50em}

\newcommand{\Q}{\hbox{Q\hskip-0.525em\lower-0.097ex\hbox{\vrule
                      height1.47ex width0.07em}}\hskip0.50em}

\newcommand{\U}{\hbox{U\hskip-0.45em\lower-0.02ex\hbox{\vrule
                height1.54ex width0.07em}}\hskip0.50em}

\newcommand{\1}{1\hskip-0.28em \text{I}}

\newcommand{\Kscr}{{\cal K}}

\newcommand{\Zscr}{{\cal Z}}

\newcommand{\argmin}{\mbox{argmin}}

\documentclass[a4paper,12pt]{article}

\usepackage[]{amstext}
\usepackage[]{amsmath}

\usepackage[]{amssymb}
\usepackage[]{graphicx}
\usepackage[]{threeparttable}
\usepackage[]{lscape}
\usepackage{xcolor}
\usepackage{ulem}
\usepackage{bbm}
\usepackage{ragged2e}

\newcommand*{\QEDB}{\hfill\ensuremath{\square}}%

\usepackage[page]{appendix}

\usepackage{caption}
\usepackage{subcaption}

\usepackage{titlesec}
\usepackage{enumerate}
\usepackage{hyperref}
\titleformat{\paragraph}
{\normalfont\normalsize\bfseries}{\theparagraph}{1em}{}
\titlespacing*{\paragraph}
{0pt}{3.25ex plus 1ex minus .2ex}{1.5ex plus .2ex}

\begin{document}

\author{}
\title{A single risk approach to the semiparametric copula competing risks model}
\vspace{1.3cm}
\author{Simon M.S. Lo\footnote{The United Arab Emirates University, United Arab Emirates, Department of Innovation in Government and Society, E--mail: losimonms@yahoo.com.hk} \\
	Ralf A. Wilke\footnote{Copenhagen Business School, Department of Economics, E--mail: rw.eco@cbs.dk}}

\maketitle
\thispagestyle{empty}

\linespread{1.5}{
\begin{abstract}
A typical situation in competing risks analysis is that the researcher is only interested in a subset of risks. This paper considers a depending competing risks model with the distribution of one risk being a parametric or semi-parametric model, while the model for the other risks being unknown. Identifiability is shown for popular classes of parametric models and the semiparametric proportional hazards model.  The identifiability of the parametric models does not require a covariate, while the semiparametric model requires at least one. Estimation approaches are suggested which are shown to be $\sqrt{n}$-consistent. Applicability and attractive finite sample performance are demonstrated with the help of simulations and data examples.
\end{abstract}}

\section{Introduction}
Duration or failure time models are routinely applied in a wide range of disciplines, including biostatistics, reliability engineering and social sciences. A common feature of these models is that not all failure times can be observed due to data restrictions or due to that there are multiple types of failures. The former corresponds to censoring, the latter to a competing risks scenario. A single risk model with censoring is therefore observationally equivalent to a competing risks model, although the latter typically gives a better understanding of the underlying data generating process. For the rest of this paper, we focus on a two risks model without loss of generality. The two latent failure times are $T$ and $C$ with survival distribution $S(t)$ and $R(c)$, respectively, and their joint distribution is $H(t,c)$. Observable are $X=\min\{T,C\}$, the minimum of the two failure times and the risk indicator $\delta = \1\{X=T\}$, but not $T$ and $C$. This incompleteness of information leads to non-identifiability and complicates the statistical modelling.

Before reviewing existing routes to identification and estimation in the literature, we briefly summarise the contributions that this paper makes. We develop a partial modelling approach that requires neither knowledge nor specification of $R(c)$. We show that $S$ and the dependence structure between $T$ and $C$ are identifiable under mild restrictions for a wide range of models for $S(t)$ including Accelerated Failure Time (AFT) model, parametric and semiparametric Proportional Hazards (PH) models. We suggest estimation procedures and show that they are $\sqrt{n}$-consistent. Extensive simulations confirm desired finite sample properties of our approaches and show that they generally outperform classical approaches that are commonly applied by empirical researchers, including maximum likelihood estimation (MLE) of parametric models with fully specified $H(t,c)$, the multivariate mixed proportional hazards model (MMPHM) for dependent risks, and models for independent risks such as the Cox proportional hazards model, the piecewise constant (PWC) hazards model, and parametric survival models (PM). Not specifying $R(c)$ avoids misspecification and improves efficiency in larger dataset as our model contains  much fewer unknown parameters. By leaving the degree of risk dependence unspecified, our approach avoids another major source of misspecification as it is well known that a misspecified risk dependence creates profound biases in the estimated $S(t)$ (Zheng and Klein, 1995; Rivest and Wells, 2001; Lo and Wilke, 2017). Our simulation results confirm this finding as the Cox, PWC, PM are shown to be sizably biased when the independent risks assumption is violated.

In the following we elaborate further how our proposed method is different from the existing approaches. A simple and commonly used restriction to resolve identifiability issues is the assumption of independent competing risks (Kalbfleisch and Prentice, 1980). In this case $S(t)$ is nonparametrically identifiable. Nonparametric estimators such as the Nelson-Aalen, Kaplan-Meier, and the Cox proportional hazards model are frequently used in empirical analysis as $S(t)$ can be estimated with independent censoring without knowing the distribution of $R(c)$. However, when risks are dependent, the competing risks model is not nonparametrically identifiable (Tsiatis, 1975) and the worst case identification bounds for $S(t)$ and $R(c)$ are wide (Peterson, 1976). Identifiability is only ensured under restrictions and three main routes to identifiability have been established.

The first approach exploits variation in latent failure times due to sufficient variation in covariates. The model is identifiable by assuming that $S(t)$ and $R(c)$ are PH, AFT, or linear transformation models  (Heckman and Honor\'{e}, 1989; Abbring and van den Berg, 2003; Lee 2006), although identifiability is weak as these approaches do not impose a parametric form of the joint distribution and partly rely on information at $t\to 0$ only. Moreover, rather restrictive conditions are required in this approach, which are difficult to verify in applications (Fermanian, 2003). Also, identifiability is only up to location and scale normalisations. Less ambitious than identifying the full model, covariate effects can be nonparametrically bounded or their sign can be identified under mild conditions (Honor\'{e} and Lleras-Muney, 2006; Lo and Wilke, 2017).

The second approach is to fully specify the joint distribution $H(t,c) = \mathcal{K}(S(t), R(c))$ by means of a known copula $\mathcal{K}$. In this case, $S(t)$ and $R(c)$ can be identified nonparametrically with the copula-graphic estimator (CGE) (Carri\`{e}re, 1995; Zheng and Klein, 1995; Rivest and Wells, 2001), which bases on the identifiable distribution of the observed $(X,\delta)$. In a regression setting with covariates $Z$, the conditional latent survival function $S(t|z)$ and $R(c|z)$ can be estimated by using different regression models for the distribution of $(X,\delta)$, including nonparametric (Braekers and Veraverbeke, 2005; Sujica and van Keilegom, 2018), semiparametric (Scheike and Zhang, 2008; Lo et al., 2017), and parametric (Lo and Wilke, 2014). Alternatively one can impose some restrictions on $S$ and $R$ such as semiparametric models (Huang and Zhang, 2008; Chen, 2010; Xu et al., 2018). Although this CGE approach allows flexibility in modeling $S(t|z)$ and $R(c|z)$ - nonparametric or semiparametric, its estimates can be sizably biased, because of an incorrect assumed degree of dependence between risks.

The third approach uses a parametric copula $\mathcal{K}_\theta$ with unknown dependence parameter $\theta$ and assumes that $S(t|\chi)$ and $R(c|\chi)$ are (semi)parametric with unknown parameters $\chi$. 
In particular, 
$\theta$ and $\chi$ are identifiable when $S(t|\chi)$ and $R(c|\chi)$ are parametric (Escarela and Carri\'{e}re, 2003; Hsu et al., 2016; Shih and Emura, 2018; Deresa and van Keilegom, 2020; Czado and van Keilegom, 2021) and semiparametric (Staplin, et al., 2015; Chen et al., 2017; Emura and Michimae, 2017). Estimation is by MLE using the known functional form of $H(t,c|\chi,\theta) = \mathcal{K}_{\theta}(S(t|\chi), R(c|\chi))$. This approach can be considered as a generalisation of another similar approach that specifies $H(t,c)$ as, for example, bivariate normal, bivariate log-normal, or bivariate Weibull (Basu and Ghosh, 1978; Emoto and Matthews, 1990; Fan and Hsu, 2012; Gupta and Gupta, 2012).

Besides these three major approaches, there is a small literature that considers models with restrictions on $S(t)$, while leaving the distribution of the other risk $R(c)$ unspecified. This approach is appealing as the distribution of other risks is often unknown in applications. Schwarz et al. (2013) show identifiability of the dependence parameter $\theta$ when $R(c)$ is unknown, although their model requires a rather restrictive independence assumption on observed duration $X$ and the risk indicator $\delta$. This assumption is violated when, for example, the two risks have different cumulative incidence functions. Without relying on this assumption, Braekers and Veraverbeke (2008) develop an estimator for unknown $R(c)$ requiring that $S(t)$, $\mathcal{K}$ as well as the dependence parameter $\theta$ are known. Under much milder restrictions, Wang (2021) shows identifiability of $S$ and $\theta$, when $T$ is exponentially distributed, while leaving $R(c)$ completely unspecified. Wang's approach bases on three observations. First, $S(t)$ can be computed from the CGE for any $\theta$. Second, the parameter of the exponential marginal, $\lambda = \log S(t) / t $, can be computed for any given $S(t)$ that is obtained by the CGE under an assumed $\theta$. Third, $\lambda$ is constant for all $t$, and a natural estimator for $\lambda$ is its sample average with different realisations of $T$. Then $S(t)$ can be computed from the exponential model using the estimated $\lambda$ for any value of $\theta$. $\theta$ is then identified by applying the Cram\'{e}r-von Mises (CvM) criterion that in $\theta$ minimises the distance between the $S(t)$ implied by the exponential model and the CGE. While Wang's approach can only identify a single parameter exponential distribution without covariates, we suggest a more general approach that is compatible with more flexible distributions with more than one parameter with or without covariates. These include the commonly used parametric regression survival models like the AFT and PH model using Weibull, log-logistic, and log-normal. We also show identifiability when $S(t)$ is a semiparametric PH model when there is at least one covariate. We therefore substantially generalise Wang's (2021) approach.

Emura et al. (2020) introduce another method that utilises a similar CvM criterion. It searches for the $\theta$ that minimises the distance between the nonparametric cumulative incidences based on the observed $(X,\delta)$ and the estimated cumulative incidences based on an assumed copula model with a semiparametric PH model for $S$ and $R$. Although this method shares similar ideas with our approach, it requires an assumed model for the marginal distribution for both risks. The CvM criterion is based on the simulated (and not closed form) cumulative incidences given the assumed models for $S$, $R$ and $\mathcal{K}$. Their simulations show that the objective function is flat in $\theta$, which corresponds to weak identification. In contrast, we formally show identifiability under weak restrictions, and our simulations and applications demonstrate non-flatness of the objective function.

We summarise the contributions of our paper as follows: (i) We show identifiability of $S$ and $\theta$ for a variety of commonly used parametric and semiparametric survival regression models, without specifying the distribution of the other risk. It therefore avoids a source of misspecification if $R$ is unknown. Other approaches that ignore $R$ enjoy great popularity among practitioners but require $\mathcal{K}$ to be the independent copula (e.g. Kaplan-Meier, Cox model). In the context of dependent competing risks this is an important practical improvement as the second risk $R(c)$ is often a dependent censoring or a pooled remainder risk (Lo and Wilke, 2010) with neither clear interpretation nor known functional form. Existing methods that do not require $R(c)$ in contrast require a fully known copula such as the CGE (Carri\`{e}re, 1995; Zheng and Klein, 1995; Rivest and Wells) or the method of Braekers and Veraverbeke (2008). Our model avoids this source of misspecification as $\theta$ can be estimated. (ii) Our identifiability results hold with minimum requirement for covariates and also hold for all $t$. This is in contrast to the general nonparametric competing risks model (Heckman and Honor\'{e}, 1989) and the MMPHM (Abbring and van den Berg, 2003), which require sufficient variation in covariates and identifiability relies on the limit point $t\to 0$ (Fermanian, 2003). Our method is therefore more practical and more stable as all durations contribute to the identifiability with (almost) no restrictions on the covariate structure. (iii) Our approach does not model the full joint distribution. By having a partial spirit it contains fewer parameters to be estimated, which leads to precise estimates with larger samples. (iv) The suggested estimation approaches are $\sqrt{n}$ consistent. (v) We show with extensive simulations that our approaches generally outperform some classical approaches that are commonly applied, including full MLE, MMPHM, COX, PWC and PM models.

The following sections introduce our model (Section \ref{s:mod}), present the identification results (Section \ref{s:ident}), suggest the estimation methods and establish their large sample properties (Section \ref{s:est}). Section \ref{s:app} illustrates the practicability using labour market duration data from economics. There is extensive supplementary material where we provide proofs, study finite sample properties with the help of simulations and show the results of an additional application. Sample code for the parametric models that are used in the empirical analysis in this paper can be downloaded from: \url{https://github.com/ralfawilke/singlerisk}.




\section{The model \label{s:mod}}
This section introduces the different model components in three subsections: competing risks, copula function, and marginal distribution.

\subsection{Competing risks}
We consider a competing risks duration model with possibly many risks. The researcher is only interested in the distribution of one risk (risk 1) and let $T \in \mathbb{R} _{0+}$ be the latent duration of this risk, where $\mathbb{R} _{0+}$ refers to the set of non-negative real number. The other risks are not of interest and they are for simplicity pooled into a second risk and denoted as censoring. Let $C \in \mathbb{R} _{0+}$ be the censoring time. The observed failure time is $X=\min(T,C) \in \mathbb{R} _{0+}$ and $\delta = \mathbbm{1}_{T<C}$ is the risk indicator function, which is equal to one when $T<C$ and zero otherwise. Let $Z\in \Zscr\subset \mathbb{R} ^k$ be a $k-$vector of covariates. 
The data are $(x_i, \delta_i, z_i)$ for randomly sampled units $i = 1, \ldots, n$. Let the joint survival distribution of $T$ and $C$ be $H(t,c|z)= \Pr(T > t, C > c;z)$
. The overall survival function is $\pi(x|z)=\Pr(X > x;z)=H(x,x|z)$. The sub-density function for $T$ is $f_t(x|z) = \lim_{\epsilon\to 0} \Pr( x \leq T < x+\epsilon \wedge \delta = 1;z)/\epsilon$. The latent marginal survival function for $T$ is $S(t|z) = \Pr(T> t; z)
$, and the latent marginal survival function for $C$ is $R(c|z) = \Pr(C> c; z)
$. Let $\nabla_{s}$ be the gradient of a function with respect to $s$, and $f^{(k)}(s)$ is the $k$-th derivative of function $f$ with respect to a scalar $s$.  

\subsection{Copula model}
In the following we discuss various conditions for the copula model.

\assumption{
\begin{enumerate}[(i)]
\item A copula generator denoted by $\phi_{\theta}(u) = \phi(u;\theta): [0,\infty) \times \Theta \to [0,1] $ is a continuous, decreasing and convex function in $u$, where $\Theta$ is a compact subset of $\mathbb{R}$; for all $\theta\in  \Theta$, $\phi_{\theta}(0)=1$, $\lim_{u\to \infty} \phi_{\theta}(u) =0$, and, by convention, $\phi_{\theta}(\infty) =0$;  for all $\theta\in  \Theta$, $u \mapsto \phi_{\theta}(u)$ is strictly decreasing on $[0, \inf\{u: \phi_{\theta}(u)=0\})$, with $(-1)^{(k)}\phi_{\theta}^{(k)}(u) \geq 0$  for $k=1,2$ and  $\phi_{\theta}^{(1)}(u) > -\infty$ for all $u\in[0, \inf\{u: \phi_{\theta}(u)=0\})$; and the quasi inverse of $\phi_{\theta}(u)$  is defined as $\phi^{-1}_{\theta}(s) = \phi^{-1}(s;\theta) : (0,1] \times \Theta \to [0,\infty)$, where, by convention,  $\phi_{\theta}^{-1}(0) = \inf \{u: \phi_{\theta}(u)= 0\}$; $\nabla_{\theta} \phi_{\theta}(u)$, $\nabla_{\theta} \phi^{-1}_{\theta}(s)$, and $\nabla_{\theta} ( \phi^{-1}_{\theta})^{(1)}(s)$ are finite for all $u\in (0,\infty)$,  $s\in (0,1)$ and  $\theta\in \Theta$;\label{a1:1}
\item $S(t|z)$ and $R(c|z)$ have joint distribution
\begin{eqnarray}
\quad H(t,c|z) &= \mathcal{K}_{\theta}[S(t|z), R(c|z)] =& \phi_{\theta}(\phi^{-1}_{\theta}[S(t|z)] + \phi^{-1}_{\theta}[R(c|z)]),\label{H}
\end{eqnarray}
where $\mathcal{K}_{\theta}(s,r) = \Pr(S \leq s, R\leq r;\theta): [0,1]^2 \times \Theta \to [0,1]$ is an Archimedean copula; by solving (\ref{H}) for $S(t|z)$,  $S(t|z)$ has a closed form solution in terms of $\pi$, $f_t$, and $\phi_{\theta}$ for any $\theta$ (Rivest and Wells, 2001):
\begin{eqnarray}
S_{\theta}(t|z) := S(t|z) = \phi_{\theta}\left[-\int_0^t (\phi^{-1}_{\theta})^{(1)}[\pi(u|z)]f_t(u|z)du\right] \label{cge}
\end{eqnarray}
\nin for all $t\in (0,\infty)$,  $z\in \Zscr$ and $\theta\in \Theta$; 
 \label{a1:3}
\item $\pi(t|z) < S(t|z)$ for all $t\in (0,\infty)$ and for all $z\in \Zscr$;\label{a1:6}
\item $(\phi^{-1}_{\theta_1})^{(1)}(s)/(\phi^{-1}_{\theta_2})^{(1)}(s)$ is strictly increasing in $s$ for any $\theta_1 <\theta_2 \in \Theta$; \label{ass1:increase}
\item $\nabla_{\theta} S_{\theta}(t|z)$ is finite for all $t\in (0,\infty)$, for all $z\in \Zscr$, and for all $\theta \in \Theta$. \label{ass1:nabla}
 \end{enumerate} \label{ass1}}

Assumption \ref{ass1}(\ref{a1:1}) is the standard definition of a 2-dimensional Archimedean copula generator (McNeil and Neslehov\'{a}, 2009), except that $\phi_{\theta}^{(1)}(u) $ is not only non-positive but also finite, i.e. $ -\infty <\phi_{\theta}^{(1)}(u) \leq 0$. This assumption implies through Lemma \ref{l:inverse} (below) that $\phi^{-1}_{\theta}(s)$ has non-zero slope at $s=1$. 
The copula generators that satisfy this condition are called non-strict Archimedean copulas (Nelson, 2006, p.122). 

\lemma{Assumption \ref{ass1}(\ref{a1:1}) implies that $\phi_{\theta}^{-1}(s)$ is a continuous and strictly decreasing function on $s\in (0,1]$ such that $(\phi_{\theta}^{-1})^{(1)}(s) < 0$  and $(\phi_{\theta}^{-1})^{(2)}(s) \geq 0$ for all $s\in (0,1]$ and for all $\theta\in  \Theta$. \label{l:inverse}}


$\theta$ is a dependence parameter that measures the rank dependence between $S$ and $R$. 
$\theta$ can generally be transformed into Kendall's $\tau$, a rank correlation coefficient, by a function of $\tau(\theta) : \Theta \to [-1,1]$. For the Clayton copula, $\tau(\theta)= \theta/(\theta+2)$ and the corresponding $\Theta = [-1,\infty)$. In this case, $\theta =-1$ corresponds to perfect negative dependence (Kendall's $\tau =-1$), and $\theta \to \infty$ approaches the case of perfect positive dependence (Kendall's $\tau = +1$). $\theta =0$ corresponds to independence (Kendall's $\tau =0$). 

Assumption \ref{ass1}(\ref{a1:3}) is a result from Sklar's theorem (Schweizer and Sklar, 1983). 
Based on (\ref{H}), $S(t|z)$ in (\ref{cge}) is a function depending on $\theta$ and we make it clear by defining $S_{\theta}(t|z)$ as an equivalent symbol for $S(t|z)$.
Assumption \ref{ass1}(\ref{a1:6})  can be ensured by transforming the duration variable $X$ by $Y= X-x^* +\epsilon$, where $x^* = \inf\{x: \pi(x|z) <S(x|z)\}  \in (0, \infty)$ and $\epsilon$ is an arbitrary number closed to zero. This transformation ignores the duration variable $X$ in the interval of $[0,x^*)$ in which $\pi(x|z)= S(x|z)$. This ignored interval is not interesting for the purpose of identifying $S(x|z)$ as it is known to be equal to $\pi(x|z)$. For  Assumption \ref{ass1}(\ref{ass1:increase}), 
Genest and MacKay (1986) and Rivest and Wells (2001) discuss the case when $\phi^{-1(1)}_{\theta_1}(s)/\phi^{-1(1)}_{\theta_2}(s)$ is increasing in $s$. We impose a stronger condition by assuming that it is strictly increasing. Together with Assumption \ref{ass1}(\ref{a1:6}) it leads to Lemma \ref{l:thetaSPHI}.


\lemma{Due to Assumption \ref{ass1}(\ref{a1:6}) and (\ref{ass1:increase}), 
$\theta \mapsto S_{\theta}(t|z)$ is strictly decreasing for all $t \in (0,\infty)$ and for any $z\in \Zscr$.  \label{l:thetaSPHI}}


Assumption \ref{ass1}(\ref{a1:1}) - (\ref{ass1:increase}) are required for identification. Assumption \ref{ass1}(\ref{ass1:nabla}) is required for the derivation of large sample properties in Section \ref{s:est}. 
The restrictions in Assumption \ref{ass1} can be checked for each copula and they  hold for Clayton copula in particular.

\lemma{\textbf{(Clayton Copula)}: Assumption \ref{ass1} holds for $\phi_{\theta}(u)= (1+u\theta)_{+}^{-1/\theta}$, where $(s)_{+} = \max\{s,0\}$.  \label{l:clayton}}

\nin The proofs of all Lemmas are given in Supplementary Material S.I.

\subsection{Marginal survival model}




We introduce two separate models for $S(t|z)$. $S(t|z)$ in (\ref{cge}) is a model of the copula under the restrictions of Assumption \ref{ass1}. Alternatively, it can be written as a marginal survival model defined by
\begin{eqnarray}
S^*(t|z)  :=  S(t|z) = \exp(-\Lambda(t|z)) \label{aft00}
\end{eqnarray}
\nin with assumed parametric or semiparametric model for $\Lambda(t|z): \mathbb{R} _{0+} \times \Zscr  \to \mathbb{R} _{0+}$. $\Lambda(t|z)$ increases in $t$ for all $z$, $\Lambda(t|z)\in (0, \infty)$ and $\lambda(t|z) = \Lambda^{(1)}(t|z) \in (0, \infty)$ for all $t \in (0,\infty)$ and for all $z$.
While models (\ref{cge}) and (\ref{aft00}) base on separate sets of restrictions without direct relationship, $S_{\theta}(t|z) = S^*(t|z) = S(t|z)$ must hold for all $t$ and $z$.

\section{Identifiability \label{s:ident}}
The non-identifiability of the competing risks model has been well explored by Cox (1962), Tsiatis (1975), and Wang (2014). According to Tsiatis (1975), a dataset of $(X,\delta)$ that is generated by a dependent competing risks model is observationally equivalent to an independent risks model. Wang (2014) proves further that two non-independent competing risks models with different $\theta$ and different marginal survival functions can generate observationally equivalent distributions of $(X,\delta)$. Therefore, without imposing further restrictions on $S$ and $R$, there exists more than one value of $\theta$ in (\ref{cge}) that is compatible with $H(t,c|z)$ in (\ref{H}). It is called the non-identifiability of $\theta$. 
Wang (2021) shows that $\theta$ is unique and identifiable if $S^*(t|z)$ in (\ref{aft00}) has an exponential distribution 
while  the distribution of $C$ is left unspecified. In this paper we generalise Wang's (2021) result by showing that $\theta$ is unique and identifiable for other parametric and semiparametric proportional hazard (PH) model of $S^*(t|z)$ while  the distribution of $C$ is left unspecified. The identifiability is established by showing that $S_{\theta}(t|z)$ in (\ref{cge}) and $S^*(t|z)$ in (\ref{aft00}) are identical for a unique $\theta$ and unique parameters of $S^*(t|z)$. This means the two models cannot match for all $t$ and $z$ for any other parameter values.




\subsection{Parametric model \label{ss:aft}}
We discuss a set of restrictions on the parametric model for $S^*(t|z)$ that are required for identifiability. 



\assumption{ (\textbf{Parametric model})  $\Lambda(t|z)$ in (\ref{aft00}) has a known parametric form with  unknown parameters $\chi \in \mathbb{R} ^p$, so that $S^*(t|z)$ in (\ref{aft00})  is denoted as
\begin{eqnarray}
 S^*(t|z;\chi) = \exp(-\Lambda(t|z; \chi)); \label{parametric}
\end{eqnarray}
\nin 
(i) $\Lambda(t|z;\chi_1) =\Lambda(t|z;\chi_2)$ for all $t \in (0,\infty)$ and for all $z\in \Zscr$ if and only if $\chi_1 = \chi_2$; (ii) $\lim_{t \to 0+} \lambda(t|z;\chi_1) / \lambda(t|z;\chi_{2})\neq 1$ for any $\chi_1 \neq \chi_2$ and for any $z\in \Zscr$. \label{assS}}

Parametric models like Exponential, Weibull,  Log-logistic and Log-normal are compatible with Assumption \ref{assS}(i) and (ii). Take the exponential model without covariates as an illustrative example: $\Lambda(t;\chi) =\chi t$, where $\chi > 0$ is a scalar. Assumption \ref{assS}(i) and (ii) are met as $\Lambda(t;\chi_1) = \Lambda(t;\chi_2)$ for all $t$ if and only if $\chi_{1} = \chi_{2}$. $\lim_{t \to 0^+} \lambda(t;\chi_{1}) / \lambda(t;\chi_{2})= \chi_{1}/ \chi_{2}$, which is a constant not equal to one for any $\chi_{1}\neq \chi_{2}$. 
A counter-example is  Gompertz model, in which $\lambda(t;\chi) = (a/b) \exp(b t)$ with $\chi = (a, b)'$ and $a>0 , b \geq 0$. Here, $\lim_{t \to 0^+} \lambda(t;\chi_{1}) / \lambda(t;\chi_{2})= (a_{1}/b_{1})/(a_{2}/b_{2})$. The latter may be equal to one even if $a_{1} \neq a_{2}$ and $b_{1} \neq b_{2}$. 

\proposition{$\theta$ in (\ref{cge}) and $\chi$ in  (\ref{parametric}) are unique under Assumptions \ref{ass1} and \ref{assS}.\label{prop1}}

The proof is given in Supplementary Material S.I. It does not require covariates. It therefore relaxes an important restriction of some related approaches, where the identifiability relies on exogenous variations in covariates. Examples are Heckman and Honor\'{e} (1989), Abbring and van den Berg (2003), Fermanian (2003), Honor\'{e} and Lleras-Muney (2006), and Lo and Wilke (2017).

Since there is no restriction on the role of $z$ in Assumption \ref{assS}, a PH regression model or an accelerated failure time (AFT) model can be chosen for $S^*(t|z;\chi)$ in (\ref{parametric}). 
We illustrate it by using the exponential regression model as an example, as it belongs to the PH and AFT model at the same time. In this case, $\Lambda(t|z;\chi) = a t\exp(z'\beta)$ and $\chi = (a, \beta)'$.
Assumption \ref{assS}(i) is satisfied as $\Lambda(t|z;\chi_{1})$ and $\Lambda(t|z;\chi_{2})$ are identical for all $t$ and $z$ if and only if $a_1=a_2$ and $\beta_1=\beta_2$. Assumption \ref{assS}(ii) is satisfied as $\lambda(t|z;\chi_{1}) / \lambda(t|z;\chi_{2}) = a_1/a_2 \exp(z'(\beta_1-\beta_2))$, which is equal to one for all $z$ if and only if $a_1=a_2$ and $\beta_1=\beta_2$. Besides the PH and AFT regression model, Assumption \ref{assS} is also valid for the general linear transformation model or the general location-scale model discussed in Lee (2006), Hsu et al. (2016), and Suijica and van Keilegom (2018). For instance, $\log(T) = - z'\beta + (1/a)\epsilon$, where $\epsilon = \log(-\log S(t;z))$. It can be rewritten as $\Lambda(t|z;\chi) = -\log S(t;z) = t^{a}[\exp(z'\beta)]^{a}$, where $\chi = (a, \beta)'$.  $\Lambda(t|z;\chi_{1})$ and $\Lambda(t|z;\chi_{2})$ are identical for all $t$ and $z$ if and only if $a_1=a_2$ and $\beta_1=\beta_2$. Assumption \ref{assS}(i) is met. Assuming without loss of generality that $a_{1} > a_{2}$, $\lambda(t|z;\chi_{1}) / \lambda(t|z;\chi_{2})= a_{1}/ a_{2} t^{a_{1}-a_{2}}\exp((z'\beta_{1})a_{1}-(z'\beta_{2})a_{2}))$, which is zero but not equal to one at $t =0$. Assumption \ref{assS}(ii) is met. 


\subsection{Semiparametric PH model \label{ss:ph}}
We consider the semiparametric PH model in this subsection, where the functional form of $\Lambda(t;z)$ is partly unknown. 

\assumption{ (\textbf{Semiparametric PH model}) (i) $\Lambda(t|z)$ in (\ref{aft00}) is a PH model, i.e.
\begin{eqnarray}
 S^*(t|z;\beta)  = \exp(-\Lambda_{0}(t)\exp(z'\beta))  \label{ph00}
\end{eqnarray}
\nin with $\beta$ is a vector of $k$ unknown parameters, $\Lambda_0(t): \mathbb{R}_{0+}  \to \mathbb{R} _{0+}$ is an unknown increasing function in $t$, such that
$\Lambda_{0}(t) \in (0, \infty)$ for all $t \in (0,\infty)$,  $\lambda_{0}(t) = \Lambda^{(1)}_{0}(t) \in (0, \infty)$ for all $t \in (0,\infty)$; 
(ii) $z$ is a vector of covariates with $z\in \Zscr \subset\mathbb{R} ^k$ and $k\geq 1$.
\label{assPH}}

\proposition{$\theta$ in (\ref{cge}), $\Lambda_0(t)$ and $\beta$ in (\ref{ph00}) are unique under Assumptions \ref{ass1} and \ref{assPH}. \label{prop2} }

The proof is given in Supplementary Material S.I. The existing literature has already shown identifiability of the semiparametric PH model (Heckman and Honor\'{e} , 1989;  Abbring and van den Berg , 2003; Fermanian, 2003), although our result is obtained under weaker restrictions. Firstly, identification of $\theta$ in (\ref{p2:eq3}) does not only rely on the information at the limit $t\to 0$  such that our estimator has the usual rate of convergence. Secondly,
$z$ is not required to be continuous to trace out the joint distribution of $T$ and $C$. In fact, a binary $z$ suffices for identification or inference purpose. Lastly, a normalising assumption on the baseline cumulative hazard function, such as $\Lambda_0(t) =1$ for some $t$ is not required. It means that $\Lambda_0(t)$ in our model is not just identified up to a scale. 

\section{Estimation \label{s:est}}
This section introduces estimation procedures for the parametric and semiparametric models of Section \ref{s:ident}. Suppose there is a random sample of $(x_i,\delta_i,z_i)$ with $i=1,\ldots,n$ observations. We denote the true value of an unknown parameter $\theta$ as $\theta_0$. For consistency of an estimator, we follow the definition as in Lemma 2.4 of Newey and McFadden (1994) and denote, say, $\hat{\theta} \overset{p}{\to} \theta_0$ as uniform convergence in probability.

The estimation is done stepwise. In the first step, $\pi(x|z)$ and $f_t(x|z)$ are estimated from the data $(x_i,\delta_i,z_i)$ by means of an existing parametric (e.g. Kalbfleisch and Prentice, 2002), semiparametric (e.g. Colvert and Boardman, 1976; Lancaster, 1990; Fine and Gray, 1999) or nonparametric (Anderson et al. 1993) model. While our approach is generally compatible with any of these estimators, we focus here the parametric case. In this case, the estimation procedure corresponds to a multiple step GMM estimator for which it is convenient to state the asymptotic properties. Semiparametric or nonparametric estimators for $\pi(x|z)$ and $f_t(x|z)$ (as in our application in Section \ref{s:app}) can be used in practice and inference can be based on the bootstrap. Suppose $\pi(x|z)$ and $f_t(x|z)$ are functions with unknown parameters $\eta$, we let  $\nu(x|z;\eta) = (\pi(x|z;\eta), f_t(x|z;\eta))'$.

\assumption{(\textbf{Estimators for the overall survival function and sub-density function}) (i)  $\pi(x|z;\eta)$ and $f_t(x|z;\eta)$ are continuous in $\eta$ for all $x$ and $z$; (ii) $E(||f_t||) <\infty$; (iii) $\hat{\eta} \overset{p}{\to} \eta_0$ and $\sqrt{N}(\hat{\eta}-\eta_0) \overset{d}{\to} N(0,\Omega_{\eta})$; (iv) $\hat{\nu}(x|z) \overset{p}{\to} \nu_0(x|z)$ and $\sqrt{N}(\hat{\nu}(x|z)-\nu_{0}(x|z)) \overset{d}{\to} N(0,\Omega_{1}(x|z;\eta))$ for all $x$.
 \label{ass2}}

\nin A maximum likelihood estimator (MLE) for $\eta$ fulfills these restrictions. See, for example, Theorem VII.2.1 in Anderson et al. (1993) for consistency and Theorem VII.2.2 in Anderson et al. (1993) for asymptotic normality. See Theorem VII.2.3 in Anderson et al. (1993) for an estimate of $\Omega_1$. 

$\hat{\pi}(x|z)$ and $\hat{f}_t(x|z)$ are then plugged into the Copula Graphic Estimator (CGE) using any $\theta\in \Theta$, which is the sample counterpart of equation $(\ref{cge})$:
\assumption{(\textbf{CGE})
(i)
\begin{eqnarray}
\hat{S}_{CGE}(x|z;\theta) = \phi_{\theta}\left[\int_0^{x} \phi_{\theta}^{-1(1)}[\hat{\pi}(u|z)]\hat{f}_t(u|z)du\right] \overset{p}{\to} S_{\theta}(x|z) \label{scge}
\end{eqnarray}
\nin for all $\theta\in\Theta$ and all $x$ and $z$; 
(ii) $\sqrt{n}[\hat{S}_{CGE}(x|z;\theta)- S_{\theta}(x|z)]\overset{d}{\to} N(0,\Omega_{S}(x|z))$ for all $x$ and $z$ and for all $\theta\in \Theta$
. \label{ass3}}

\nin For more details of these assumptions, see Rivest and Wells (2001). To sum up, $\hat{S}_{CGE}(x|z;\theta)$ is a consistent estimator for $S_{\theta}(x|z)$ for all values of $\theta \in \Theta$, including but not limited to $\theta_0$. The estimated CGE will be used in the subsequent steps, where we distinguish between the parametric models of Subsection \ref{ss:aft} and the semiparametric PH model of Subsection \ref{ss:ph}.

\subsection{Accelerated Failure Time model \label{eAFT}}
Take the AFT model which can be written as
\begin{eqnarray}
\log([\alpha x_i \exp(z_i'\beta)]^{\sigma})& =&  w_i  \label{aft111}
\end{eqnarray}
\nin with unknown parameters $\alpha, \sigma \in \mathbb{R} _+$, $\beta\in \mathbb{R} ^k$. $w_i$ is a nuisance term with a known survival distribution $S_{W}(w_i)$. The latent marginal survival function  in $(\ref{aft00})$ is related with $S_W(w_i)$ by $S^*(x_i|z_i) = \Pr(X>x_i;z_i) = \Pr(W>w_i)=S_W(w_i)$. Hence, we have
\begin{eqnarray}
w_i &=& S_W^{-1}[S^*(x_i|z_i)] \label{aft112}.
\end{eqnarray}

\nin By writing  (\ref{aft111}) as a linear function and substitute  (\ref{aft112}) into it, we obtain
\begin{eqnarray}
\log(x_i) &=& -\log(\alpha) - z_i'\beta + (1/\sigma) S_W^{-1}[S^*(x_i|z_i)].  \label{aft1}
\end{eqnarray}
\nin Due to Assumptions \ref{ass1} and \ref{assS}, $S^*(x_i|z_i)=S_{\theta}(x_i|z_i)$ for all $x_i$ and $z_i$ when the latter is evaluated at $\theta = \theta_0$. We can therefore replace  $S^*(x_i|z_i)$ in (\ref{aft1}) by its estimate $\hat{S}_{CGE}(x|z;\theta_0)$ to obtain the following regression model
\begin{eqnarray}
\log(x_i) &=& -\log(\alpha) - z_i'\beta + (1/\sigma)S_W^{-1}[\hat{S}_{CGE}(x_i|z_i;\theta_0)] + \epsilon_i \non \\
          &=& \hat{w_i}'\chi+\epsilon_i \label{logt2}
\end{eqnarray}
with $\hat{w}_i = (-1,-z_i',S_W^{-1}[\hat{S}_{CGE}(x_i|z_i;\theta_0)])'$, and  $\chi = (\log \alpha ,\beta' ,1/\sigma )'$. 
$\epsilon_i$ is an unknown error term. Due to Assumption \ref{ass3}, i.e. $\hat{S}_{CGE}(x_i|z_i;\theta_0)\overset{p}{\to}S_{\theta_0}(x_i|z_i)$, it follows that
$\epsilon_i$ diminishes as $n\rightarrow \infty$. 
Let $\log(x)$, $\hat{w}$ and $\epsilon$ be the vectors or matrices with column dimension $n$ that stack all observations of $\log(x_i)$, $\hat{w_i}$ and $\epsilon_i$. 
The feasible generalised least squares estimator for $\chi$ is
\begin{eqnarray}
\hat{\chi} = (\hat{w}\hat{\Omega}_{\epsilon}^{-1}\hat{w}')^{-1}(\hat{w}\hat{\Omega}_{\epsilon}^{-1}\log(x)'). \label{chithe}
\end{eqnarray}
\nin The following assumption ensures that $\hat{\chi}$ is consistent and efficient.

\assumption{(i) $E(\epsilon|\hat{w})=0$; (ii) $E(\epsilon'\epsilon)=\Omega_{\epsilon}$, which is $n\times n$ positive definite; (iii) $\hat{\Omega}_{\epsilon}\overset{p}\to\Omega_{\epsilon}$; (iv) $E(\hat{w}\hat{\Omega}_{\epsilon}^{-1}\hat{w}')$ is a non-singular matrix. \label{ass4}}


\nin Given $\hat{\chi}$, the estimated marginal survival for the AFT model is
\begin{eqnarray}
S_{AFT}(x_i|z_i;\hat{\chi}) &=& S_W(\log([\hat{\alpha} x_i \exp(z_i'\hat{\beta})]^{\hat{\sigma}})).\label{aft01}
\end{eqnarray}
\nin Due to Assumptions \ref{ass1}, \ref{assS}, \ref{ass3} and \ref{ass4}, we have
\begin{eqnarray}
S_{AFT}(x_i|z_i;\hat{\chi}) - \hat{S}_{CGE}(x_i|z_i;\theta_0) \overset{p}{\to} 0. \label{aftcge}
\end{eqnarray}
\nin This reasoning fails if any $\theta\neq \theta_0$ is used for $\hat{S}_{CGE}$ in (\ref{aftcge}). To emphasise that $\hat{\chi}$ depends on $\theta$, we denote it as $\hat{\chi}(\theta)$. Given (\ref{aftcge}), $\theta_0$ is estimated by minimising the CvM criterion as in Emura et al. (2020)
\begin{eqnarray}
\hat{\theta}_0 = \argmin_{\theta\in \Theta} \frac{1}{n}\sum_{i=1}^n\{S_{AFT}(x_i|z_i;\hat{\chi}(\theta))  -  \hat{S}_{CGE}(x_i|z_i;\theta)\}^2, \label{estaft}
\end{eqnarray}
which is a minimum distance estimator for $\theta_0$. 

We illustrate the above arguments by simulating a model with $\theta_0=8$ (Kendall's $\tau_0 = 0.8$). Figure \ref{fig:ident}(a) in Supplementary Material S.III shows that the estimated $S_{AFT}$ (black line) and $\hat{S}_{CGE}$ (black circles) are identical when $\theta=8$ is used in computing the CGE. Panel (b) shows these estimates  when $\theta=0$  (Kendall's $\tau = 0$) is used in computing the CGE and it is apparent that the two curves are different.  Hence, (\ref{estaft}) is minimized at $\theta =\theta_0$.

After $\hat{\theta}_0$ is obtained from (\ref{estaft}), $\hat{S}_{CGE}(x_i|z_i;\hat{\theta}_0)$ is obtained from (\ref{scge}). Putting it into $\hat{w}$, $\hat{\chi}_0 = (\hat{\alpha}_0,\hat{\beta}_0',\hat{\sigma}_0)'$ is obtained by (\ref{chithe}).

We consider in detail three popular AFT models in Supplementary Material S.II: Weibull, Log-logistic and Log-normal models. By writing the corresponding functional forms of $\Lambda(t|z;\chi)$, we discuss how the restrictions of Assumption \ref{assS} are fulfilled to ensure their identifiability as stated in Proposition \ref{prop1}.


\subsection{ Other parametric models \label{ss:pph}}
We briefly outline in the following how the parametric PH model and the general linear transformation model can be estimated. The parametric PH model is
\begin{eqnarray}
S^*(t_i|z_i) &=& \exp(-\Lambda(t_i;\chi)\exp(z_i'\beta)), \non
\end{eqnarray}
\nin where $\Lambda$ is a known function with unknown parameters $\chi$. After rearranging, we have
\begin{eqnarray}
\log \Lambda(t_i;\chi)  &=&   z_i'\beta - \log ( - \log S^*(t_i|z_i)) , \non
\end{eqnarray}
\nin which is equivalent to the general linear transformation model. By replacing $S^*(t_i|z_i)$ by $\hat{S}_{CGE}(t_i|z_i;\theta_0)$, $\chi$ and $\beta$ can be consistently estimated by a linear regression. The equivalence of (\ref{aft01}) for the PH model is
\begin{eqnarray*}
S_{PH}(x_i|z_i;\hat{\chi}(\theta)) &=& \exp(-\Lambda(x_i;\hat{\chi}(\theta))\exp(z_i'\hat{\beta})).
\end{eqnarray*}
The equivalent Cramer-von Mises criterion of (\ref{estaft}) is used to estimate $\theta$:
\begin{eqnarray}
\hat{\theta} = \argmin_{\theta\in \Theta} \frac{1}{n}\sum_{i=1}^n\{S_{PH}(x_i|z_i;\hat{\chi}(\theta))  -  \hat{S}_{CGE}(x_i|z_i;\theta)\}^2. \label{estpara}
\end{eqnarray}

\subsection{Semiparametric model \label{esemi}}
We consider the semiparametric PH model in (\ref{ph00})
\begin{eqnarray}
S^*(x_i|z) &=& \exp(-\Lambda_{0}(x_i)\exp(z_i'\beta)).  \label{ph00a}
\end{eqnarray}
\nin To ease readability, we start with the case $k=1$. The case of $k>1$ is considered later.
By evaluating (\ref{ph00a}) for  two arbitrarily different values of $z$, say $z_1 \neq z_2$, we define
\begin{eqnarray}
b_i:=\log\left[\frac{\log[S^*(x_i|z_{2})]}{\log[S^*(x_i|z_{1})]}\right]/(z_{2}-z_1).   \label{beta} 
\end{eqnarray}
\nin Let $B$ be a random variable with realisations $b_1, ..., b_n$ and variance $\sigma_b^2$. By definition (\ref{ph00a}), $b_i = \beta$ for all $i$ and $\sigma^2_b=0$. For other models than (\ref{ph00a}) this is not true. This observation forms the basis for estimation.

$S^*(x_i|z_i)=S_{\theta_0}(x_i|z_i)$ for all $x_i$ and $z_i$ due to Assumption \ref{ass1} and \ref{assS}. For estimation replace $S^*(x_i|z_i)$ in (\ref{beta}) by  $\hat{S}_{CGE}(x|z;\theta_0)$. $b_i$ is estimated by the sample analogue of (\ref{beta})
\begin{eqnarray}
\hat{b}_i  = \log\left[\frac{\log[\hat{S}_{CGE}(x_i|z_{2};\theta_0)]}{\log[\hat{S}_{CGE}(x_i|z_{1};\theta_0)]}\right]/(z_{2}-z_1) . \label{ebeta}
\end{eqnarray}
\nin By Assumption \ref{ass3}, $\hat{S}_{CGE}(x_i|z;\theta_0) \overset{p}{\to} S_{\theta_0}(x_i|z)$, we therefore have $\hat{b}_i \overset{p}{\to} b_i=\beta$ for all $i$ and the sample variance of $\hat{b}_i$ is zero, i.e.
\begin{eqnarray}
\hat{\sigma}^2(\hat{b}_i)  = \frac{1}{n-1}\sum_{i=1}^{n} (\hat{b}_{i} - \hat{b}^*)^2& \overset{p}{\to}& 0 ,\label{sigmab}
\end{eqnarray}
\nin where $\hat{b}^* = n^{-1}\sum_{i=1}^{n}\hat{b}_{i}$. Note that (\ref{sigmab}) does not hold whenever $\hat{S}_{CGE}(x_i|z;\theta)$ in (\ref{ebeta}) is evaluated at any $\theta \neq \theta_0$. To emphasise that $\hat{b}_{i}$ in (\ref{ebeta}) depends on $\theta$, we write $\hat{b}_{i}(\theta)$. Because of (\ref{sigmab}) for $\theta=\theta_0$, we obtain $\hat{\theta}$ by
\begin{eqnarray}
\hat{\theta} = \argmin_{\theta\in \Theta}  \frac{1}{n-1}\sum_{i=1}^{n} [\hat{b}_{i}(\theta) - \hat{b}^*(\theta)]^2, \label{thebeta1}
\end{eqnarray}
\nin where $\hat{b}^*(\theta) = n^{-1}\sum_{i=1}^{n}\hat{b}_{i}(\theta)$.
When there are $k$ covariates, $\hat{b}_{i}(\theta)$ becomes a $k$-vector of coefficients. Using the above argument as for $k=1$, the sample covariance matrix for $\hat{b}(\theta_0)$ converges in probability to a matrix of zeros, while it does not for $\theta\neq \theta_0$. $\theta$ is then estimated by
\begin{eqnarray}
\hat{\theta} = \argmin_{\theta\in \Theta}  \frac{1}{n-1}\sum_{i=1}^{n} I_k'[\hat{b}_{i}(\theta) - \hat{b}^*(\theta)][\hat{b}_{i}(\theta) - \hat{b}^*(\theta)]'I_k, \label{thebeta}
\end{eqnarray}
\nin where $I_k$ is the $k\times 1$ unit vector. Two practical remarks in relation to the implementation of the estimation procedure are given in Supplementary Material S.IV. These are useful for enhancing the numerical stability in an application.

\subsection{Large sample properties \label{s:asymp}}
We establish $\sqrt{n}$-consistency and asymptotic normality of our estimators by showing that the multiple-step estimation procedures correspond to GMM estimators. For this purpose we define the estimator in different steps as a series of estimation equations. We closely follow Newey and McFadden's (1994) framework for stepwise estimation to establish the properties.

We let $l_i(x_i, \delta_i, z_i;\eta)$ be the log-likelihood function that estimates $\eta$ in the first step. $\hat{\eta}$ solves the following estimation equation with probability approaching one:
\begin{eqnarray}
m_{1n}(\eta) = \frac{1}{n}\sum_{i=1}^n m_1(x_i, \delta_i, z_i; \eta)= \frac{1}{n}\sum_{i=1}^n \nabla_{\eta} l_i(x_i, \delta_i, z_i;\eta) =  0. \label{m1}
\end{eqnarray}
\nin $\hat{\eta}$ is consistent and asymptotically normal because of Assumption \ref{ass2}.


\subsubsection*{Parametric models}
This is for the models given in Subsections \ref{eAFT} and \ref{ss:pph}. The estimating equation for $\hat{\chi}$ in the second step is
\begin{eqnarray}
m_{2n}(\chi;\eta) = \frac{1}{n}\sum_{i=1}^n m_2(x_i,  \delta_i, z_i; \chi, \eta)  =  \frac{1}{n}\sum_{i=1}^n \sum_{j=1}^n w_i \Omega_{\epsilon[i,j]}^{-1}(\log(x_j)' - w_j'\chi] = 0,  \label{m2AFT}
\end{eqnarray}
\nin where $\Omega_{\epsilon[i,j]}^{-1}$ is the $(i,j)$-th element in $\Omega_{\epsilon}^{-1}$.

\lemma{Let $\hat{\chi}$ be the solution to $m_{2n}(\chi;\hat{\eta})$ and  $\hat{\chi}^*$  be the solution to $m_{2n}(\chi;\eta_0)$. (i) $\hat{\chi}\overset{p}{\to}  \chi_{0}$. 
(ii) $\sqrt{n} (\hat{\chi}^* - \chi_{0}) \overset{d}{\to} N(0,\Omega_2)$.\label{propchi}}
\nin It means that (i) $\hat{\chi}$ is a consistent estimator for $\chi_0$ when $\hat{\eta}$ is used in (\ref{m2AFT}); (ii) $\hat{\chi}^*$ as the estimator for $\chi_0$ is asymptotically normal distributed when the true $\eta_0$ is used in (\ref{m2AFT}). The proof is given in Supplementary Material S.I.


The estimating equation for $\theta$ in the third step is defined as
\begin{eqnarray}
m_{3n}(\theta;\eta,\chi) = \frac{1}{n}\sum_{i=1}^n m_3(x_i,  \delta_i, z_i; \theta, \eta, \chi) = \frac{1}{n}\sum_{i=1}^n [S_{AFT}(x_i|z_i;\chi) - \hat{S}_{CGE}(x_i|z_i;\theta)] = 0. \label{m3AFT}
\end{eqnarray}
\lemma{Let $\hat{\theta}$  be the solution to $m_{3n}(\theta; \hat{\eta},\hat{\chi})$ and $\hat{\theta}^*$ be the solution to $m_{3n}(\theta;\eta_0,\chi_0)$. (i) $\hat{\theta}\overset{p}{\to}\theta_0$. (ii) $\sqrt{n} (\hat{\theta}^* - \theta_0) \overset{d}{\to} N(0,\Omega_3)$. \label{proptheta}}
\nin The proof is also given in Supplementary Material S.I.


The estimation procedure has three steps and we denote it as the parametric three stage estimator (3SE). We define a vector for the three estimating equations in (\ref{m1}) - (\ref{m3AFT}) as
\begin{eqnarray*}
g_1(x_i, \delta_i, z_i ;  \eta, \chi, \theta) = [m'_1(x_i, \delta_i, z_i;\eta), m'_2(x_i,\delta_i, z_i; \chi, \eta), m_3(x_i, \delta_i, z_i; \theta,\eta,\chi)]'.
\end{eqnarray*}
\nin The estimator for $\varphi = (\eta', \chi',\theta)'$ is a GMM estimator, which solves the following system of estimating equations with probability approaching one:
\begin{eqnarray}
m_{4n}(\varphi) 
= \frac{1}{n}\sum_{i=1}^n g_1(x_i, \delta_i, z_i; \eta, \chi, \theta) = 0. \label{m5}
\end{eqnarray}
\nin The resulting $\hat{\varphi}$ hat the following properties.
\proposition{(i) $\hat{\varphi}  \overset{p}{\to}  \varphi_0$. (ii) $\sqrt{n} (\hat{\varphi} - \varphi_0) \overset{d}{\to} N(0,\Omega_4)$. \label{prop5}}
\nin Proposition \ref{prop5} follows from Assumptions \ref{ass2}, \ref{ass3} and $\ref{ass4}$, Lemmas \ref{propchi} and \ref{proptheta} and Theorem 6.1 of Newey and McFadden (1994). The rate of convergence for the parametric estimates is $\sqrt{n}$. 
$\Omega_4$ is the variance matrix of the estimator and depends on the chosen log-likelihood function in the first step and the copula generator in the last two steps. Given its complexity, estimation by the bootstrap should be attractive in applications.

\lemma{$\Omega_4$ can be consistently estimated by the bootstrap.\label{propbt}}
\textbf{Proof:} We follow Mammen (1992). For $\hat{\varphi}(x_i,z_i, \delta_i)$ obtained from (\ref{m5}), let $L_n=(\hat{\varphi}-\varphi_0)'\Omega_4^{-1}(\hat{\varphi}-\varphi_0)$. Define $\hat{\varphi}^*(x^{*}_i,z^{*}_i, \delta^{*}_i)$ as the estimates obtained from the bootstrap sample $(x^{*}_i,z^{*}_i, \delta^{*}_i)$ and  $L^{*}_n=(\hat{\varphi}^*-\hat{\varphi})'\Omega_4^{-1}(\hat{\varphi}^*-\hat{\varphi})$. Let $G_n(l)=\Pr(L_n\leq l)$ and $G^*_n(l)=\Pr^*(L^*_n\leq l)$, where $\Pr^*$ is the probability distribution induced by bootstrap sampling, conditional on the original data $\{x_i,z_i, \delta_i\}$. Then $G^*_n$ consistently estimates $G_n$ as $G_n \overset{d}\to N(0,1)$ because of Proposition \ref{prop5}. \QEDB

\subsubsection*{Semiparametric model}
In the semiparametric model, $\hat{\theta}$ is estimated in the second step by solving  the following estimation equation with probability approaching one:
\begin{eqnarray}
m_{5n}(\theta;\eta) = \frac{1}{n-1}\sum_{i=1}^n m_5(x_i, \delta_i, z_i; \theta, \eta) =  \frac{1}{n-1}\sum_{i=1}^n I_k'\left(\hat{b}_{i}(\theta)-\hat{b}^*(\theta)\right)= 0. \label{m4}
\end{eqnarray}
\nin Since $\hat{b}_{i}$ in  (\ref{ebeta}) is a deterministic function of $\theta$, no extra step is needed to estimate it.  The estimation procedure has two steps and we denote it as the semiparametric two stage estimator (2SE).

\lemma{Let $\hat{\theta}$ be the solution to $m_{5n}(\theta;\hat{\eta})$ and $\hat{\theta}^*$ be the solution to $m_{5n}(\theta;\eta)$. (i) $\hat{\theta}\overset{p}{\to}  \theta_0$. (ii) $\sqrt{n} (\hat{\theta}^* - \theta_0) \overset{d}{\to} N(0,\Omega_5)$.
\label{propsemi}}
\nin The proof is given in Supplementary Material S.I.
We define a vector for the two estimating equations in (\ref{m1}) and (\ref{m4}) as
\begin{eqnarray*}
g_2(x_i, \delta_i, z_i ;  \eta, \theta) = [m'_1(x_i, \delta_i, z_i;\eta), m_5(x_i,\delta_i, z_i; \theta, \eta)]'.
\end{eqnarray*}
\nin The estimator for $\kappa = (\eta', \theta)'$ is a GMM estimator solving with probability one
\begin{eqnarray}
m_{6n}(\kappa) = \frac{1}{n}\sum_{i=1}^n g_2(x_i, \delta_i, z_i; \eta, \theta) = 0. \label{mg}
\end{eqnarray}

The resulting $\hat{\kappa}$ has the equivalent properties as in Proposition \ref{prop5}, i.e. it is consistent and asymptotically normal because of Assumptions \ref{ass2} and \ref{ass3}, Lemma \ref{propsemi} and Theorem 6.1 of Newey and McFadden (1994). The rate of convergence is $\sqrt{n}$. The bootstrap is also applicable (Lemma \ref{propbt}).

\subsection{Finite Sample Performance and Robustness}
We conduct a series of Monte Carlo simulations to investigate the finite sample performance of the suggested parametric and semiparametric estimation procedures under correct and incorrect model specification. The simulations also include comparisons with existing methods such as full MLE, MMPHM, semiparametric Cox PH model with independent risks (Cox), PWC PH model with independent risks, and PM with independent risks. The simulation results are given in Supplementary Material S.V. They confirm nice finite sample properties of the suggested approaches, which often outperform existing methods, in particular under misspecification of the latter. Estimation of the competing risks model is generally
sensitive to the assumed model for the latent marginals and the assumed dependency. For this
reason, it is desirable to work with milder parametric restrictions if they are not known to
hold. In this regard, our three-step estimator has a clear advantage over existing methods in
competing risk models that require assumptions on both risks or an independence assumption.

\section{Application \label{s:app}}
We conduct two real data applications to illustrate the applicability of the suggested approaches. We focus here on an anlaysis of unemployment duration with the semiparametric model. The results of an application to employment duration with the parametric model are given in Supplementary Material S.VII  To avoid any source of misspecification, we use nonparametric first stage estimators for $\pi(t|z)$ and $f_t(t|z)$.

In this example we put semiparametric models into practice to analyse a sample of unemployment benefit durations for seasonally laid-off workers. The sample is extracted from the sample of the integrated labour market biographies (SIAB) 1975-2014 of the institute for Employment Research (IAB), Germany. For more information on the SIAB see Antoni et al. (2016). We use a sample that is extracted using the same criteria as in Lo et al. (2020), although we restrict it to the first observed unemployment period for each individual in the data. The resulting sample contains around 3,800 single spells.

We study three exit routes out of unemployment benefits: (1) job at a new seasonal employer, (2) job in another business sector, and (3) receiving other benefits like unemployment assistance, or public training measures for the unemployed. All other reasons for terminating their benefit claims are pooled into unknown/other risks. We apply our semiparametric model and use an age dummy variable as the identifying covariate, in particular whether the unemployed is aged less than 30 (young) or not (old). We use age because it is well known to be a major determinant for the length of unemployment in Germany.

To analyse the multiple competing risks model using a bivariate competing risks framework, we use the pooling method suggested by Lo and Wilke (2010). Namely, when we consider risk (1), i.e. starting a job at a new seasonal employer, all other risks than risk (1) are pooled as dependent censoring. Similarly, for risk (2), i.e., obtaining a job in a new business sector, all other risks than risk (2) are pooled. Note most observations terminate with exits to jobs of different types, while risk (3) "other benefits" is less frequent. It means that when we estimate the dependence between risk (1) and the pooled other risks, the estimated $\tau$ mostly measures the dependence between exits to different jobs types. The same is the case when we estimate the dependence between risk (2) and the combined other risks. In contrast, when we estimate the dependence between risk (3) "other benefits" and the combined other risks, the estimated $\tau$ measures the dependence between remaining unemployed (receiving other benefits) and the exiting to one of the various job types. We therefore expect $\tau$ to be positive and similar in the first two cases, while its anticipation is difficult in the last.

\begin{landscape}
\begin{figure}[!htbp]
	\hspace{2.5cm} (a) \hspace{5.5cm}   (b) \hspace{6cm}    (c) \\
    \centering
    \includegraphics[scale=0.30]{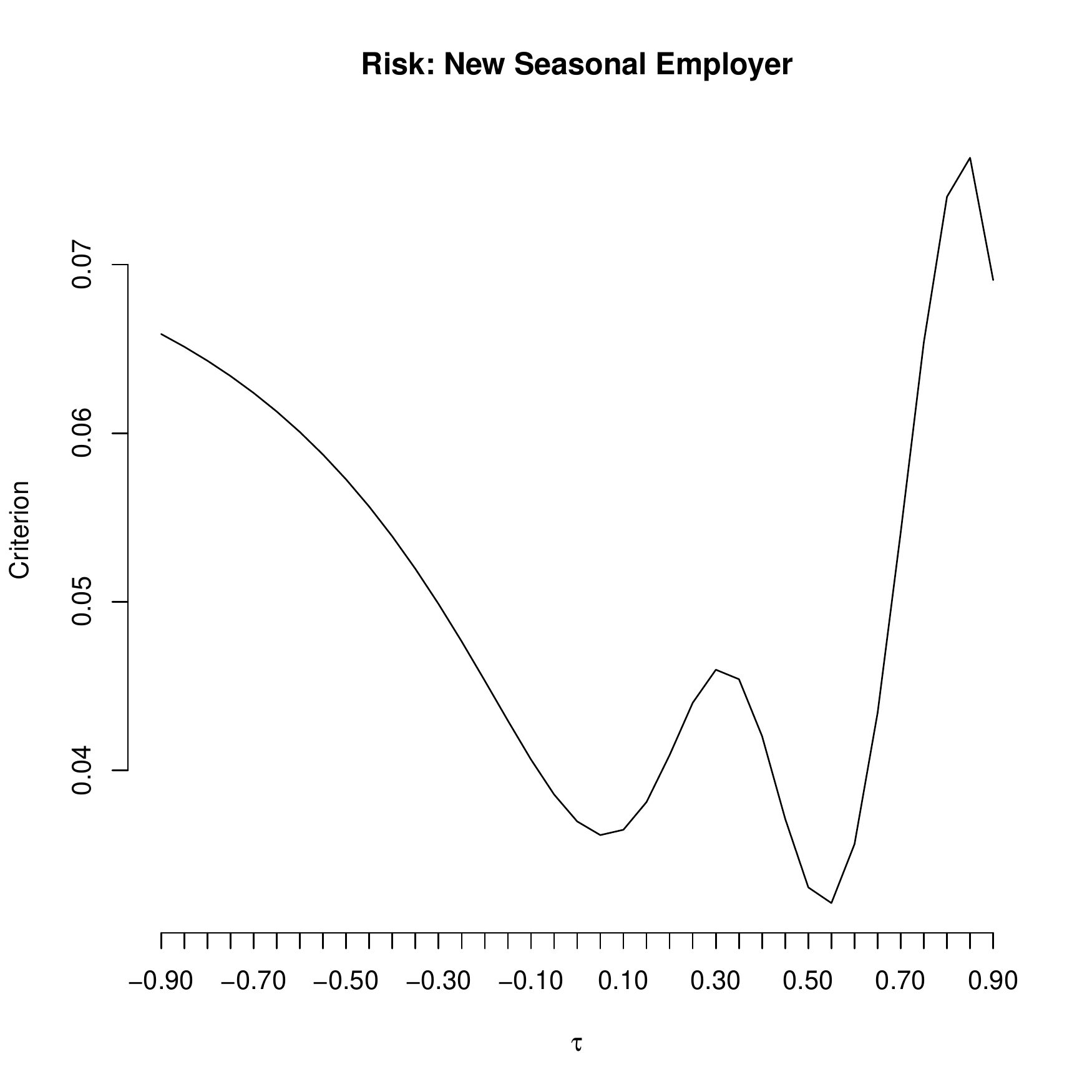}\hspace{1cm}
    \includegraphics[scale=0.30]{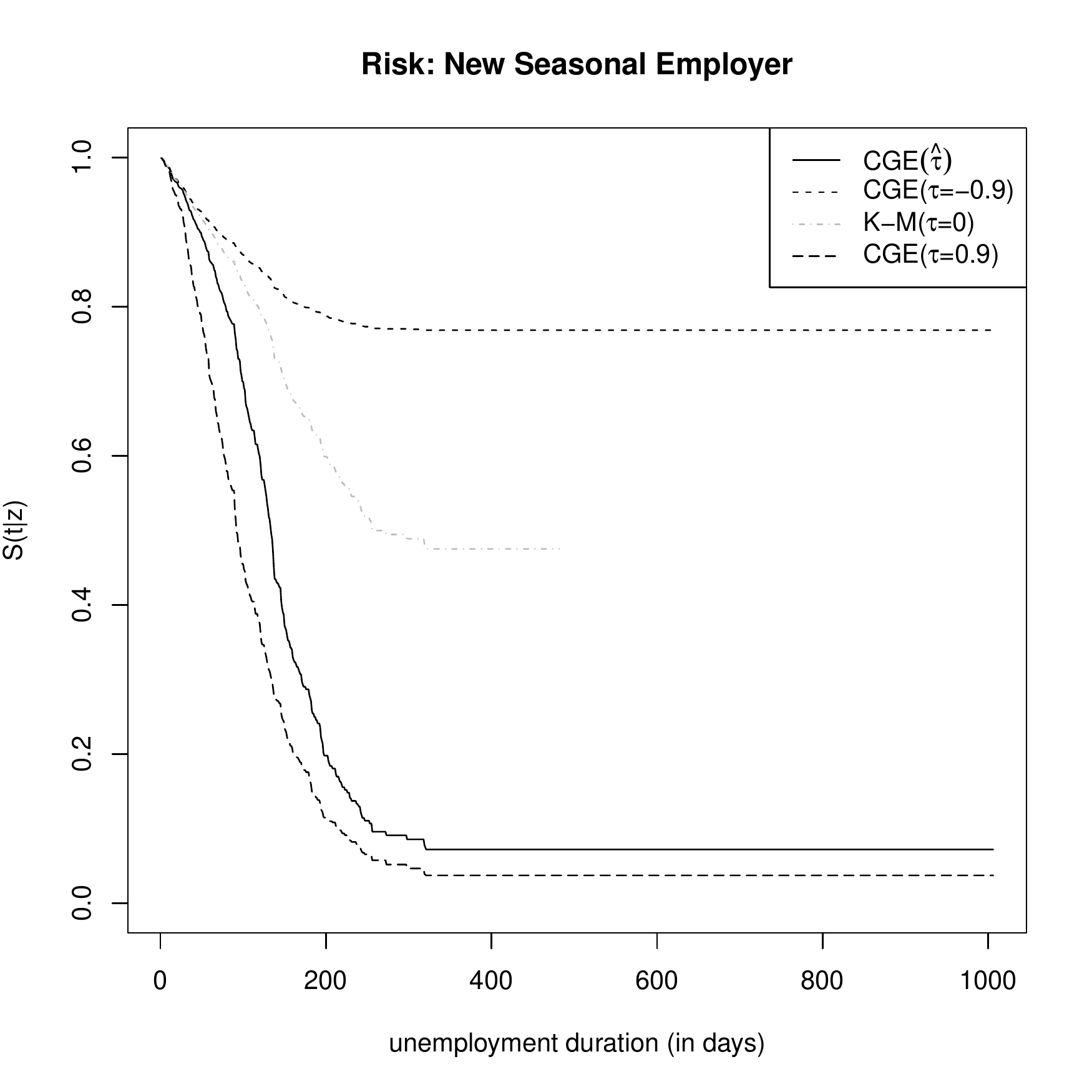}\hspace{1cm}
    \includegraphics[scale=0.30]{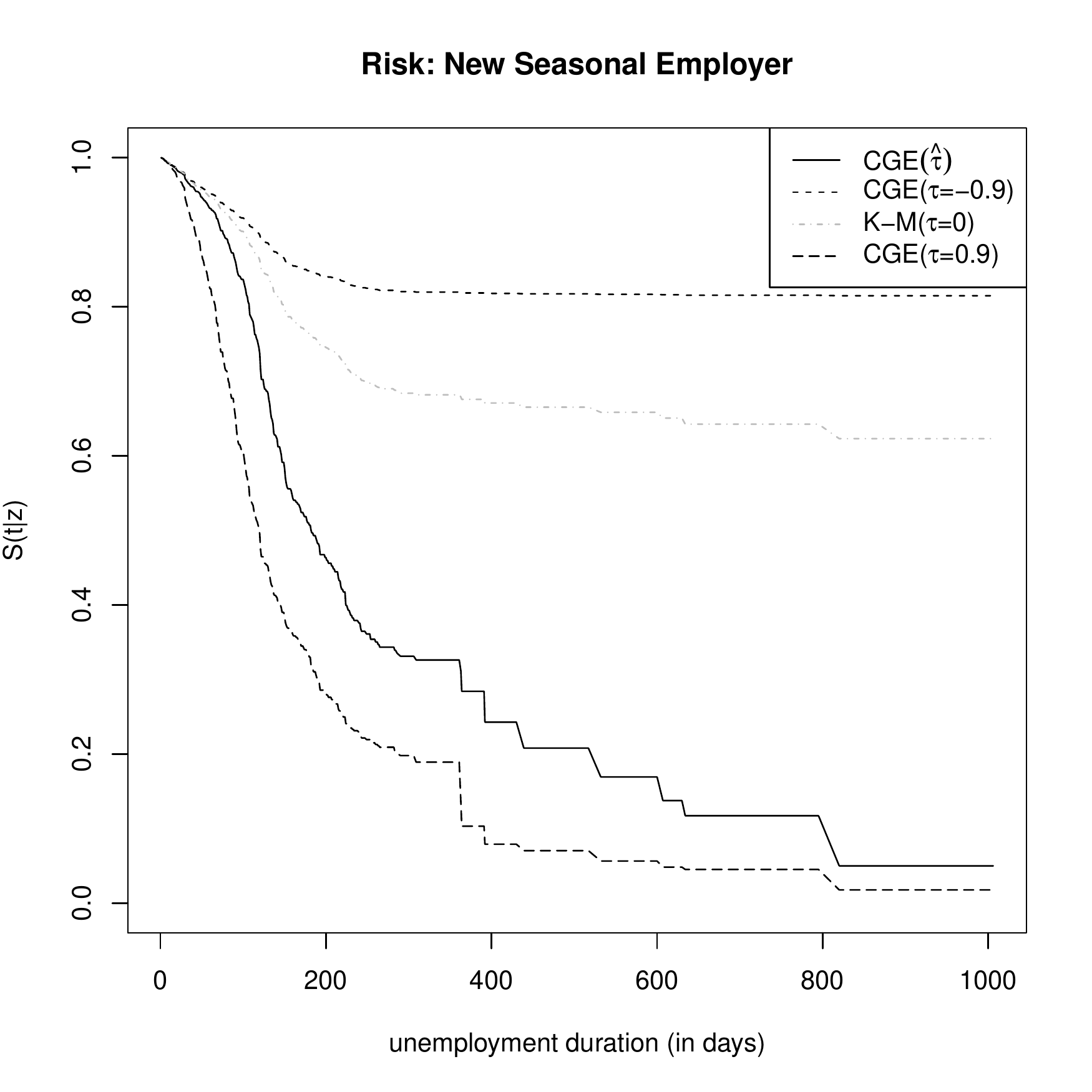}\hspace{1cm}\\

	\includegraphics[scale=0.30]{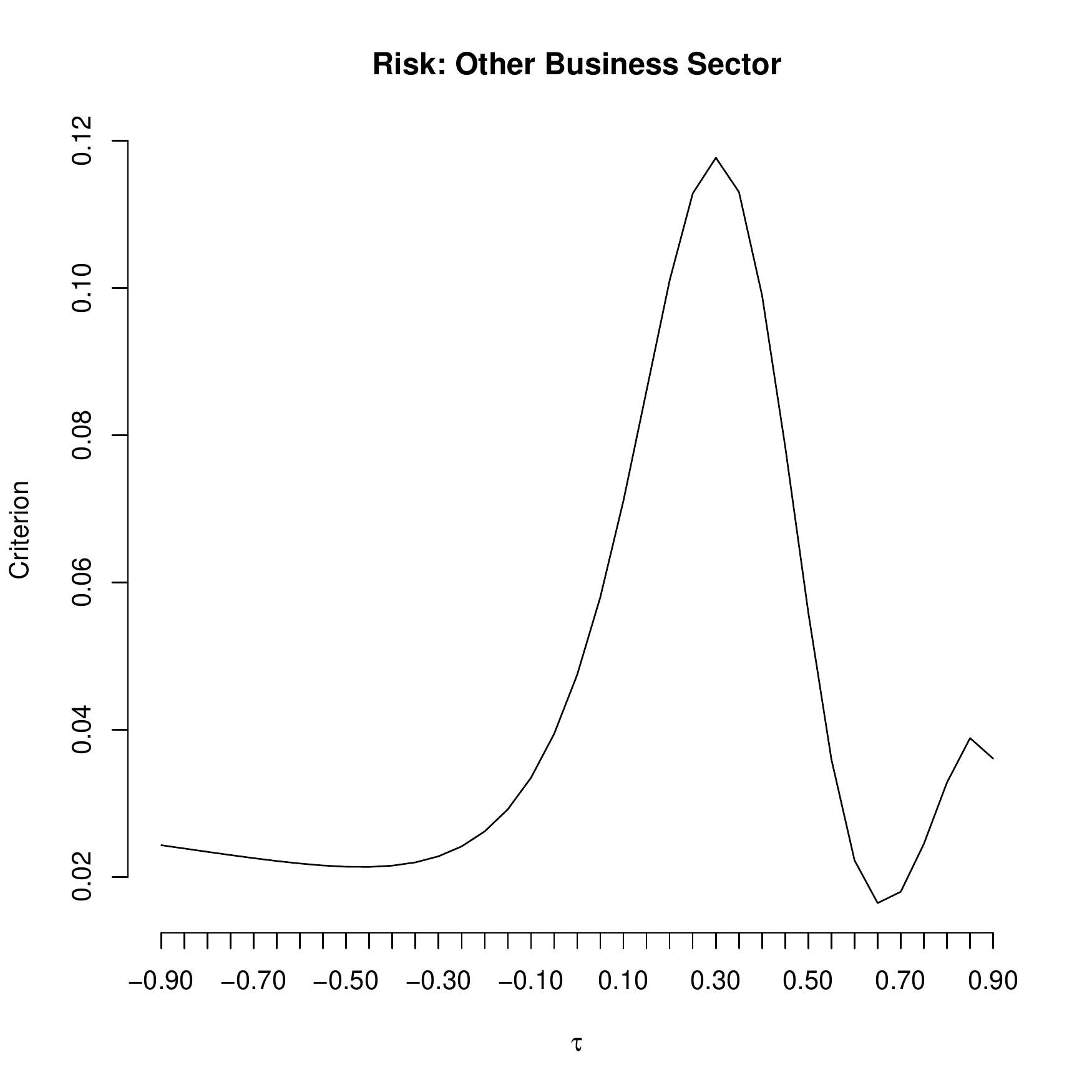}\hspace{1cm}
    \includegraphics[scale=0.30]{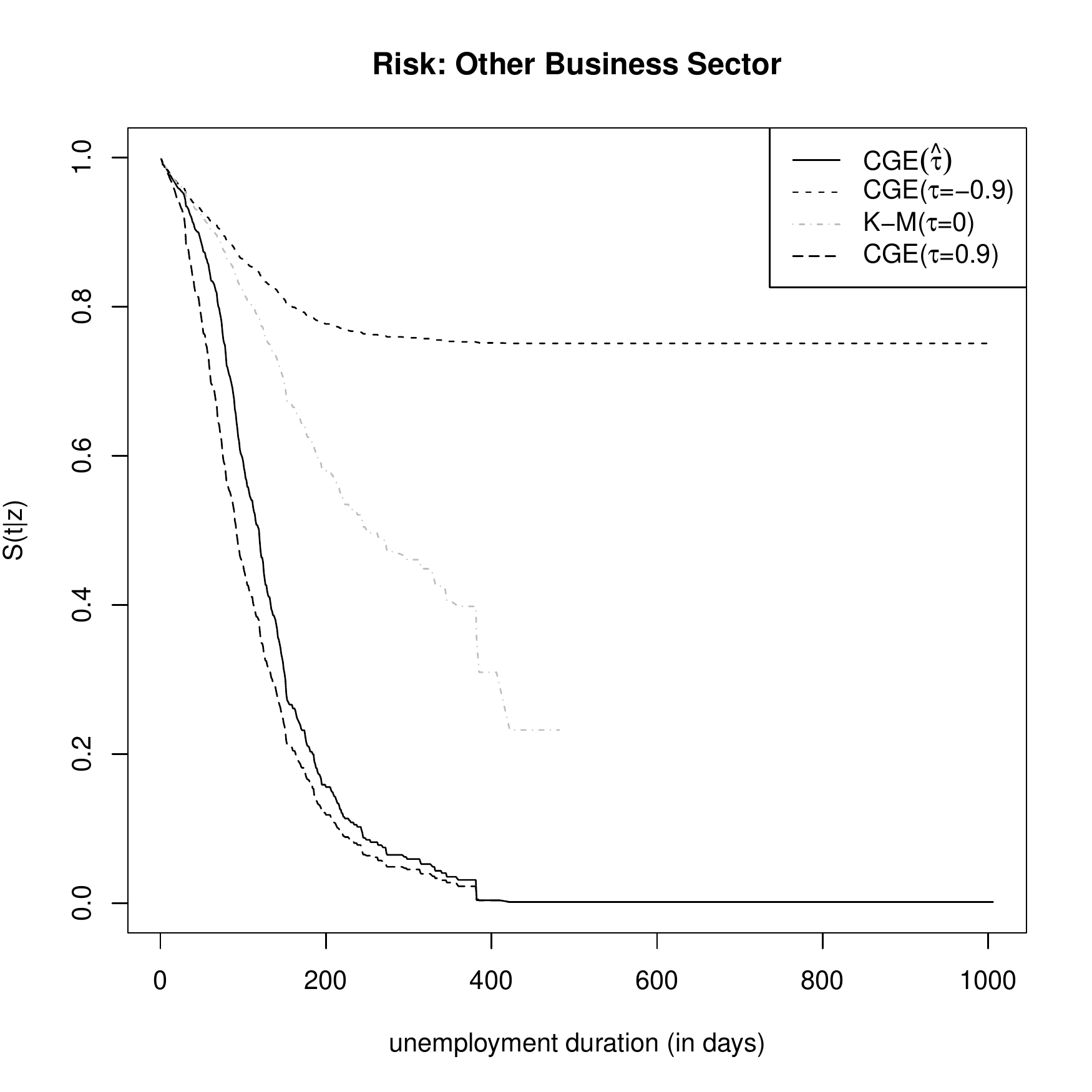}\hspace{1cm}
    \includegraphics[scale=0.30]{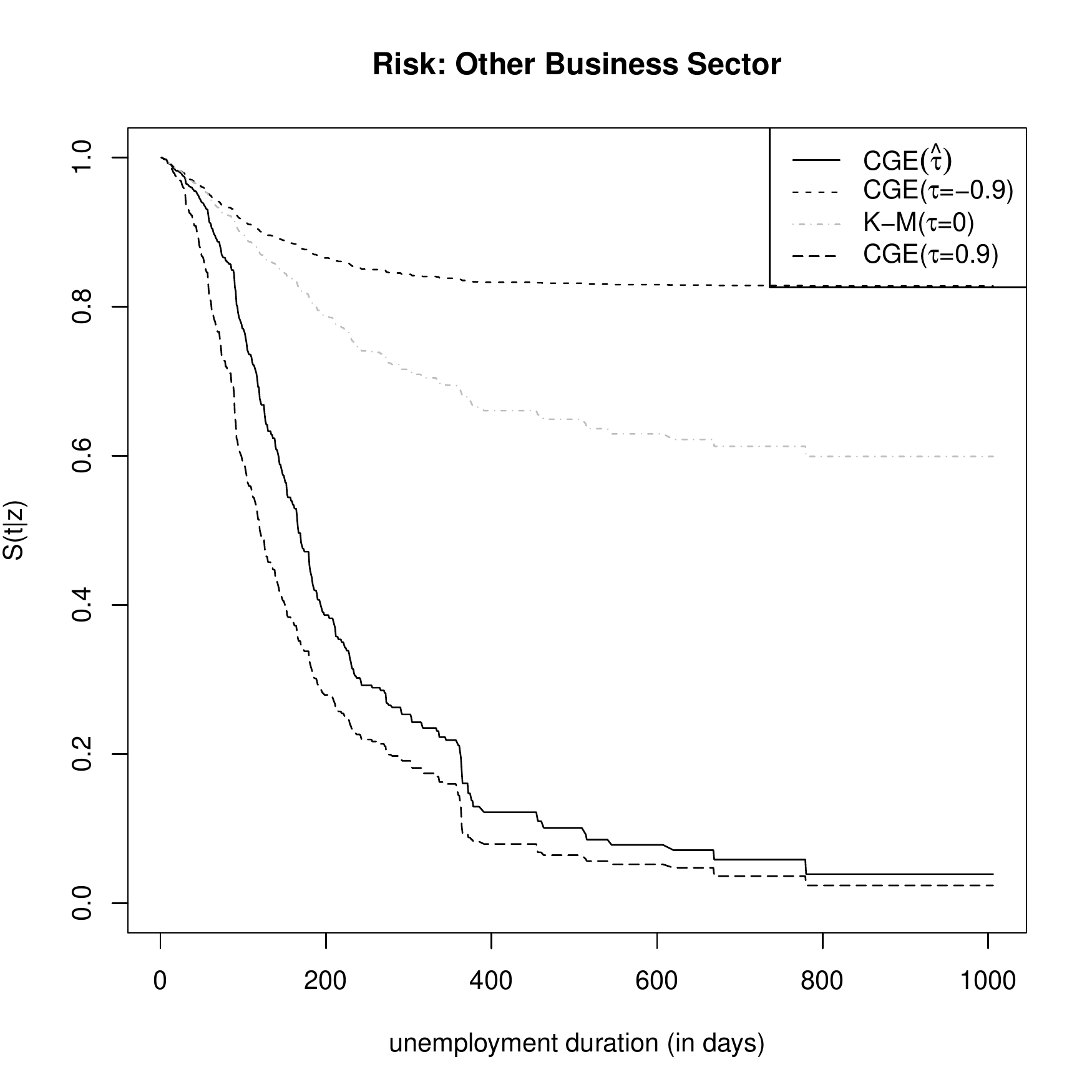}\hspace{1cm}\\

	\includegraphics[scale=0.30]{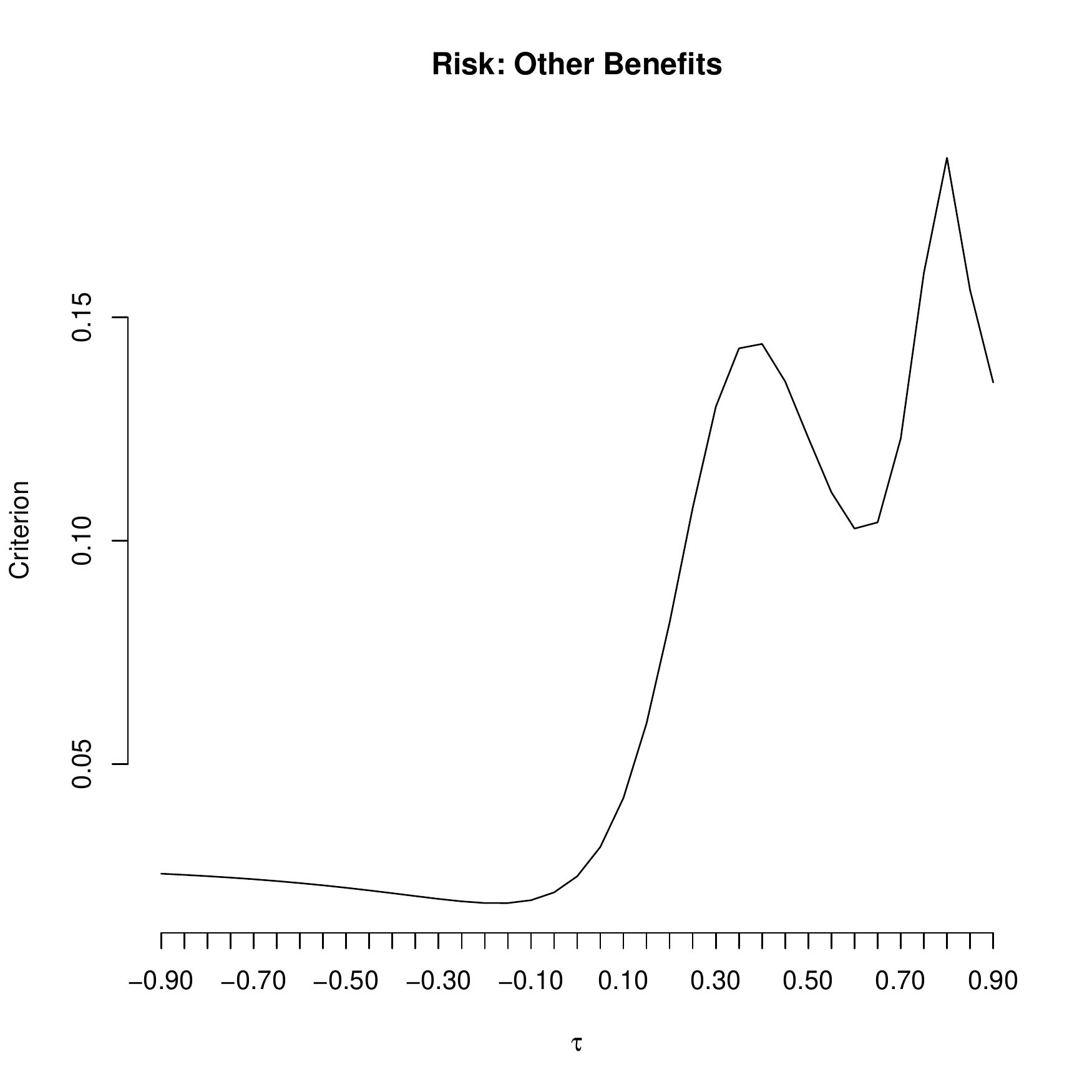}\hspace{1cm}
	\includegraphics[scale=0.30]{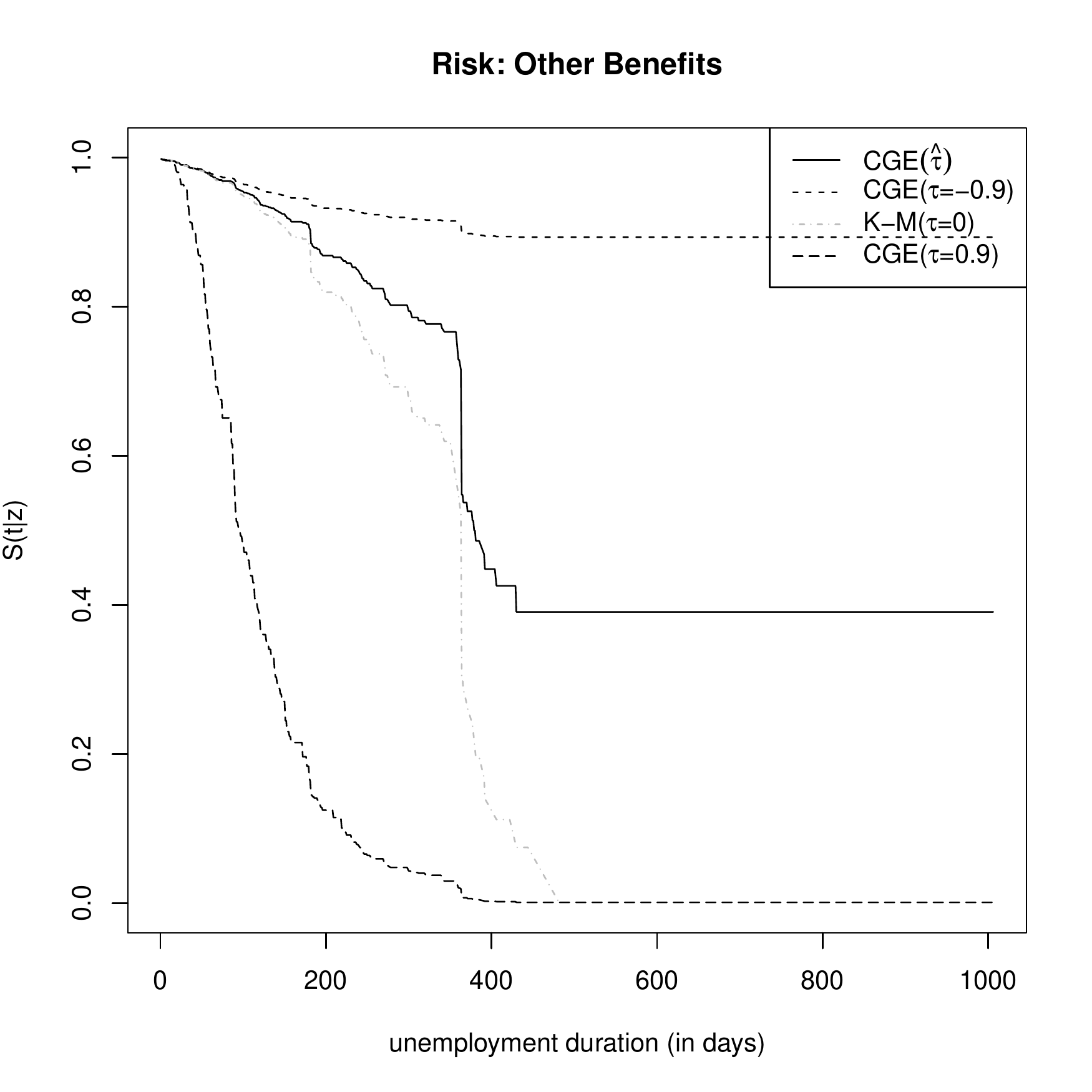}\hspace{1cm}
    \includegraphics[scale=0.30]{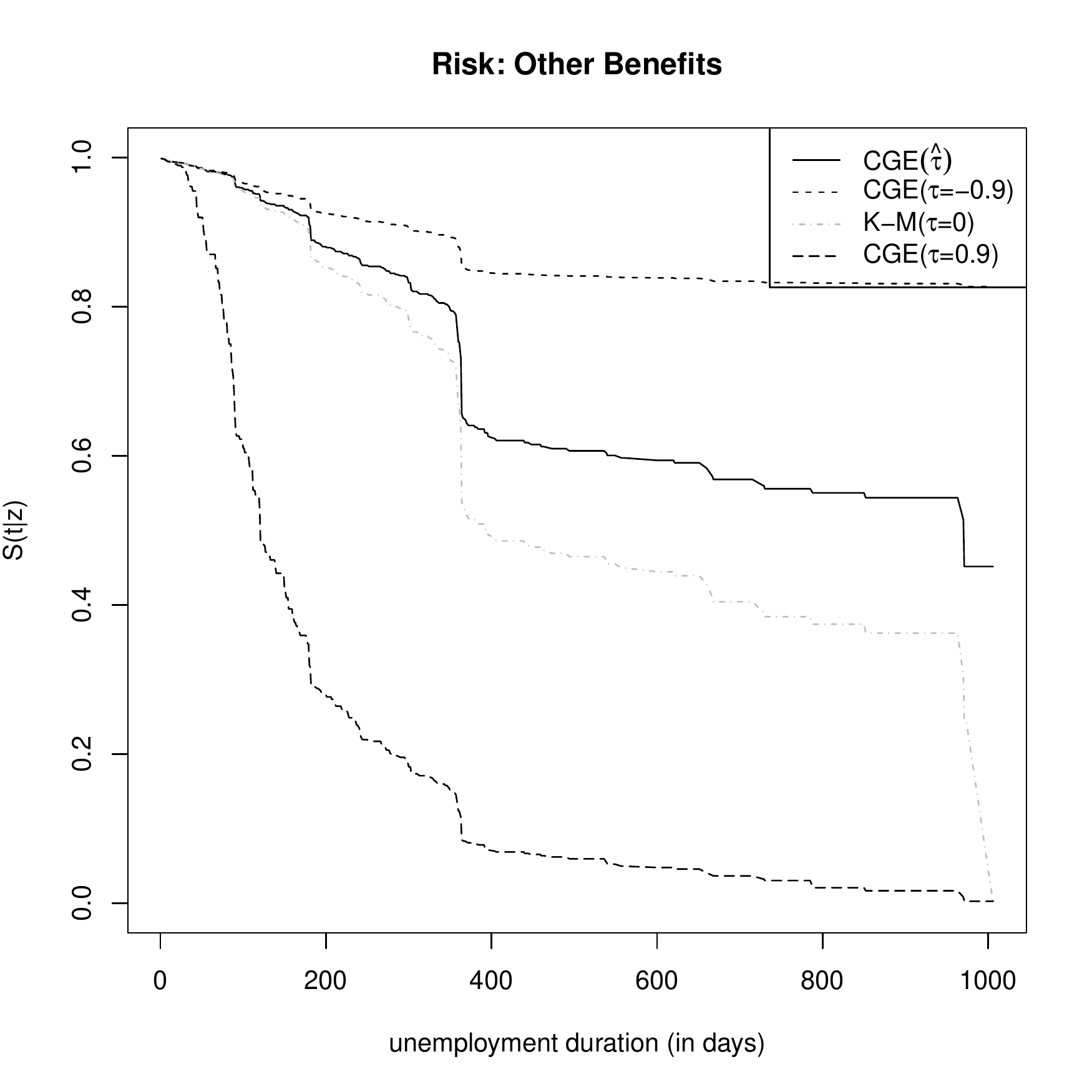}

	\caption{Criterion (\ref{thebeta}) (a) and estimated conditional survival curves using the unemployment benefits duration data (young: (b), old: (c).}
	\label{F:app}
\end{figure}
\end{landscape}

Figure \ref{F:app} shows in panel (a) the shape of the objective function in $\tau$ for the three risks of interest. It is apparent that there is a unique minimum in all three cases, although the objective function is very flat to the left for risk "other benefits". The estimated $\tau$ is positive and significantly different from 0 for the first two risks. In particular, it is 0.5367 with 95\% bootstrap C.I. [0.0805,0.8233] for new seasonal employer and 0.6628 with 95\% bootstrap C.I. [0.0238,0.9000] for other business sectors. In contrast, the estimated $\tau$ is -0.1725 with 95\% bootstrap C.I. [-0.900,0.1481] for risk (3) vs. starting a job. The negative point estimate is therefore not significantly different from zero, which implies a different dependence than for risks 1 and 2. The bootstrap distributions for the estimated $\tau$ are given in Figure \ref{fig:boot} in the supplementary material. These findings can be explained as follows. A person who is more motivated to search for and start a job or can be more successful in obtaining a job, is expected to more quickly receive an offer for any type of job, which is reflected by a positive tau in the first two cases. In fact, the point estimates suggest a rather strong positive dependence ($\hat{\tau}=0.6$), suggesting that this pattern could be very pronounced (considering the uncertainty in the estimates). In contrast, an individual who is less motivated or is less successful in finding a new job is expected to remain longer in unemployment and to transfer to less generous benefit types after the unemployment benefit entitlement has expired. As the maximum entitlement period for unemployment benefit is regulated by laws (depending on employment history and laws), it is difficult to anticipate the direction of the dependence, which is somehow reflected by the imprecise point estimate.

Figure \ref{F:app} also reports the estimated survival curves for younger individuals in panel (b) and for older individuals in panel (c). The estimates are stratified nonparametric estimates of the CGE for different values of $\tau$. We report these instead of the marginal survivals implied by the semiparametric model, because they are more flexible. Therefore, the only ingredient that comes from the semiparametric model is $\hat{\tau}$. The figure clearly demonstrates the width of the possible ranges implied by the different $\tau$. Therefore, by having an estimate of $\tau$ allows the researcher to develop a much clearer understanding of the location of the survival. For comparison the Kaplan-Meier estimator for $\tau=0$ is shown. It is apparent that it is often far off the estimate given $\hat{\tau}$.

Figure \ref{F:app} also confirms the well known fact of the German labour market that older unemployed have much longer unemployment benefit duration than the younger unemployed. This is also reflected by the $\hat{\beta}$s of the 2SE as show in column (1) of Table \ref{T:semipara}, which are positive for the three risks. The hazard rate for a young person to start a job in the same or another business sector is around twice ( hazard ratio $=\exp(0.6962)\approx 2.01)$ as high than for an old person. On the other hand, the hazard rate for remaining in unemployment and receiving another benefit is only slightly higher (hazard ratio $=\exp(0.1653)\approx 1.18)$ for the two groups and the coefficient is not significant. It is remarked that there are no younger unemployed claiming unemployment benefits after a bit more than one year ($>$ 400 days). Therefore, no nonparametric estimates are available beyond this point, which restricts the objective functions of 2SE to this range of $x_i$.

Column (2) of Table \ref{T:semipara} contains the bootstrap standard error for the 2SE when $\tau$ in the first step is not estimated but assumed to be $\hat{\tau}$. This exercise provides insights about how the standard error is affected. For the first risk, the standard error in column (2) is only slightly smaller than that in column (1), which indicates that the main source of variance comes from the second step estimation when $\beta$ is estimated. For the other two risks, however, the standard error in column (2) is 25-40\% smaller, which points to estimation of $\tau$ causing the precision of estimates to decrease. Next, we compare 2SE with COX model results which are reported in column (3). The results are similar for the first two risks but quite different for the third risk, whereby any differences can only be explained by the incorrect assumption about $\tau$. As expected, the standard errors for the COX model are smaller than for 2SE as the latter involves estimation of $\tau$. A fairer comparison is Cox with 2SE results for assumed $\hat{\tau}$, where the difference is rather small for the first two risks but sizable for the third. We explain the differences by the fact that the Cox estimated is based on partial MLE that ignores the nonparametric baseline, while our approach bases on a nonparametric $\hat{S}(t)$ obtained by the CGE. Columns (4) to (6) show the estimated $\beta$ by using the CGE when $\tau$ is assumed to have different values: -0.9, 0 and 0.9. The results suggest that the $\hat{\beta}$ can sizably differ when an incorrect assumption about $\tau$ is being made. This suggests that 2SE with estimated $\tau$ is preferable as it avoids misspecification bias.

\begin{table}
\centering
\caption{Comparison of $\hat{\beta}$ for younger individuals for semiparametric unemployment duration models.}
\hspace{12 pt}
\begin{tabular}{lcccccc}
		\hline\hline
										& (1) 			& (2) 			& (3) 	& (4)	& (5) & (6) \\
Estimator							& 2SE 			& 2SE 			 	& Cox & CGE & CGE& CGE \\
Assumed $\tau$				& - 			 & $\hat{\tau}$ & 0	& 0 & -0.9 & 0.9 \\
\hline
	                            &  \multicolumn{6}{c}{Risk: New Seasonal Employer}\\
    $\hat{\beta}$			   &   	0.6701           &		0.6701	                           & 	0.5569    &      0.5765           &    0.4930           &   0.5200     \\
	bootstrap S.E.	           &	0.0883	        & 		0.0794                              & 	0.0692     &   0.0887           &    0.0954           &   0.0581      \\
\hline
	                            &  \multicolumn{6}{c}{Risk: Other Business Sector}\\
    $\hat{\beta}$			   &   	0.6962           &		0.6962	                           & 	  0.7773     &     0.6777    &     0.5721           &   0.5197        \\
	bootstrap S.E.	           &	0.1298	      & 	0.0756	                            & 	 0.0737      &     0.0908         &   0.0970            &    0.0577       \\
\hline
	                            &  \multicolumn{6}{c}{Risk: Other Benefits}\\
    $\hat{\beta}$			   &   	0.1653       &		0.1653	                               & 	 0.4539    &      0.2195          &    0.0705           &     0.5427  \\
	bootstrap S.E.	           &	0.2201	     & 		0.1655                             & 	0.1003      &      0.1616       &     0.1758          &    0.0862         \\
 \hline	
 \hline	
	\end{tabular}
\label{T:semipara}
\end{table}

\section{Summary and Conclusion}
We consider a model with one latent marginal of interest that is either parametric or semiparametric, while the other risk is entirely unspecified. The dependence structure is modelled by a copula with unknown degree of dependence. We show that the  parametric marginal and the risk dependence are identifiable with and without covariates, while it requires at least one covariate in the case of the semiparametric marginal. We suggest a parametric three step and a semiparametric two step estimation procedure, respectively, and demonstrate through simulations and applications the applicability and nice properties of our approach. Estimates are shown to have nice properties for data sets of a couple of thousand observations and quickly converge. Competing classical models such as parametric MLE, the multivariate mixed proportional hazards model, the Cox model and MLE under assumed risk independence are shown to likely suffer more from misspecification bias as they operate under stronger restrictions. Our application illustrates that the use of our method gives plausible and more insightful results compared to the classical approaches. We therefore consider our approach as an interesting extension of the portfolio of analysis methods for competing risks models. Of the existing models, we only found MLE to be better in smaller samples, despite possible misspecification of $R(c)$ and the Cox model when the dependence structure between risks is (close enough) to independence.


	\newpage

	\setcounter{page}{1}
    \setcounter{figure}{1}
    \setcounter{table}{1}


\centering\LARGE{\textbf{A single risk approach to the semiparametric copula competing risks model}}\\
\LARGE{\textbf{SUPPLEMENTARY MATERIAL}\\
\normalsize
\vspace{1.3cm}
\justifying
\author{\nin Simon M.S. Lo\footnote{The United Arab Emirates University, United Arab Emirates, Department of Innovation in Government and Society, E--mail: losimonms@yahoo.com.hk} \\
	Ralf A. Wilke\footnote{Copenhagen Business School, Department of Economics, E--mail: rw.eco@cbs.dk}}

\maketitle


\thispagestyle{empty}
			
\renewcommand{\thetable}{S\arabic{table}}
\renewcommand{\thefigure}{S\arabic{figure}}

\linespread{1.3}{
\normalsize
\section*{S.I: Proofs}

\nin \textbf{Proof of Lemma \ref{l:inverse}:} $\phi_{\theta}(u)$ is continuous and strictly decreasing in $u$ on $[0, \inf\{u: \phi_{\theta}(u)=0\})$  by Assumption \ref{ass1}(i). It is therefore a bijection for which the inverse always exists and is also a bijection. The quasi inverted function $u=\phi^{-1}_{\theta}(s)$ is therefore also continuous and strictly decreasing in $s$ on $(0, 1]$ with $\phi^{-1}_{\theta}(1) =0$ and $\phi^{-1}_{\theta}(0) = \inf\{u:\phi_{\theta}(u) =0\}$. According to the inverse function theorem, the derivative of the inverse function is the reciprocal of the derivative of a function. We can therefore compute  $(\phi^{-1}_{\theta})^{(1)}(s)$ from  $\phi^{(1)}_{\theta}(u)$. Because of Assumption \ref{ass1}(i),  $\phi_{\theta}^{(1)}(u) \in (-\infty, 0]$ and $\phi_{\theta}^{(1)}(u)$ is increasing in $u$, its reciprocal  $ (\phi^{-1}_{\theta})^{(1)}(s) = 1/\phi_{\theta}^{(1)}(u) <0$. Similarly, by Assumption \ref{ass1}(i), $\phi^{(2)}_{\theta}(u) \geq 0$  on $u\in [0,\inf\{u:\phi_{\theta}(u) =0\})$, its reciprocal  $(\phi^{-1}_{\theta})^{(2)}(s)  \geq 0$ for all $s \in (0,1]$. 

\vspace{12pt}



\nin \textbf{Proof of Lemma \ref{l:thetaSPHI}:} Let $S_{\theta_1}(t;z)$ and $S_{\theta_2}(t;z)$ be obtained from (\ref{cge}) for any $\theta_2 > \theta_1 \in \Theta$.  Proposition 2 of Rivest and Wells (2001) suggests $S_{\theta_2}(t;z) \leq S_{\theta_1}(t;z)$ for all $t$ and $z$, if $(\phi^{-1}_{\theta_1})^{(1)}(s)/(\phi^{-1}_{\theta_2})^{(1)}(s)$ is increasing  in $s$. Their proposition can be carried over to the case $S_{\theta_2}(t;z) < S_{\theta_1}(t;z)$ when Assumption \ref{ass1}(\ref{a1:6})  holds. The proof follows Rivest and Wells (2001) with some modifications. We ignore $z$ for the proof. By differentiating (\ref{cge}) with respect to $t$ at $\theta_2$, we obtain
\begin{eqnarray}
S_{\theta_2}^{(1)}(t) &=& -\frac{(\phi^{-1}_{\theta_2})^{(1)}[\pi(t)] f_t(t)}{(\phi^{-1}_{\theta_2})^{(1)}[S_{\theta_2}(t)]}.\non
\end{eqnarray}
\nin By integrating the derivative of $\phi_{\theta_1}^{-1}[S_{\theta_2}(t)]$ with respect to $t$, where $\theta_2>\theta_1$, we have
\begin{eqnarray}
\phi_{\theta_1}^{-1}[S_{\theta_2}(t)] &=& \int_0^{t} (\phi^{-1}_{\theta_1})^{(1)}[S_{\theta_2}(u)] S_{\theta_2}^{(1)}(u) du \non\\
&=& -\int_0^{t} (\phi^{-1}_{\theta_1})^{(1)}[S_{\theta_2}(u)] \frac{(\phi^{-1}_{\theta_2})^{(1)}[\pi(u)] f_t(u)}{(\phi^{-1}_{\theta_2})^{(1)} [S_{\theta_2}(u)]} du . \label{lsthe}
\end{eqnarray}
\nin Next, we have $S_{\theta}(t;z) > \pi(t;z)$ for all $t\in (0, \infty)$ by Assumption \ref{ass1}(\ref{a1:6}) and $(\phi^{-1}_{\theta_1})^{(1)}(s)/(\phi^{-1}_{\theta_2})^{(1)}(s)$ is strictly increasing in $s$ by Assumption \ref{ass1}(\ref{ass1:increase}) and $(\phi_{\theta}^{-1})^{(1)}(s) <0$ for all $s \in (0,1]$ by Lemma \ref{l:inverse}. These observations together imply for all $u \in (0, \infty)$
\begin{eqnarray}
\frac{(\phi_{\theta_1}^{-1})^{(1)}[S_{\theta_2}(u)]}{(\phi_{\theta_2}^{-1})^{(1)}[S_{\theta_2}(u)]} >\frac{(\phi_{\theta_1}^{-1})^{(1)}[\pi(u)]}{(\phi_{\theta_2}^{-1})^{(1)}[\pi(u)]},  \non \\
(\phi_{\theta_1}^{-1})^{(1)}[S_{\theta_2}(u)]\frac{(\phi_{\theta_2}^{-1})^{(1)}[\pi(u)]}{(\phi_{\theta_2}^{-1})^{(1)}[S_{\theta_2}(u)]} <(\phi_{\theta_1}^{-1})^{(1)}[\pi(u)]. \non
\end{eqnarray}
\nin Put it into (\ref{lsthe}), we have
\begin{eqnarray}
\phi_{\theta_1}^{-1}[S_{\theta_2}(t)] > -\int_0^{t} (\phi_{\theta_1}^{-1})^{(1)}[\pi(u)] f_t(u) du = \phi_{\theta_1}^{-1}[S_{\theta_1}(t)].  \non
\end{eqnarray}
\nin Because that $\phi^{-1}_{\theta_1}$ is strictly decreasing on $(0,1]$ (see Lemma \ref{l:inverse}), we conclude that $S_{\theta_2}(t) < S_{\theta_1}(t)$ for all $t \in (0, \infty)$. Therefore, $\theta \mapsto S_{\theta}(t)$ is strictly decreasing for any $t\in (0, \infty)$. \QEDB

\paragraph*{Proof of Lemma \ref{l:clayton}:} We show that the Clayton copula is compatible with Assumption \ref{ass1}(i), (iii), and (v) in what follows.

Assumption \ref{ass1}(i): (a) We show that $\phi^{(1)}_{\theta}(u) > -\infty$ for all $u\in [0, \inf\{u:\phi_{\theta}(u)=0\})$ and for all $\theta\in \Theta = [-1,\infty)$. Since $\phi_{\theta}(u) = [u\theta+1]^{-1/\theta}_+$,  $\phi^{(1)}_{\theta}(u) = - [u\theta+1]^{-(1/\theta +1)}_+$. For $\theta \in [-1,0)$, $-(1/\theta+1) \in [0, \infty)$, $[u\theta+1]_+ \in [0,1]$. Hence, when $u$ increases, $[u\theta+1]_+$ decreases, and $[u\theta+1]^{-(1/\theta +1)}_+$ decreases, $\phi^{(1)}_{\theta}(u)$ is non-positive and increasing in $u$. For $\theta \in (0, \infty)$, $1/\theta +1>1$, and $[u\theta+1]_+ \geq 1$. When  $u$ increases, $[u\theta+1]_+$ increases, and $[u\theta+1]^{-(1/\theta +1)}_+$ decreases, $\phi^{(1)}_{\theta}(u)$ is again non-positive and increasing in $u$. We can conclude that $\phi^{(1)}_{\theta}(u)$ attains its minimum at $u = 0$, when $\theta \neq 0$. It is enough to prove that  $\phi^{(1)}_{\theta}(0) > -\infty$ in this case. We consider the value of $\phi^{(1)}_{\theta}(0) = - 1^{-(1/\theta +1)}$ for different value of $\theta$.   For $\theta \in (-1,0)$ and $(0,\infty)$, $\phi^{(1)}_{\theta}(0) = -1$. For $\theta = -1$, $\phi_{-1}(u) = 1-u$ and $\phi^{(1)}_{-1}(u) = -1$ so $\phi^{(1)}_{-1}(0) = -1$. Finally, when $\theta=0$, by L'H\^{o}pital's rule, $\lim_{\theta \to 0} \phi_{0}(u) = \exp(-u)$, and by convention, $\phi_{0}(u) = \exp(-u)$. Hence, $\phi^{(1)}_{0}(0) = - \exp(-0) = -1$. In all cases $\phi^{(1)}_{\theta}(0) = -1 >-\infty$.


Assumption \ref{ass1}(i): (b) We show that $\nabla_{\theta}\phi_{\theta}(u)$ is bounded for all $u\in(0,\infty)$ and for all $\theta \in \Theta = [-1,\infty)$. We can show that $\nabla_{\theta}\phi_{\theta}(u) = - A(u) - B(u)$, where $A(u) = u\theta^{-1}\phi_{\theta}(u)^{(\theta+1)}$ and $B(u)=  \theta^{-1}\phi_{\theta}(u)\log(\phi_{\theta}(u))$. Since  $\theta$ is bounded below by a number greater than $-1$,  we have $\theta +1\geq 0$. Moreover, $\phi_{\theta}(u) \in (0,1)$ for all $u\in (0,\infty)$, we have $\phi_{\theta}(u)^{(\theta+1)} \in (0,1]$. Note also that $|\log(x)|<C(|x|^{-\epsilon}+|x|^{\epsilon})$ for any $\epsilon >0$ and for any constant $C$ big enough, $\log(\phi_{\theta}(u))$ is therefore bounded when $\phi_{\theta}(u)$ is bounded. Put them together, $|A(u)| \leq | u | |\theta|^{-1} |\phi_{\theta}(u)|^{|\theta+1|} <\infty$ and $ |B(u)| \leq  |\theta|^{-1}|\phi_{\theta}(u)| |\log(|\phi_{\theta}(u)|)| <\infty$, and thus  $|\nabla_{\theta}\phi_{\theta}(u)|\leq |A|+|B| < \infty$, except at the point of $\theta =0$. But, by convention, $\phi^{(1)}_{0}(u) = - \exp(-u)$, $\nabla_{\theta}\phi_{0}(u) = 0 <\infty$. To summarise, $\nabla_{\theta}\phi_{\theta}(u)$ is bounded.

Assumption \ref{ass1}(i): (c) We show that $\nabla_{\theta}\phi^{-1}_{\theta}(s)$ is bounded for all $s\in(0,1)$ and for all $\theta \in \Theta = [-1,\infty)$. We can show that $\nabla_{\theta}\phi^{-1}_{\theta}(s) = - A(s) - B(s) + C(s)$, where $A(s) = \theta^{-1}\log(s) s^{-\theta}$, $B(s) = \theta^{-2}s^{-\theta}$ and  $C(s) = \theta^{-2}$. Since $s\in(0,1)$, $s^{-\theta}$ is bounded in $(0,1)$ if $\theta \in [-1,0)$, and $s^{-\theta}$ is bounded in $(1,\infty)$ if $\theta \in (0, \infty)$. For the same reason discussed above, $\log(s)$ is bounded when $s$ is bounded. Put them together, we have $|\nabla_{\theta}\phi^{-1}_{\theta}(s)| \leq  |A(s)| + |B(s)| + |C(s)| \leq |\theta|^{-1}|\log(|s|)| |s|^{-|\theta|} + |\theta|^{-2} |s|^{-|\theta|} +  |\theta|^{-2} <\infty$ , except at the point of $\theta =0$. But, by convention, $\phi^{-1}_{0}(s) = - \log(s)$, $\nabla_{\theta}\phi^{-1}_{0}(u) = 0<\infty$. To summarise, $\nabla_{\theta}\phi^{-1}_{\theta}(s)$ is bounded.

Assumption \ref{ass1}(i): (d) We show that $\nabla_{\theta}(\phi^{-1}_{\theta})^{(1)}(s)$ is bounded. We can show that $\nabla_{\theta}(\phi^{-1}_{\theta})^{(1)}(s) =  \log (s) s^{-(1+\theta)}$. Since $1+\theta >0$ for $\theta\in [-1,\infty)$, $s^{-(1+\theta)}$ is bounded at $s^{-|1+|\theta||}$, which is $(0,1)$. We therefore have $|\nabla_{\theta}(\phi^{-1}_{\theta})^{(1)}(s)| \leq  |\log (|s|)| |s|^{-|1+|\theta||} <\infty$.

Assumption \ref{ass1}(iv): We show that $(\phi^{-1}_{\theta_1})^{(1)}(s)/(\phi^{-1}_{\theta_2})^{(1)}(s)$ is strictly increasing in $s$. Since $(\phi^{-1}_{\theta})^{(1)}(s) = -s^{-(\theta+1)}$, $(\phi^{-1}_{\theta_1})^{(1)}(s)/(\phi^{-1}_{\theta_2})^{(1)}(s) = s^{(\theta_2-\theta_1)}$. For any $\theta_2>\theta_1$, and $s\in (0,1]$,  $(\phi^{-1}_{\theta_1})^{(1)}(s)/(\phi^{-1}_{\theta_2})^{(1)}(s)$ must be strictly increasing with $s$.

Assumption \ref{ass1}(v): We show boundedness of $\nabla_{\theta}S(t)$ for $t\in (0, \infty)$ and $\theta\in \Theta$ in five steps.

Step (a): We show that $\nabla_{\theta}(\phi^{-1}_{\theta})^{(1)}(\pi(t))$ is bounded. $\nabla_{\theta}(\phi^{-1}_{\theta})^{(1)}(s)$ is bounded if $s$ is bounded  as it has been shown in the context of the proof for Assumption  \ref{ass1}(i)(d) above. Since $\pi(t)$ is bounded in $(0,1)$ for all $t\in (0,\infty)$, $\nabla_{\theta}(\phi^{-1}_{\theta})^{(1)}(\pi(t))$ is bounded.

Step (b): We show that $I(t) = - \int_0^t \nabla_{\theta}(\phi^{-1}_{\theta})^{(1)}(\pi(s))f_t(s)ds$ is bounded at all $t\in(0,\infty)$. It is because $||I(t)|| \leq \int_0^t ||-\nabla_{\theta}(\phi^{-1}_{\theta})^{(1)}(\pi(s))||\times ||f_t(s)||ds$ is bounded, as $||\nabla_{\theta}(\phi^{-1}_{\theta})^{(1)}(\pi(s))||$ in step (a) is bounded, $||f_t(s)||$ is bounded by Assumption \ref{ass2}(ii), and a definite integral of a bounded function is also bounded.

Step (c): We show that $J(t) = - \int_0^t (\phi^{-1}_{\theta})^{(1)}(\pi(s))f_t(s)ds$ is bounded, as $||J(t)|| \leq  \int_0^t ||-(\phi^{-1}_{\theta})^{(1)}(\pi(s))||\times ||f_t(s)||ds$ is bounded  at all $t\in(0,\infty)$. It is because $\log(\pi(t))$ and $f_t(t)$ are bounded at all $t\in(0,\infty)$ as mentioned above. $\phi^{-1(1)}_{\theta}(s) = -s^{-(\theta+1)} $  is also bounded, as it has been discussed in the context of the proof for Assumption \ref{ass1}(i)(d) above.  Note also that $J(t)>0$ as $(\phi^{-1}_{\theta})^{(1)}(s) = -s^{-(\theta+1)}<0$  by definition.

Step (d): We show that $\nabla_{\theta}\phi_{\theta}(J(t))$ is bounded. $\nabla_{\theta}\phi_{\theta}(s)$ is bounded when $s$ is bounded as shown in the context of Assumption \ref{ass1}(i)(b). From step (c) above, it is shown that $J(t)$ is bounded. Hence,  $\nabla_{\theta}\phi_{\theta}(J(t))$ is bounded.

Step (e): We can show from equation (\ref{cge}) that, $\nabla_{\theta}S(t) = \nabla_{\theta}\phi_{\theta}(J(t)) \times I(t)$ and $||\nabla_{\theta}S(t)|| \leq  ||\nabla_{\theta}\phi^{-1}_{\theta}(J(t))|| \times ||I(t)||$, which is bounded in $t \in (0,\infty)$ from steps (b) and (d). \QEDB


\paragraph*{Proof of Proposition \ref{prop1}:} 
We omit $z$ in the proof to ease readability as covariates are not required. Uniqueness of $\chi$ follows directly from Assumption \ref{assS}(i) as $\Lambda(t;\chi_1) = \Lambda(t;\chi_2)$ for all $t$ if and only if $\chi_1 = \chi_2$. $\chi$ is therefore unique for a given $S(t)$.

Next we show that $\theta$ is unique by contradiction. Suppose the same observable distribution of $(X,\delta)$ was generated by two different values of $\theta$, i.e. $\theta_1 \neq \theta_2$. Due to equation (\ref{cge}) and Lemma \ref{l:thetaSPHI}, there were then two different $S_{\theta_1}(t)$ and $S_{\theta_2}(t)$ that were compatible with the observed distribution $f_t(t)$ and $\pi(t)$. Additionally suppose that both $S_{\theta_1}(t)$ and $S_{\theta_2}(t)$ have the same parametric form as given in (\ref{parametric}) such that
\begin{eqnarray}
S_{\theta_1}(t)  = S^*(t;\chi_1) = \exp(-\Lambda(t;\chi_1)) \label{sss}\\
S_{\theta_2}(t) = S^*(t;\chi_2) = \exp(-\Lambda(t;\chi_2)) \non
\end{eqnarray}
\nin for all $t$. $\chi_1\neq \chi_2$ due to Assumption \ref{assS}(i). We show that this is impossible.

By inverting $(\ref{cge})$ and differentiating it with respect to $t$, we have
\begin{eqnarray}
(\phi^{-1}_{\theta})^{(1)}[S_{\theta}(t)]S^{(1)}_{\theta}(t) = -(\phi^{-1}_{\theta})^{(1)}[\pi(t)]f_t(t).\label{pf0}
\end{eqnarray}
Solve this equation for $f_t(t)$, equate it for $\theta_1 \neq \theta_2$, and solve $S^{(1)}_{\theta}(t)$ using (\ref{sss}) to obtain
\begin{eqnarray}
\frac{(\phi^{-1}_{\theta_2})^{(1)}[\pi(t)]}{(\phi^{-1}_{\theta_1})^{(1)}[\pi(t)]}\times  \frac{(\phi^{-1}_{\theta_1})^{(1)}[S_{\theta_1}(t)]}{(\phi^{-1}_{\theta_2})^{(1)}[S_{\theta_2}(t)]}
 &=& \frac{S_{\theta_2}(t) \lambda(t;\chi_{2})}{S_{\theta_1}(t) \lambda(t;\chi_{1})} . \non
\end{eqnarray}


\nin By noting $S_{\theta}(t) \to 1$ and $\pi_{\theta}(t) \to 1$ as $t\to 0^+$, it follows
\begin{eqnarray}
\frac{(\phi^{-1}_{\theta_2})^{(1)}[1]}{(\phi^{-1}_{\theta_1})^{(1)}[1]}\times  \frac{(\phi^{-1}_{\theta_1})^{(1)}[1]}{(\phi^{-1}_{\theta_2})^{(1)}[1]} &=& \lim_{t\to 0^+}\frac{\lambda(t;\chi_{2})}{\lambda(t;\chi_{1})} . \label{pf2}
\end{eqnarray}
\nin The LHS is one as $(\phi^{-1}_{\theta})^{(1)}[1]$ is non-zero and finite due to Lemma \ref{l:inverse}. By Assumption \ref{assS}(ii), $\lambda(t;\chi_{1}) / \lambda(t;\chi_{2})$ is a constant unequal to $1$ at $t\to 0^+$ when $\chi_{1}\neq \chi_{2}$. It leads to contradiction. It is therefore not possible that the same distribution of $(X,\delta)$ can be generated by two different $\theta$ and at the same time $S_{\theta}(t)$ has the same parametric model implied by $S^*(t;\chi)$ with two different values of $\chi$. \QEDB

\paragraph*{Proof of Proposition \ref{prop2}:}
To ease readability, we consider the case $k=1$. Suppose $\theta_1\in\Theta$ and $\theta_2\in\Theta$ are two candidates for $\theta$ such that $S_\theta(t|z)=S^*(t|z)$ for all $t$ and $z$. Due to equation (\ref{cge}) and Lemma \ref{l:thetaSPHI}, there were then two different $S_{\theta_1}(t)$ and $S_{\theta_2}(t)$ that were compatible with the observed distribution $f_t(t)$ and $\pi(t)$. Also, suppose that both $S_{\theta_1}(t)$ and $S_{\theta_2}(t)$ have the same semiparametric form given in (\ref{ph00}) such that the following equalities hold for all $t \in (0,\infty)$ and $z \in \Zscr$:
\begin{eqnarray}
S_{\theta_1}(t|z) = S^*(t|z;\beta_1) =  \exp(-\Lambda_{0}(t)\exp(z\beta_{1})) \label{p2:eq11}, \\
S_{\theta_2}(t|z) = \tilde{S}^*(t|z;\beta_2) =   \exp(-\tilde{\Lambda}_{0}(t)\exp(z\beta_{2})), \label{p2:eq12}
\end{eqnarray}
\nin where $\tilde{\Lambda}_{0}(t)$ can be different from $\Lambda_{0}(t)$ for all $t$ and $\beta_2$ can be different from $\beta_1$ generally.


In the first step, we show that $\beta$ in (\ref{ph00}) is unique, i.e. $\beta_{1} = \beta_{2}$. We evaluate (\ref{pf0}) at  $\theta=\theta_1$ and at $z_1$ and $z_2$ with $z_1\neq z_2$ and take their ratio. By taking the derivative of (\ref{p2:eq11}), $\Lambda_0^{(1)}(t)$ cancels out as by Assumption \ref{assPH}(i) it is non-zero and finite. This gives
\begin{eqnarray}
\frac{f_t(t|z_1)}{f_t(t|z_2)} = \frac{\phi^{-1(1)}_{\theta_1}[\pi(t|z_2)]}{\phi^{-1(1)}_{\theta_1}[\pi(t|z_1)]}\times \frac{\phi^{-1(1)}_{\theta_1}[S_{\theta_1}(t|z_1)]}{\phi^{-1(1)}_{\theta_1}[S_{\theta_1}(t|z_2)]} \frac{S_{\theta_1}(t|z_1)}{S_{\theta_1}(t|z_2)}  \frac{\exp(z_1\beta_{1})}{\exp(z_2\beta_{2})}. \non
\end{eqnarray}
\nin
\nin At the limit $t\to 0^+$, we have $S_{\theta_1}(t|z_j) \to 1$, $\pi(t|z_j) \to 1$ for $j=1,2$. Without loss of generality $z$ can be recoded to take on the value of $0$. By choosing $z_2=0$, the above equation simplifies to
\begin{eqnarray}
\lim_{t\to 0^+} \frac{f_t(t_1|z_1)}{f_t(t_1|z_2)}  & = &  \exp(z_1\beta_{1}). \label{p2:eq21}
\end{eqnarray}
\nin We repeat the same steps for the case of $\theta_2$ in (\ref{p2:eq12}) and obtain
\begin{eqnarray}
\lim_{t\to 0^+} \frac{f_t(t_1|z_1)}{f_t(t_1|z_2)}  & = & \exp(z_1\beta_{2}). \label{p2:eq22}
\end{eqnarray}
\nin By equating (\ref{p2:eq21}) and  (\ref{p2:eq22}), we  have $\beta_{1}=\beta_{2} = \beta$.  There is a unique $\beta$ in model (\ref{ph00}).

In the second step we prove that $\theta$ is unique. Since $\beta$ is unique, we can rewrite (\ref{p2:eq11}) as
\begin{eqnarray}
S_{\theta_1}(t|z) = S^*(t|z;\beta) =  \exp(-\Lambda_0(t)\exp(z\beta)). \label{p2:eq16}
\end{eqnarray}
\nin Assumption \ref{assPH}(i)  guarantees that there exists a unique  $t_1 \in (0,\infty)$ such that $\Lambda_0(t_1) = c$ for some positive and finite constant $c$. 
From (\ref{p2:eq16})  we have for all $z$
\begin{eqnarray}
S_{\theta_1}(t_1|z) = \exp(-c\exp(z\beta)). \label{p2:eq3}
\end{eqnarray}
\nin The RHS, which is implied by the marginal survival model in Assumption \ref{assPH}, is a fixed value at $t_1$ for any $z \in \Zscr$. The LHS, which is implied by the copula model in Assumption \ref{ass1}, is a strictly decreasing in $\theta$ at $t_1$ for any $z \in \Zscr$ according to Lemma \ref{l:thetaSPHI}. Therefore, $\theta_1$ in the copula model is unique, i.e.  $\theta_1=\theta_2$.

We prove $\Lambda_0(t)$ is unique by taking $z=0$ in (\ref{p2:eq16}) so that $S_{\theta_1}(t|z=0)= \exp(-\Lambda_0(t))$. Since $\theta_1$ is unique, the LHS is unique for all $t\in(0,\infty)$. Therefore, the RHS, in essence $\Lambda_0(t)$, must also be unique for all $t\in(0,\infty)$: $\Lambda_0(t)=\tilde{\Lambda}_0(t)$. \QEDB

\paragraph*{Proof of Lemma \ref{propchi}:} Part (i) requires two ways: (a) $\hat{w}_{i}$ converges in probability to $w_{i} = (-1, -z_i, S_W^{-1}[S_{\theta}(x_i|z_i)])$ with any $\theta$, such that the estimated regressor $\hat{w}_{i}$ is a consistent estimate for the true value of $w_{i}$. Therefore
the estimator $\hat{\chi}$ relying on the estimated $\hat{w}_{i}$ converges in probability to the estimator $\hat{\chi}^{*}$ using the true $w_{i}$, i.e. $\hat{\chi}\overset{p}{\to} \hat{\chi}^{*}$; and (b) $\hat{\chi}^{*}$ is a consistent estimator for the true $\chi_{0}$, i.e. $\hat{\chi}^{*} \overset{p}{\to} \chi_{0}$. When (a) and (b) hold, $\hat{\chi}\overset{p}{\to} \hat{\chi}^{*} \overset{p}{\to} \chi_{0}$. We first prove part (a). We have $\hat{S}_{CGE}(x_i|z_i;\hat{\eta};\theta) \overset{p}{\to}  S_{\theta}(t_i|z_i)$ for any $\theta$ under Assumption \ref{ass3}, which implies that $\hat{w}_{i} \overset{p}{\to} w_{i}$ and $\hat{\chi}\overset{p}{\to} \hat{\chi}^{*}$ from Theorem D.16 in Greene (2012). For part (b), Assumption \ref{ass4} meets the conditions in Theorem 7.3 of Wooldridge (2010), and hence $\hat{\chi}^{*} \overset{p}{\to} \chi_0$. This completes the proof of Lemma \ref{propchi}(i). Part (ii) requires that $\hat{\chi}^{*}$ is asymptotically normal distributed given the known $w_{i}$, which is a direct result of part (i), Assumption \ref{ass3} and Theorem 7.3 of Wooldridge (2010).\QEDB

\paragraph*{Proof of Lemma \ref{proptheta}:} From Theorem D.16 in Greene (2012), $S_{AFT}(x|z;\hat{\chi}) \overset{p}{\to} S_{AFT}(x|z;\chi_{0})$, since $\hat{\chi} \overset{p}{\to} \chi_{0}$ by Lemma \ref{propchi}(i). Similarly $\hat{S}_{CGE}(x|z;\hat{\eta}; \theta) \overset{p}{\to} \hat{S}_{CGE}(x|z;\eta_0; \theta)$ for any $\theta \in \Theta$, since $\hat{\eta} \overset{p}{\to} \eta_0$ by Assumption \ref{ass2}. Therefore, for Lemma \ref{proptheta} to hold, it suffices to show (i) consistency and (ii) asymptotically normality of $\hat{\theta}^*$ that solves $m_{3n}(\theta;\eta_0,\chi_0)$.

\nin I.) We prove part (i) first. From Theorem 2.1 of Newey and McFadden (1994), $\hat{\theta}^*  \overset{p}{\to}  \theta_0$, if (a) the objective function $Q_0(\theta)= -E\{S_{AFT}(x|z;\chi(\theta))-\hat{S}_{CGE}(x|z;\theta,\eta)\}^2$ is uniquely maximised at $\theta_0$; (b) $\Theta$ is compact; (c) $Q_0(\theta)$ is continuous; and (d) $\hat{Q}_n(\theta)= -\frac{1}{n}\sum_{i=1}^n \{S_{AFT}(x_i|z_i;\hat{\chi}(\theta))-\hat{S}_{CGE}(x_i|z_i;\theta,\eta)\}^2$ converges uniformly to $Q_0(\theta)$. We prove these four conditions subsequently.

\nin (a) follows from Proposition \ref{prop1}.  
(b) follows from Assumption \ref{ass1}(\ref{a1:1}).
(c) and (d) follow from Lemma 2.4 of Newey and McFadden (1994), which requires (1) $g(x_i,\theta) = S_{AFT}(x_i|z_i;\chi(\theta))-\hat{S}_{CGE}(x_i|z_i;\eta,\theta)$ is continuous at each $\theta\in \Theta$, and (2) $||g(x,\theta)|| \leq d(x)$ for all $\theta\in \Theta$, and $E[d(x)]<\infty$. Due to the fact that the sum of a finite number of continuous functions is a continuous function, Lemma 2.4(1) requires that $S_{AFT}(x_i|z_i;\chi(\theta))$ and $\hat{S}_{CGE}(x_i|z_i;\theta,\eta)$ are continuous. From (\ref{aft01}), it can be seen that $S_{AFT}(x|z;\chi(\theta))$ is continuous in all $x>0$ and for all $\theta \in \Theta$ as $S_W(w)$ is continuous in $w$ and $w=\log(\lambda x \exp(z'\beta))^{\sigma}$ is continuous in $x$ for $x>0$. From (\ref{scge}), $\hat{S}_{CGE}(x_i|z_i;\eta,\theta)$ is continuous given that $\phi_{\theta}$, $(\phi^{-1}_{\theta})^{(1)}$, $\pi$ and $f_t$ are all continuous functions, as stated in Assumption \ref{ass1}(\ref{a1:1}) and Assumption \ref{ass2}(i). These observations prove Lemma 2.4(1). Lemma 2.4(2)holds, because $S_{AFT}(x_i|z_i;\chi(\theta))$ and $\hat{S}_{CGE}(x_i|z_i;\eta,\theta)$ are both bounded in $[0,1]$ by definition. Hence, $||g(x_i,\theta)|| = ||S_{AFT}(x_i|z_i;\chi(\theta) - \hat{S}_{CGE}(x_i|z_i;\eta, \theta)|| \leq ||d(x)||$, with $d(x) = 1$. To summarise, conditions (c) and (d) are met.

\nin II.) Next, we prove part (ii). From Theorem 3.4 of Newey and McFadden (1994), $\sqrt{n}(\hat{\theta}-\theta_0) \overset{d}{\to} N(0,\Omega_3)$, if (a) $\hat{\theta}  \overset{p}{\to} \theta_0$; (b) $\theta_0 \in$ interior of $\Theta$; (c) $m_{3n}(\theta;\eta_0,\chi_0)$ is continuously differentiable in a neighborhood $\mathcal{N}$ of $\theta_0$; (d) $E(m_{3n}(\theta;\eta_0,\chi_0)=0$ and $E(||m_{3n}(\theta;\eta_0,\chi_0)||^2)$ is finite; (e) $E[\sup_{\theta \in \mathcal{N}}||\nabla_{\theta}m_{3n}(\theta;\eta_0,\chi_0)||]$ is finite; (f) for $M =E[\nabla_{\theta}m_{3n}(\theta;\eta_0,\chi_0)]$, $M'M$ is nonsingular. We prove these conditions subsequently.

\nin (a) comes from part (i). (b) is fulfilled by Assumption \ref{ass1}(\ref{a1:1}). (c) is met when both $S_{AFT}$ and $\hat{S}_{CGE}$ are differentiable w.r.t. $\theta$. For $\hat{S}_{CGE}$ in (\ref{scge}), it is differentiable w.r.t. $\theta$ according to Assumption \ref{ass1}(\ref{a1:1}). For $S_{AFT} = S_W(\log([\lambda x_i\exp(z_i\beta)]^{\sigma}))$ in (\ref{aft01}), it is a continuous function of the parameter $\chi=(\alpha,\beta,\sigma)'$ where $\chi= E(w_{i}'\Omega_{\epsilon}^{-1}w_{i})^{-1}(w_{i}'\Omega_{\epsilon}^{-1}\log(x_i))$, which is also continuous in $w_{i}$. And, $w_{i}=(-1, -z_i, S_W^{-1}(\hat{S}_{CGE}(x_i|z_i;\eta,\theta)))$ is continuous in $\hat{S}_{CGE}(x_i|z_i;\eta,\theta)$, which is differentiable w.r.t. $\theta$ for the reason mentioned above. To summarise, $S_{AFT}$ is differentiable w.r.t. $\theta$.

\nin (d) $E(m_{3n}(\theta;\eta_0,\chi_0))=0$ is the result of Proposition \ref{prop1}. $E(||m_{3n}(\theta;\eta_0,\chi_0)||^2)$ is obvious as $||S_{AFT}(x_i;\theta_0)||<1$ and $||\hat{S}_{CGE}(x_i|z_i;\eta,\theta)||<1$ for all $i$.

\nin (e) To prove $E||\nabla_{\theta}m_{3n}(\theta;\eta_0,\chi_0)||$ is finite for all $\theta$, one can show that all elements in  $\nabla_{\theta}S_{AFT}(x_i|z_i;\chi_0)$ and $\nabla_{\theta} \hat{S}_{CGE}(x_i|z_i;\eta,\theta)$ are finite. For $\hat{S}_{CGE}$, it follows from Assumption \ref{ass1}(\ref{a1:1}). For $S_{AFT}$, it requires that $S_W$ is differentiable, which holds for AFT models, and $E[\nabla_{\theta}\chi] = E[\nabla_{\theta}\lambda, \nabla_{\theta}\beta, \nabla_{\theta}\sigma]'$ exist and are bounded, which can be checked by expanding $E[\nabla_{\theta} \chi] = E[A+B]$, where $A= \nabla_{\theta}[(w_{i}'\Omega_{\epsilon}^{-1}w_{i})^{-1}][w_{i}'\Omega_{\epsilon}^{-1}\log(x_i)]$ and $B= [w_{i}'\Omega_{\epsilon}^{-1}w_{i}]^{-1}\nabla_{\theta}[w_{i}'\Omega_{\epsilon}^{-1}\log(x_i)]$. Because $\nabla_{\theta}(G^{-1}) = - G^{-1}(\nabla_{\theta}G)G^{-1}$, and $\nabla_{\theta}[w_{i}'\Omega_{\epsilon}^{-1}w_{i}]$ = $[0, 0, D]'$ with $D =2\sum_i\sum_j S_W^{-1'}(\nabla_{\theta} \hat{S}_{CGE,i})S_W^{-1'}(\nabla_{\theta} \hat{S}_{CGE,j})]'$, each element in $E[\nabla_{\theta}\chi]$ is a linear combination of $E[D$], $E[w_{i}'\Omega_{\epsilon}^{-1}w_{i}]$ and $E[w_{i}'\Omega_{\epsilon}^{-1}\log(x_i)]$. Under Assumption \ref{ass4}, $\Omega_{\theta}$ is non-singular, together with the assumption that $z$, $x$, $\hat{S}_{CGE}(x_i|z_i;\eta,\theta)$ and $\nabla_{\theta}\hat{S}_{CGE}(x_i|z_i;\eta,\theta)$  are finite, we know that $E[\nabla_{\theta}\chi]$ is finite.

\nin (f) $S_{AFT}(x|\cdot)$ and $\hat{S}_{CGE}(x|\cdot)$, and thus $\nabla_{\theta}S_{AFT}(x|\cdot)$ and $\nabla_{\theta}\hat{S}_{CGE}(x|\cdot)$, are not linearly dependent for all $x$. Therefore $M'M = E(\nabla_{\theta}S_{AFT}(x|\cdot)-\nabla_{\theta}\hat{S}_{CGE}(x|\cdot))'(\nabla_{\theta}S_{AFT}(x|\cdot)-\nabla_{\theta}\hat{S}_{CGE}(x|\cdot))$ is non-singular. \QEDB

\paragraph*{Proof of Lemma \ref{propsemi}:} The proof follows closely the proof of Lemma \ref{proptheta}. It suffices to prove the consistency and asymptotic normality of $\hat{\theta}^*$ that solves the corresponding moment function $m_{5n}(\theta; \eta_0)$ with probability one.

\nin I.) To prove part (i), we check the four conditions of Theorem 2.1 of Newey and McFadden (1994). For (a), the objective function $Q_0(\theta)= -E[I_k'(\beta_{i}(\theta)- \beta^*(\theta))(\beta_{i}(\theta)- \beta^*(\theta))'I_k$ is uniquely maximised at $\theta_0$ due to Proposition \ref{prop2}.  Condition (b) follows from Assumption  \ref{ass1}(\ref{a1:1}). Condition (c) and (d) follows from the conditions (1) and (2) of Lemma 2.4 of Newey and McFadden (1994). For Lemma 2.4(1), $m_{5n}(\theta, \eta_0)$ is continuous, as $\hat{S}_{CGE}$ is continuous in $x_i$ for all $x_i >0$ and due to the fact that the product of a finite number of continuous functions is a continuous function, $\hat{\beta}_{i}(\theta)$ in (\ref{ebeta}) is also continuous in $x_i$. By being a linear combination of $\hat{\beta}_{i}(\theta)$, $m_{5n}(\theta; \eta_0)$ is also continuous. For Lemma 2.4(2), $m_{5n}(\theta; \eta_0)$ is bounded by some $d(x)$ for all $t\in(0,\infty)$. It is fulfilled as in the data sample the minimum $x_i$ is greater than zero while the maximum $x_i$ is less than infinity. Specifically, let $x_{min} = \min\{X_i\}$ and $x_{max} = \max\{X_i\}$, $\hat{\beta}_{i}(\theta)$ in (\ref{ebeta}) is bounded by $d_{i}(x) = \max\{||\log[\log(S(x_{min}))/\log(S(x_{max}))]||, ||\log[\log(S(x_{max}))/\log(S(x_{min}))]||\} + ||z_1-z_2||$. By being a linear combination of $d_{i}(x)$, $m_{5n}(\theta; \eta_0)$ is also bounded. To sum up, conditions (c) and (d) are met.

\nin II.) To prove part (ii), we check the conditions of Theorem 3.4 of Newey and McFadden (1994). (a) $\hat{\theta}^*  \overset{p}{\to} \theta_0$ is the result of Lemma \ref{propsemi}(i); (b) is fulfilled by Assumption \ref{ass1} (\ref{a1:1}); (c) $m_{5n}(\theta,\eta_0)$ is continuously differentiable in a neighborhood $\mathcal{N}$ of $\theta_0$. It can be shown by taking differentiation of $\beta$, which is $\nabla_{\theta}\beta = \nabla_{\theta}[\hat{S}_{CGE}(t|z_1)]/[\hat{S}_{CGE}(t|z_1)\log(\hat{S}_{CGE}(t|z_1))]- \nabla_{\theta}[\hat{S}_{CGE}(t|z_2)]/[\hat{S}_{CGE}(t|z_2)\log(S_{CGE}(t|z_2))]$. First note that $\nabla_{\theta}[S_{CGE}(t|z_1)]$ and $\nabla_{\theta}[S_{CGE}(t|z_2)]$ are both differentiable as is mentioned above. Next, $||S_{CGE}(t|z_k)\log(S_{CGE}(t|z_k))|| < \infty$ for $t\in (0,\infty)$ for $k=1,2$ as $\hat{S}_{CGE}$ and $\log(\hat{S}_{CGE})$ are bounded; (d) $E(m_{5n}(\theta; \eta_0)=0$ because of Proposition \ref{prop2}. $E(||m_{5n}(\theta; \eta_0)||^2)$ is finite as $||z_1||$ and  $||z_2||$ are finite, $\hat{S}_{CGE}(t|z_k)$ is bounded and $\log(\hat{S}_{CGE}(t|z_k))$ is also bounded for $t\in(0,\infty)$. (e) $E[\sup_{\theta \in \mathcal{N}}||\nabla_{\theta}m_{5n}(\theta; \eta_0)||]$ is finite when $\nabla_{\theta} \hat{S}_{CGE}(x_i,z_i;\theta)$ is bounded for all $x_i$ and for all $\theta \in \Theta$. (f)
For $M =E[\nabla_{\theta}m_{5n}(\theta; \eta_0)]$, $M'M$ is nonsingular  as long as $z_1\neq z_2$. \QEDB

\newpage

\section*{S.II: AFT model estimation: Examples}

\example{\textbf{(Weibull Model)}: $W$ has an extreme value distribution, $S_{W}(w)= \exp(-\exp(w))$ and $S_W^{-1}(s) = \log(-\log(s))$, and (\ref{aft1}) becomes
\begin{eqnarray}
\log(t) &=& -\log(\alpha) - z'\beta + (1/\sigma)\log(-\log(S^*(t|z))). \non
\end{eqnarray}
\nin Rewrite it as
\begin{eqnarray}
- \log S^*(t|z)  = & \Lambda(t|z;\chi) & = (\alpha t \exp{(z'\beta)})^{\sigma}, \non
\end{eqnarray}
\nin with $\alpha, \sigma>0$, $\beta\in \mathbb{R}$ and $\chi=(\alpha, \sigma, \beta')'$. Assumption \ref{assS}(i) is satisfied as $\alpha_{1}^{\sigma_{1}}t^{\sigma_{1}}[\exp(z'\beta_{1})]^{\sigma_{1}}$ and $\alpha_{2}^{\sigma_{2}}t^{\sigma_{2}}[\exp(z'\beta_{2})]^{\sigma_{2}}$ are identical for all $t$ and $z$ if and only if $\alpha_1 = \alpha_2$, $\beta_1=\beta_2$ and $\sigma_1=\sigma_2$. Next,  $\lambda(t|z;\chi) = \sigma \alpha^{\sigma} t^{\sigma-1} [\exp(z'\beta)]^{\sigma}$. Assume without loss of generality $\sigma_{2} > \sigma_{1}$. This gives
\[\frac{\lambda(t|z;\chi_2)}{\lambda(t|z;\chi_1)} = \frac{\sigma_{2} \alpha_{2}^{\sigma_{2}}[\exp(z'\beta_{2})]^{\sigma_{2}}}{\sigma_{1} \lambda_{1}^{\sigma_{1}}[\exp(z'\beta_{1})]^{\sigma_{1}}} t^{\sigma_{1}-\sigma_{1}}.\]
\nin which is zero for $t=0$. Therefore, the restriction of Assumption \ref{assS}(ii) is satisfied.}

\example{\textbf{(Log-logistic Model)}: $W$ has logistic distribution $S_W(w) = 1/(1+\exp(w))$ and $S_W^{-1}(s) = \log(\frac{1-s}{s})$. Equation
(\ref{aft1}) becomes
\begin{eqnarray}
\log(t) &=& -\log(\alpha) - z\beta + (1/\sigma) \log\left(\frac{1-S^*(t|z)}{S^*(t|z)}\right), \non
\end{eqnarray}
\nin which can be rewritten as
\begin{eqnarray}
- \log S^*(t|z)  = & \Lambda(t|z;\chi) & = \log [1+ (\alpha t \exp{(z'\beta)})^{\sigma}], \non
\end{eqnarray}
\nin with $\alpha, \sigma>0$, $\beta\in \mathbb{R}$ and $\chi=(\alpha, \sigma, \beta')'$. It is clear that Assumption \ref{assS}(i) is satisfied. Assume without loss of generality $\sigma_{2} > \sigma_{1}$
\begin{eqnarray}
\frac{\lambda(t|z;\chi_2)}{\lambda(t|z;\chi_1)} &=& \left(\frac{\sigma_{2} (\alpha_{2}\exp{(z'\beta_{2}))^{\sigma_{2}}}}{1+ (\alpha_{2}t\exp{(z'\beta_{2})})^{\sigma_{2}} } \left/ \frac{\sigma_{1}(\alpha_{1}\exp{(z'\beta_{1}))^{\sigma_1}}}{1+ (\alpha_{1}t\exp{(z'\beta_{1})})^{\sigma_{1}}}\right) t^{\sigma_{2}-\sigma_{1}} \right. , \non
\end{eqnarray}
which is zero for $t=0$. Assumption \ref{assS}(ii) therefore holds.}


\example{\textbf{(Log-normal)}:  $W$ has standard normal distribution with $S_W(w) = 1-\Phi(w)$ and  $S^{-1}_W(s) = \Phi^{-1}(1-s)$. Equation
(\ref{aft1}) becomes
\begin{eqnarray}
\log(t) &=& -\log(\alpha) - z'\beta + (1/\sigma) \Phi^{-1}[1-S^*(t|z)],\non
\end{eqnarray}
\nin which can be rewritten as
\begin{eqnarray}
- \log S^*(t|z)  = & \Lambda(t|z;\chi) & = - \log [ 1- \Phi (\log (\alpha t \exp{(z'\beta)})^{\sigma})] \non
\end{eqnarray}
\nin with $\alpha, \sigma>0$, $\beta\in \mathbb{R}$ and $\chi=(\alpha, \sigma, \beta')'$. It is clear that Assumption \ref{assS}(i) is satisfied.
\[\lambda(t|z;\chi) = \frac{ \phi( \log (\alpha t \exp{(z'\beta)})^{\sigma})}{ 1 -\Phi (\log (\alpha t \exp{(z'\beta)})^{\sigma})}\sigma t^{-1} = \varsigma[\log (\alpha t \exp{(z'\beta)})^{\sigma}] \sigma t^{-1},\]
\nin where $\varsigma(s)$ is the inverse-mill ratio. At $s\to -\infty$, $\varsigma(s) \to 0$ and $\varsigma^{(1)}(s) \to 1$. Take the ratio of the hazards for any $\chi_1\neq \chi_2$, and by L'H\^{o}pital's rule,
\[\lim_{t\to 0^+} \frac{\lambda(t|z;\chi_2)}{\lambda(t|z;\chi_1)} =\lim_{t\to 0^+} \frac{\varsigma[\log (\alpha_{2} t \exp{(z'\beta_{2})}^{\sigma_{2}})] \sigma_{2}}{\varsigma[\log (\alpha_{1} t \exp{(z'\beta_{1})}^{\sigma_{1}})] \sigma_{1}} \to  \frac{\sigma^2_{2}}{\sigma^2_{1}}.\]
\nin It is not equal to one for any $\sigma_{1}\neq \sigma_{2}$.}



\newpage

\section*{S.III: Figures}
\begin{figure}[h]
     \centering
     \begin{subfigure}[b]{0.45\textwidth}
         \centering
         \includegraphics[width=\textwidth]{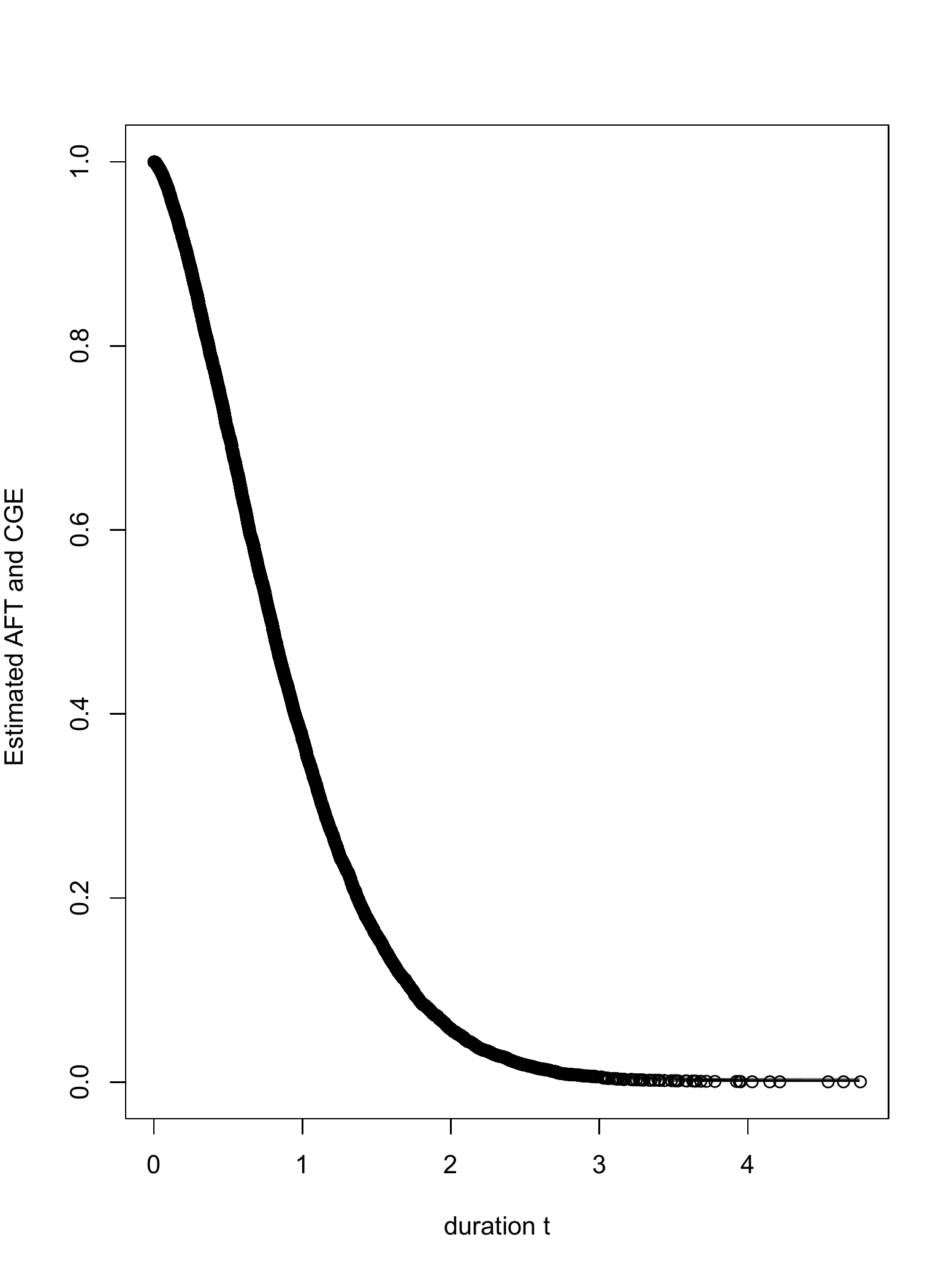}
         \caption{$\theta = 8$}
         \label{fig:obj03}
     \end{subfigure}
     \begin{subfigure}[b]{0.45\textwidth}
         \centering
         \includegraphics[width=\textwidth]{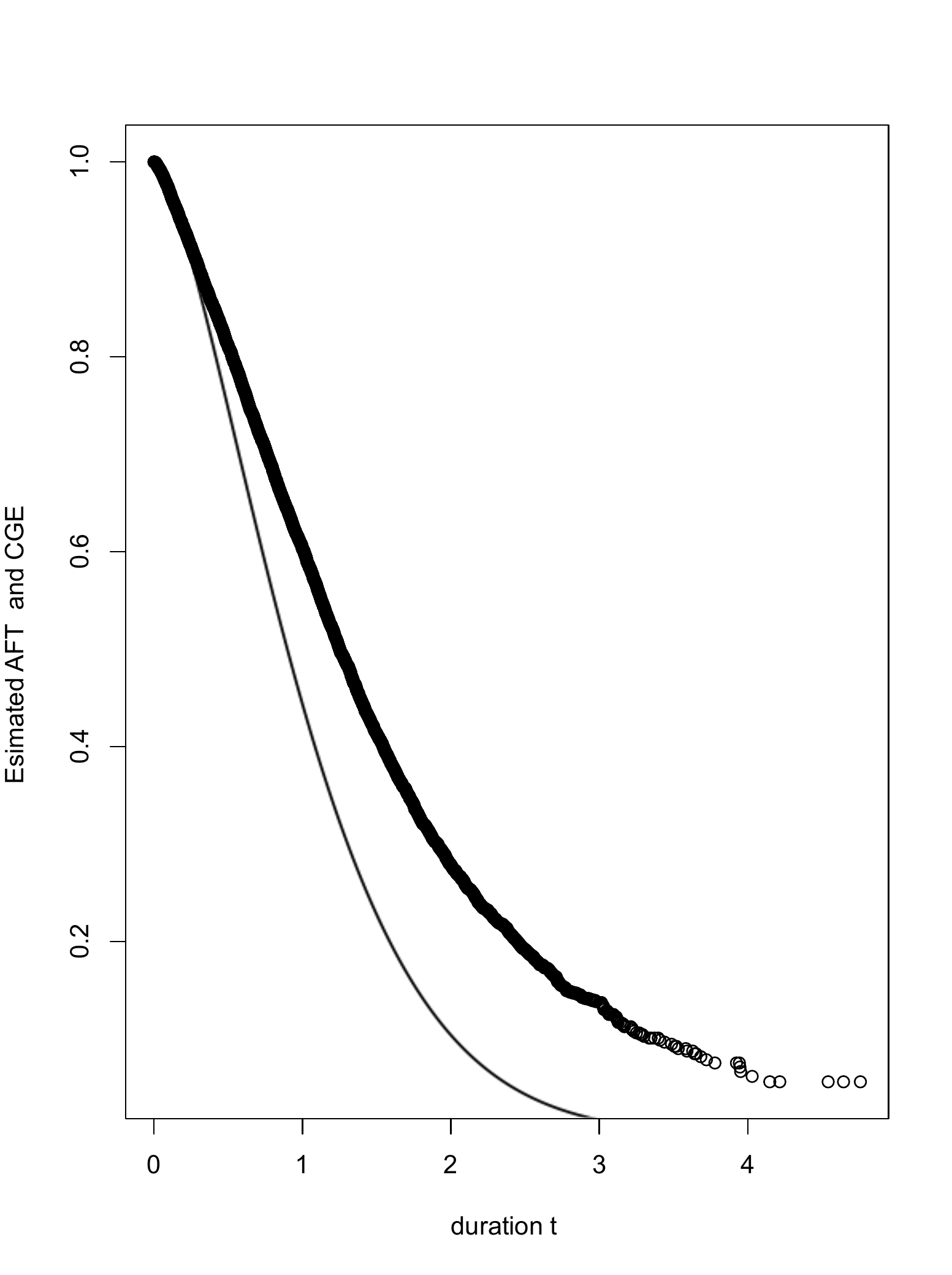}
         \caption{$\theta = 0$}
         \label{fig:obj08}
     \end{subfigure}
        \caption{The estimated $S_{AFT}$ (black line) and $\hat{S}_{CGE}$ (black circles) with  $\theta_0 = 8$.}
        \label{fig:ident}
\end{figure}

\begin{figure}[h!]
     \centering
     \begin{subfigure}[b]{0.3\textwidth}
         \centering
         \includegraphics[width=\textwidth]{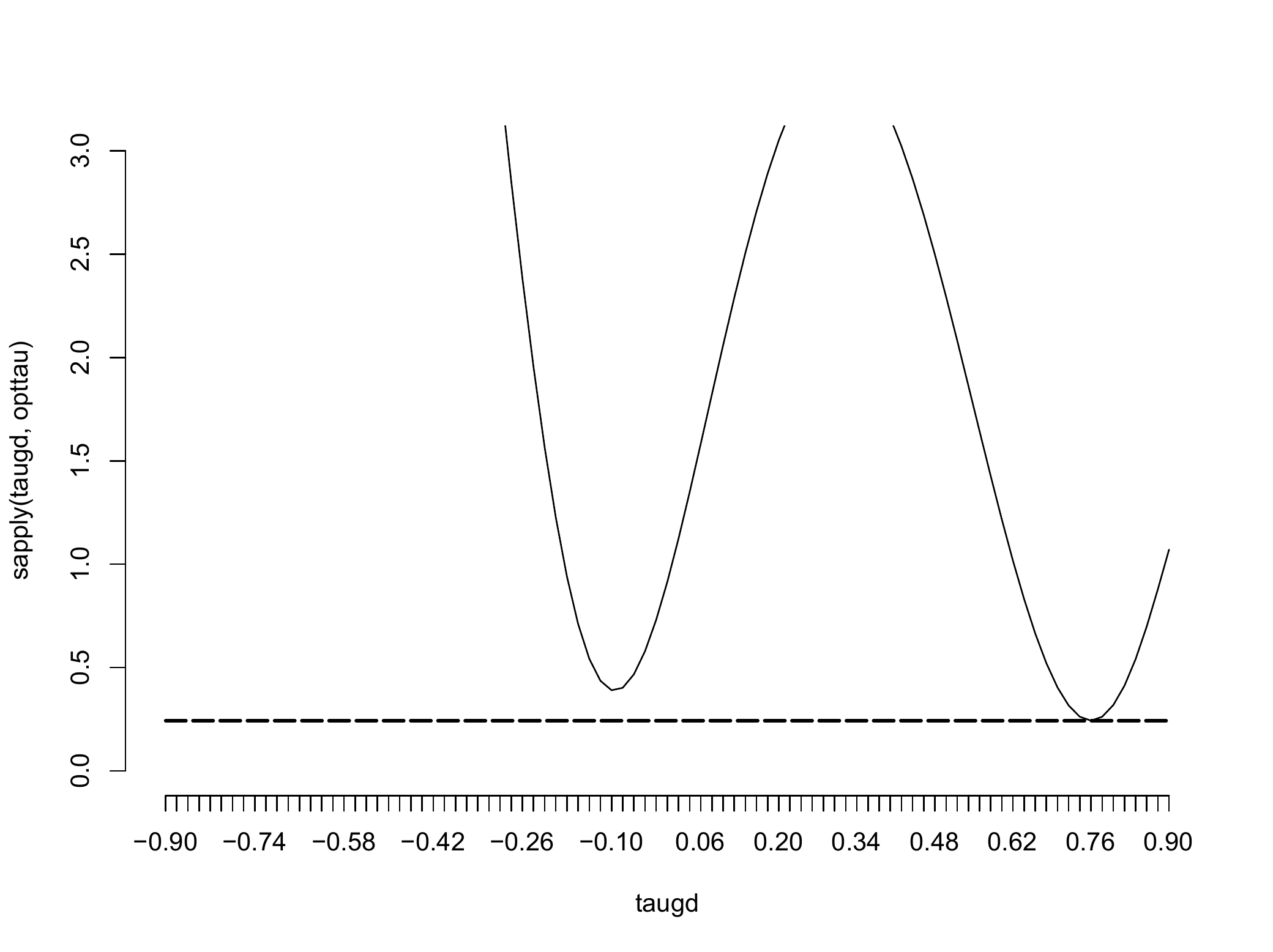}
         \caption{$\tau_0$ = 0.8}
         \label{a}
     \end{subfigure}
     \hfill
     \begin{subfigure}[b]{0.3\textwidth}
         \centering
         \includegraphics[width=\textwidth]{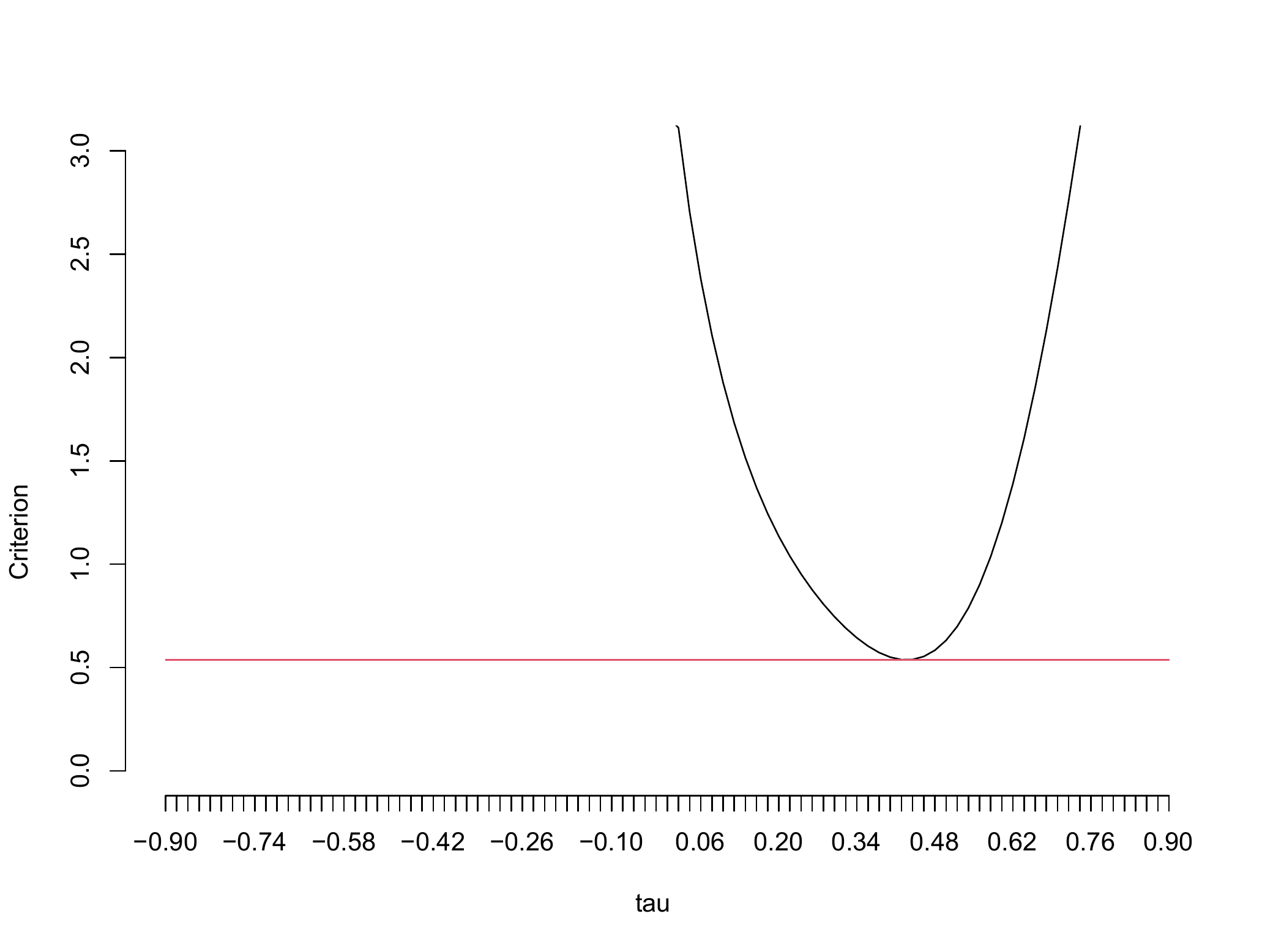}
         \caption{$\tau_0$ = 0.4}
         \label{b}
     \end{subfigure}
     \hfill
     \begin{subfigure}[b]{0.3\textwidth}
         \centering
         \includegraphics[width=\textwidth]{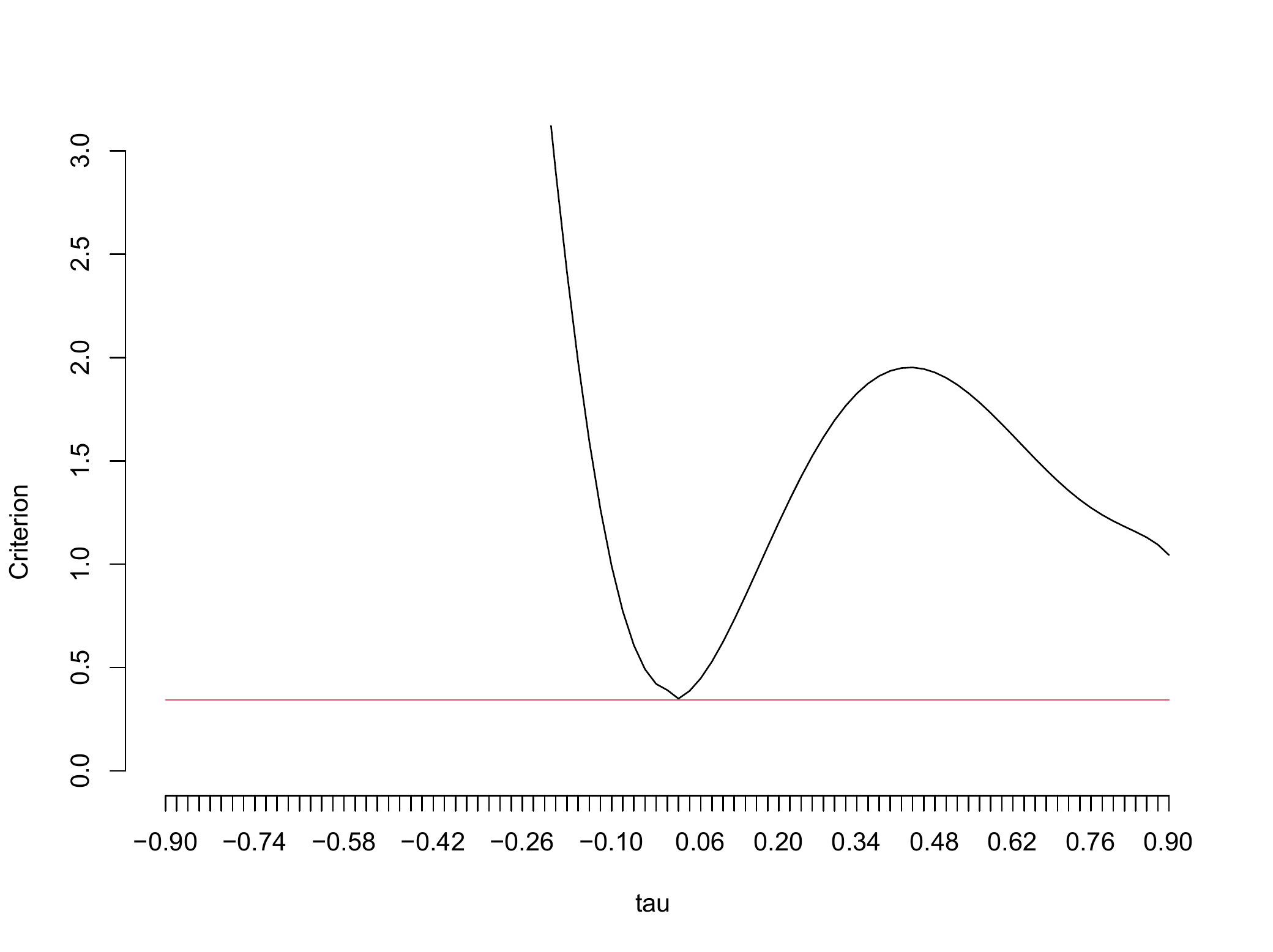}
         \caption{$\tau_0$ = 0}
         \label{c}
     \end{subfigure}
		   \hfill
		\begin{subfigure}[b]{0.3\textwidth}
         \centering
         \includegraphics[width=\textwidth]{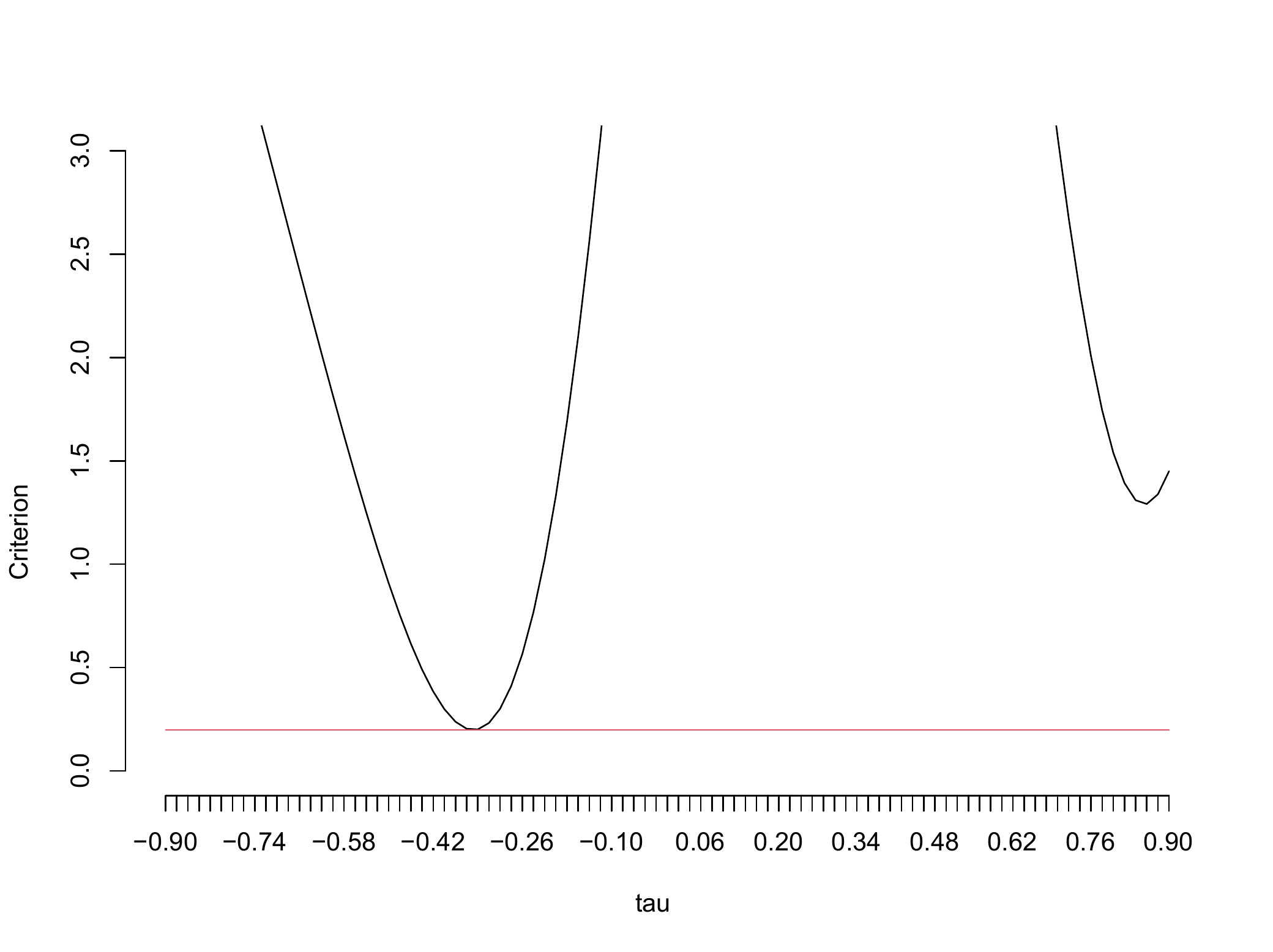}
         \caption{$\tau_0$ = -0.4}
         \label{a}
     \end{subfigure}
		   \hfill
		   \begin{subfigure}[b]{0.3\textwidth}
         \centering
         \includegraphics[width=\textwidth]{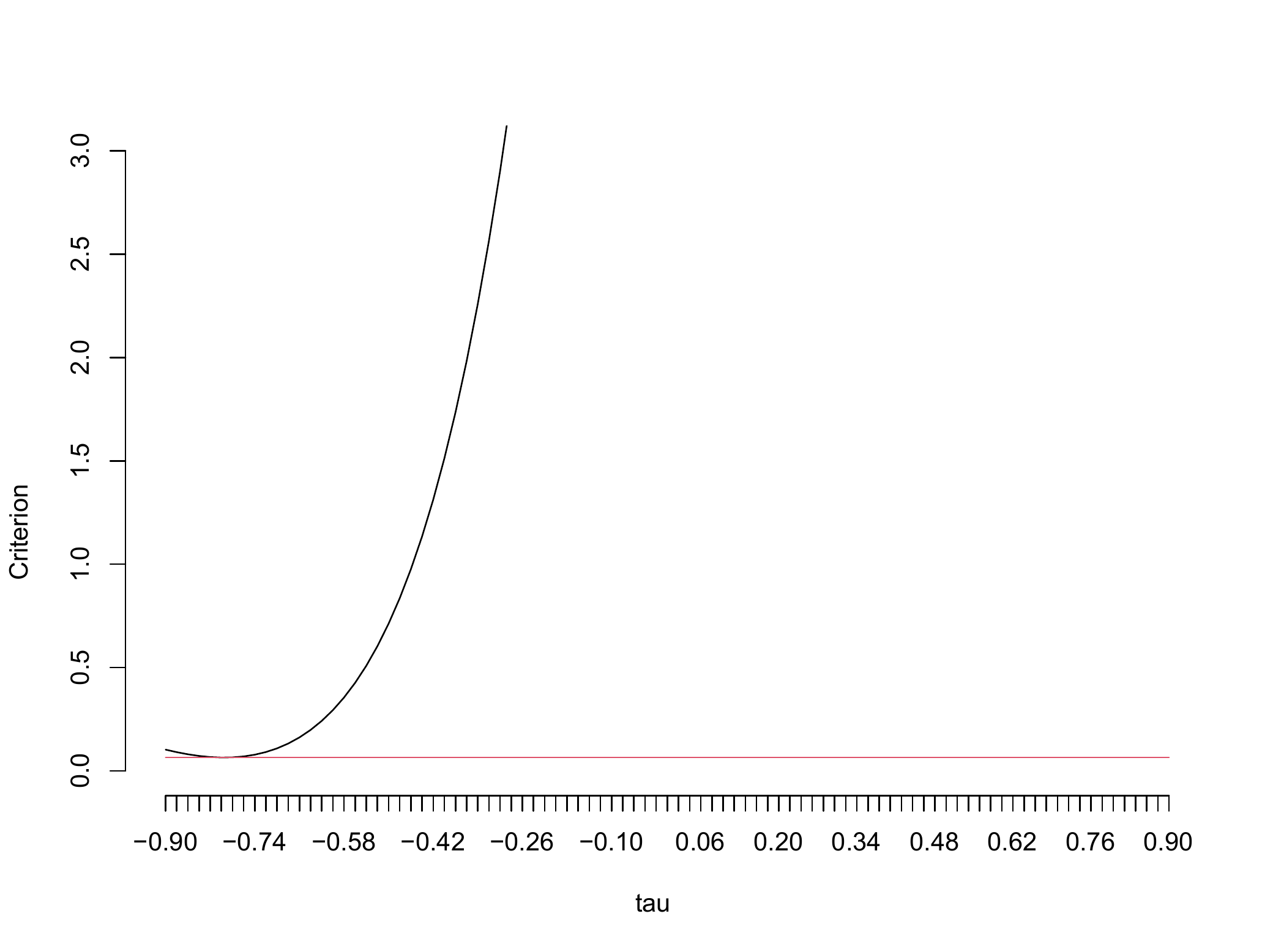}
         \caption{$\tau_0$ = -0.8}
         \label{b}
     \end{subfigure}
        \caption{Objective function (\ref{estaft}) for the Weibull model with different Kendall's $\tau_0$.}
        \label{fig:sen1}
\end{figure}


\begin{figure}[h!]
     \centering
     \begin{subfigure}[b]{0.3\textwidth}
         \centering
         \includegraphics[width=\textwidth]{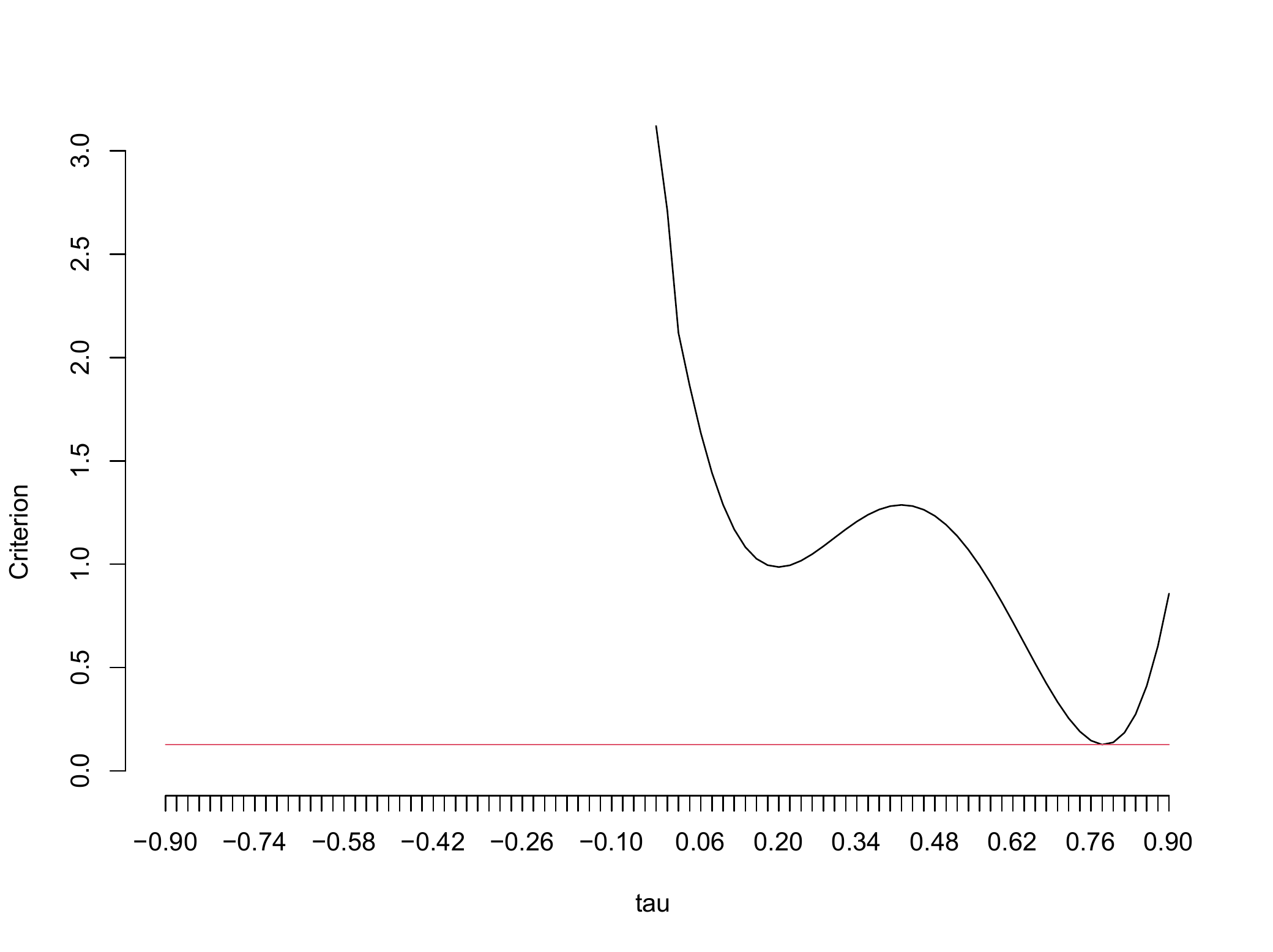}
         \caption{$\tau_0$ = 0.8}
         \label{a}
     \end{subfigure}
     \hfill
     \begin{subfigure}[b]{0.3\textwidth}
         \centering
         \includegraphics[width=\textwidth]{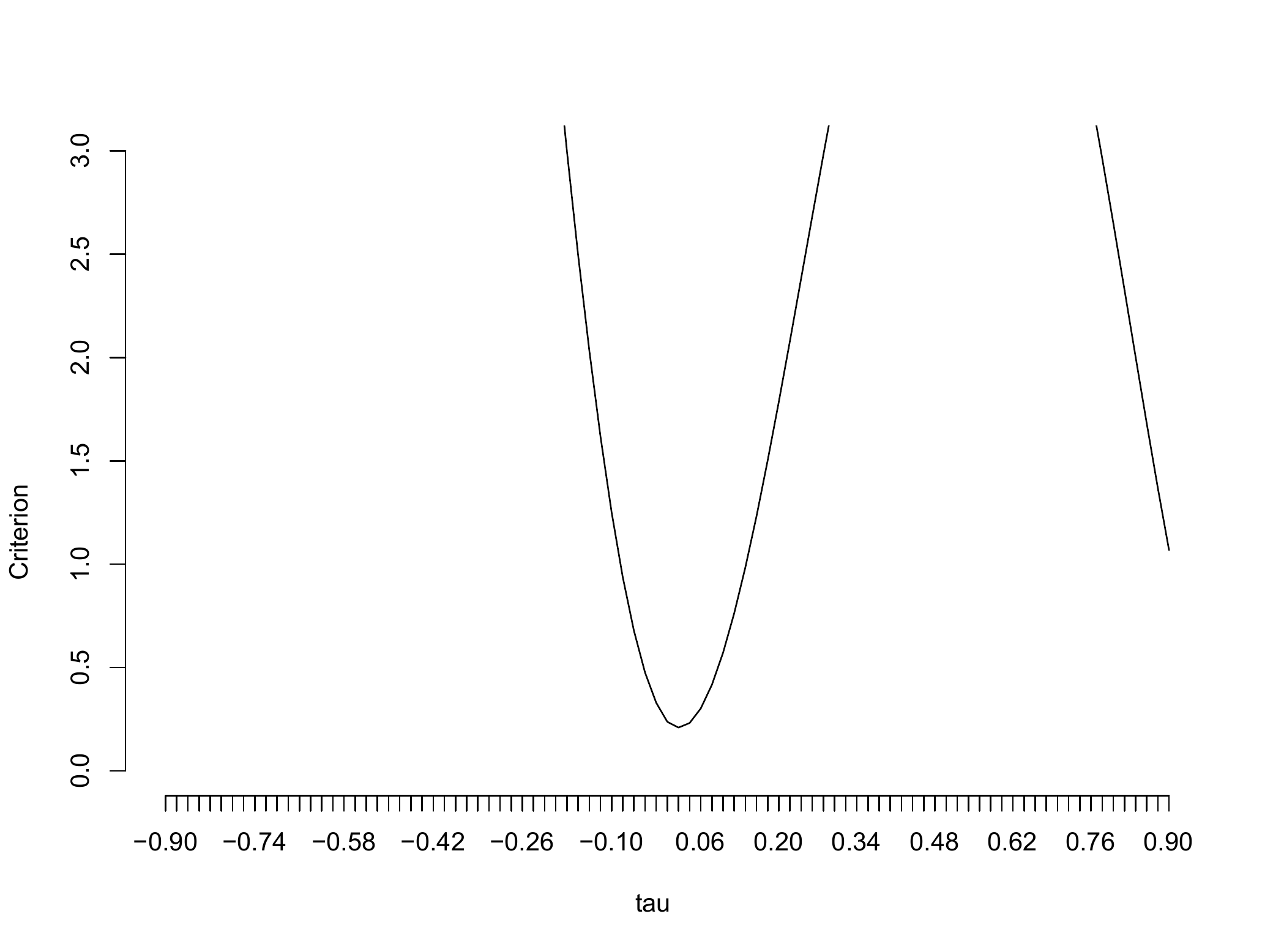}
         \caption{$\tau_0$ = 0.4}
         \label{b}
     \end{subfigure}
     \hfill
     \begin{subfigure}[b]{0.3\textwidth}
         \centering
         \includegraphics[width=\textwidth]{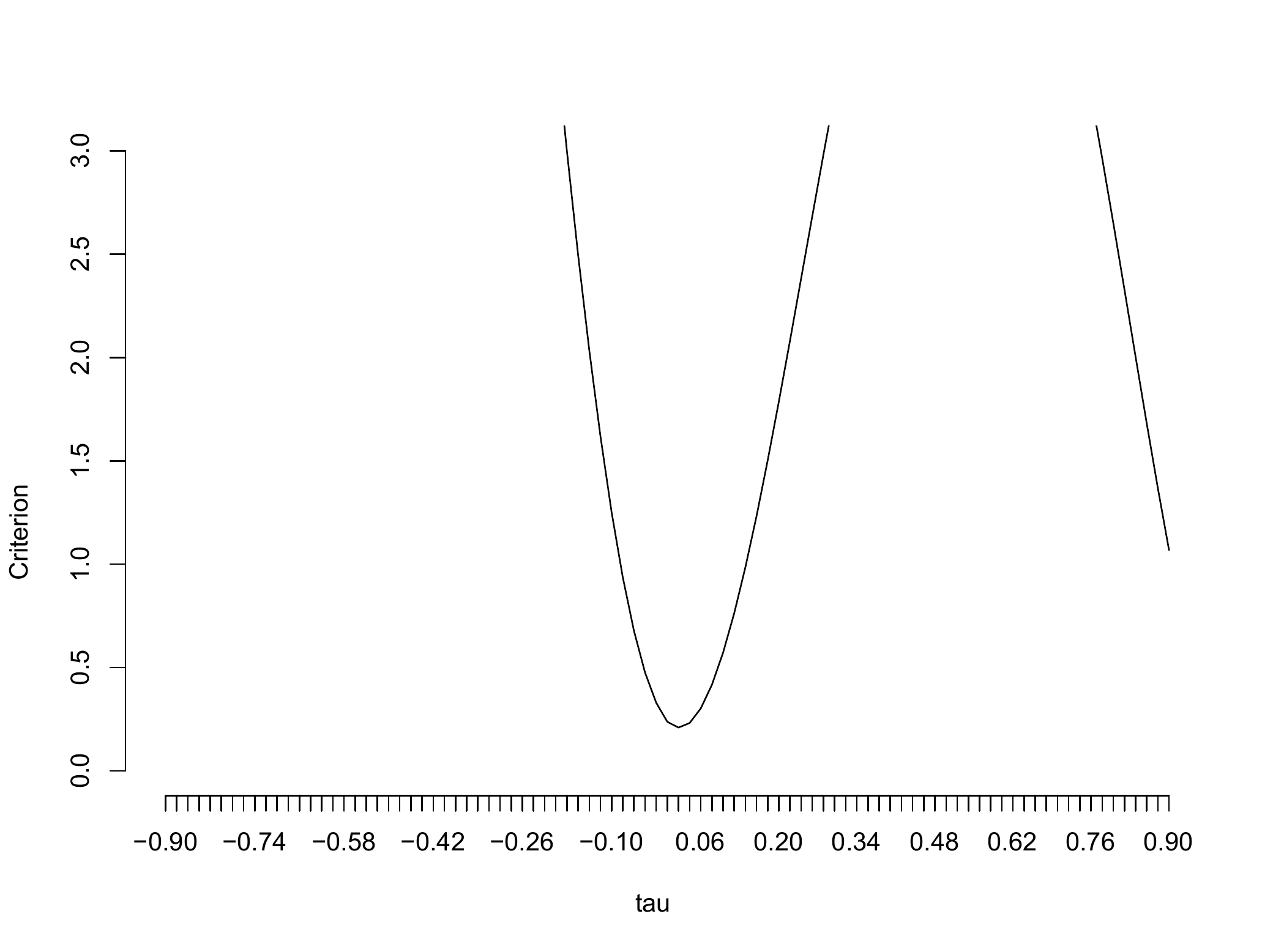}
         \caption{$\tau_0$ = 0}
         \label{c}
     \end{subfigure}
		   \hfill
		\begin{subfigure}[b]{0.3\textwidth}
         \centering
         \includegraphics[width=\textwidth]{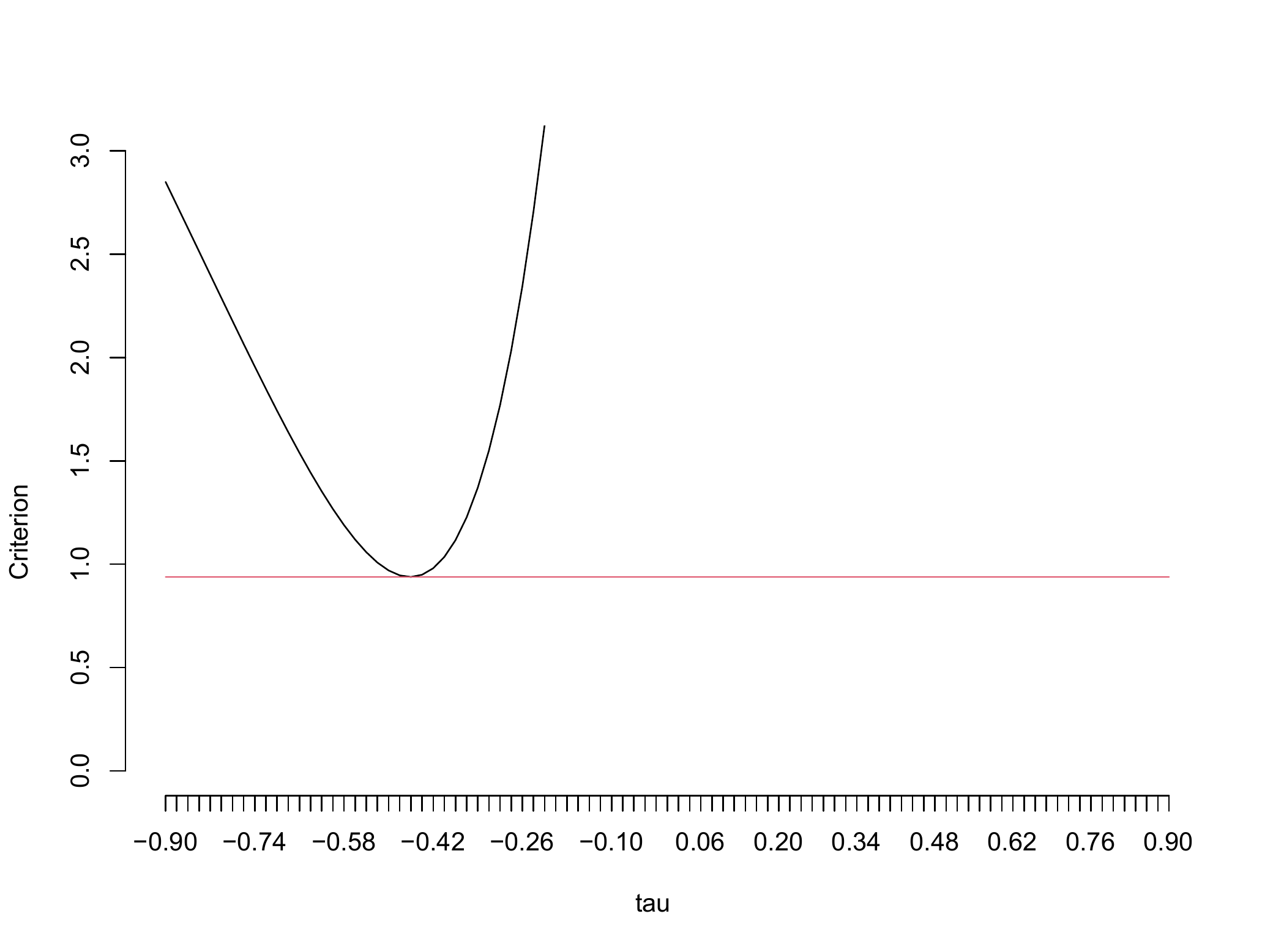}
         \caption{$\tau_0$ = -0.4}
         \label{a}
     \end{subfigure}
		   \hfill
		   \begin{subfigure}[b]{0.3\textwidth}
         \centering
         \includegraphics[width=\textwidth]{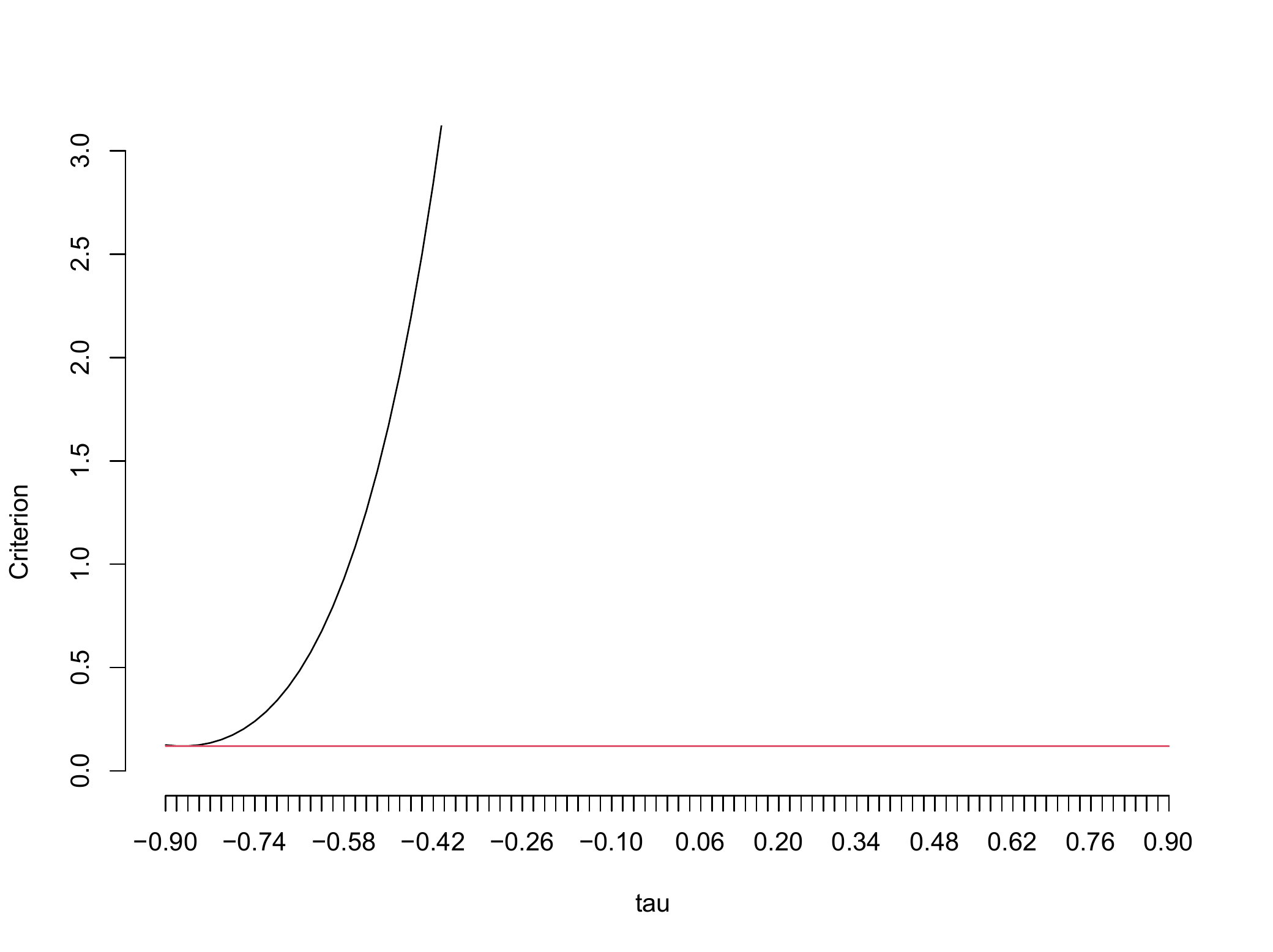}
         \caption{$\tau_0$ = -0.8}
         \label{b}
     \end{subfigure}
        \caption{Objective function (\ref{estaft}) for the Log-logistic model with different Kendall's $\tau_0$.}
        \label{fig:sen3}
\end{figure}

\begin{figure}[h!]
     \centering
     \begin{subfigure}[b]{0.3\textwidth}
         \centering
         \includegraphics[width=\textwidth]{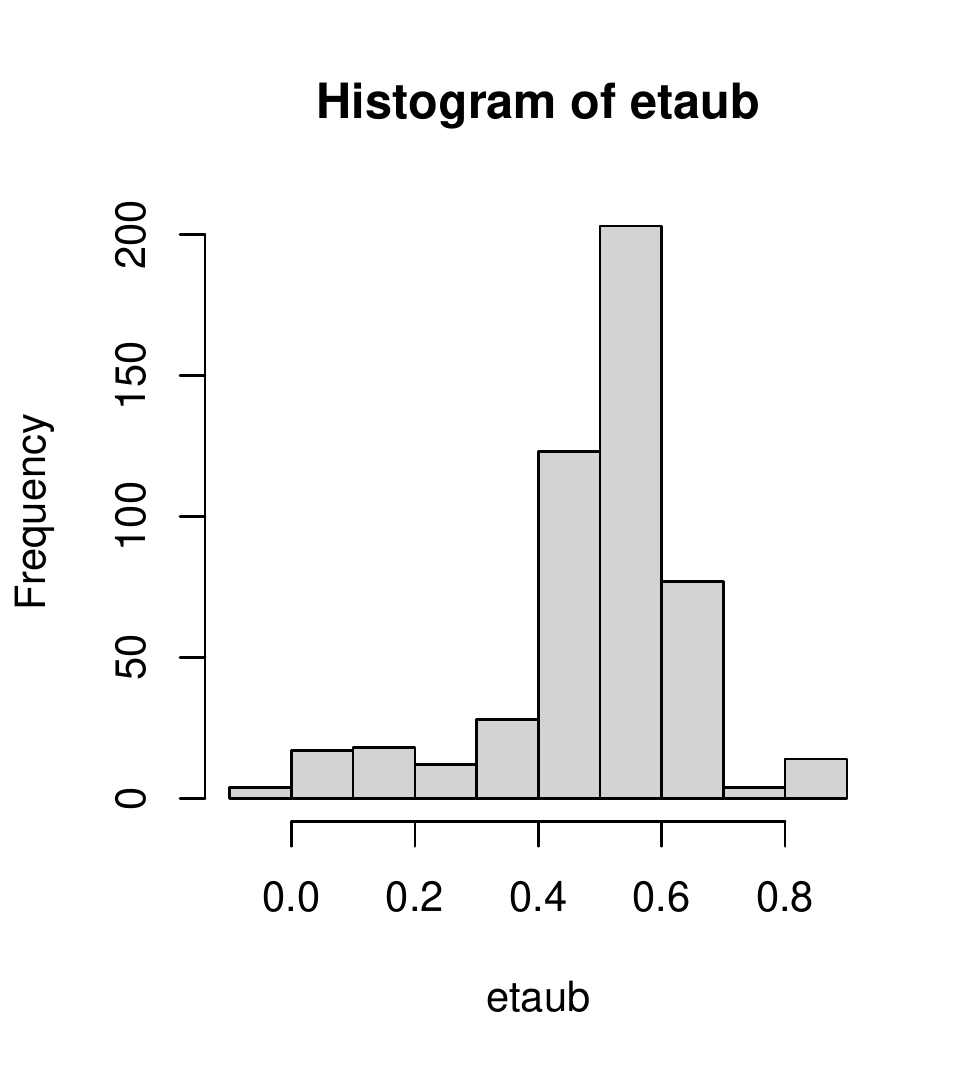}
         \caption{Risk: New Seasonal Employer}
         \label{a}
     \end{subfigure}
     \hfill
     \begin{subfigure}[b]{0.3\textwidth}
         \centering
         \includegraphics[width=\textwidth]{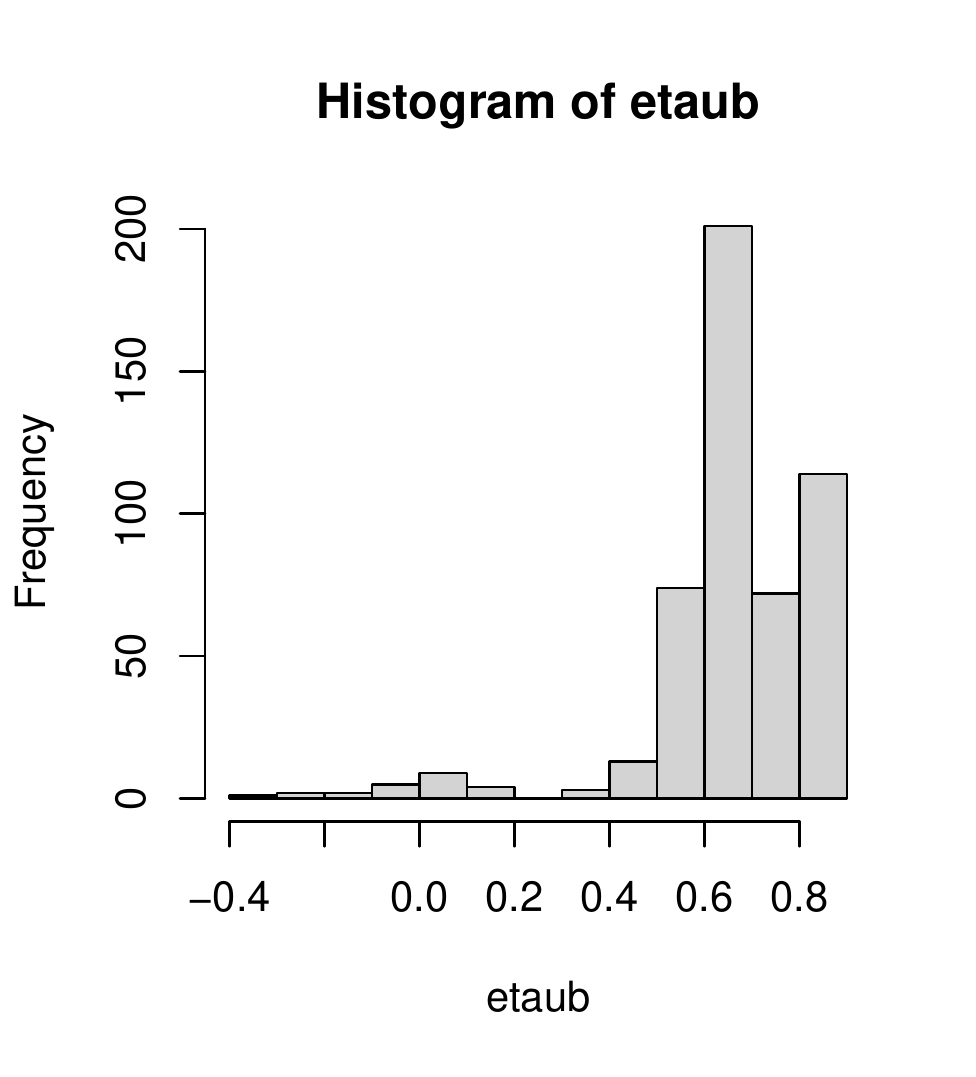}
         \caption{Risk: Other business sector}
         \label{b}
     \end{subfigure}
     \hfill
     \begin{subfigure}[b]{0.3\textwidth}
         \centering
         \includegraphics[width=\textwidth]{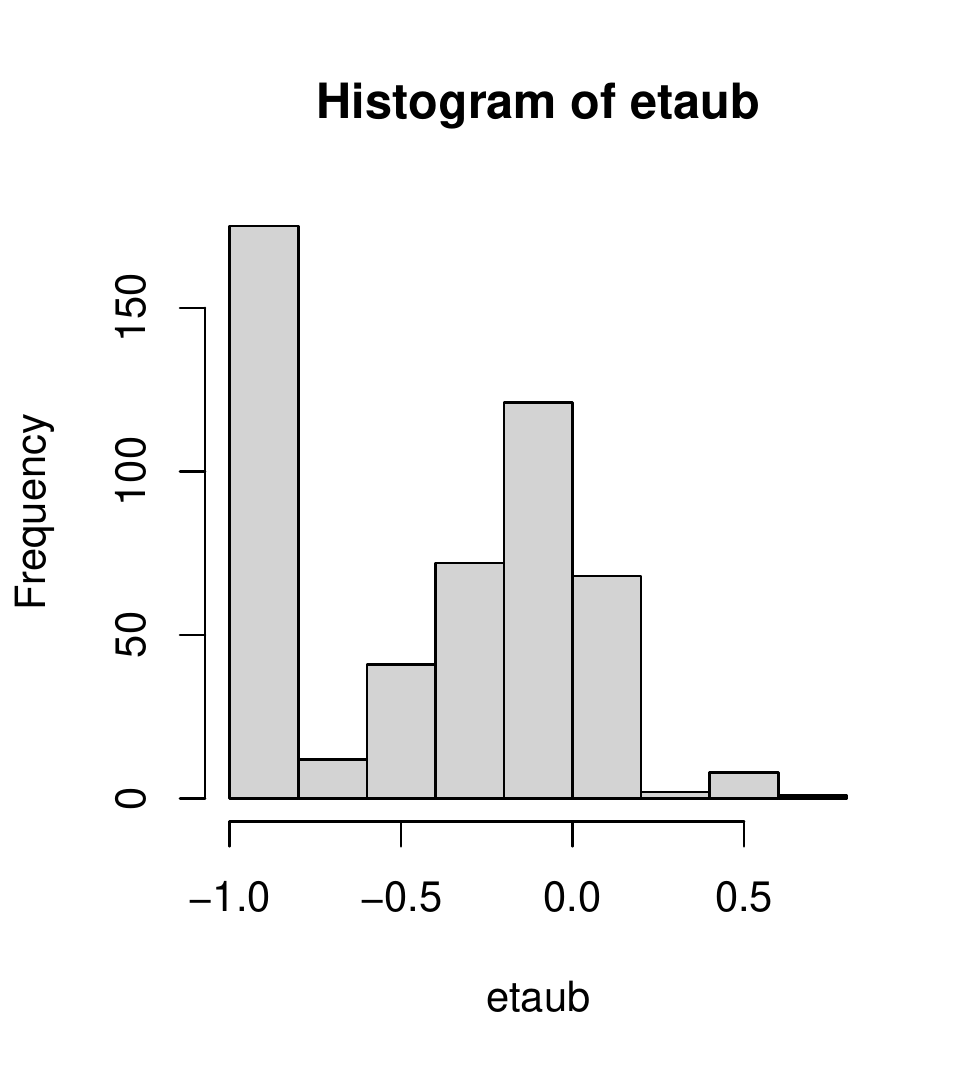}
         \caption{Risk:\\ Other benefits}
         \label{c}
     \end{subfigure}
        \caption{Bootstrap distribution of estimated $\tau$ in the unemployment duration distribution.}
        \label{fig:boot}
\end{figure}


\newpage
\section*{S.IV: Practical remarks for the implementation of (30) in the semiparametric model.}

There are two practical remarks in relation to the implementation of (\ref{thebeta}) in the semiparametric model. First, the estimated $\hat{S}_{CGE}(x_i|z;\theta)$ can be an improper survival distribution function for some data sets. For instance, $\hat{S}_{CGE}(x_i|z_{1};\theta)$ does not descend further to zero but stay on a plateau value for all $x_i>x^*$ at some large enough $x^*$, while, $\hat{S}_{CGE}(x_i|z_{2};\theta)$ continues to decrease to $0$ for $x>x^*$. This issue is common in estimating latent marginal survival function in a competing risks model, when at some $x$ there are no more observed failures for observations with $z=z_{1}$, while there are still failures for $z=z_{2}$. As can be seen from (\ref{ebeta}), $\log(\hat{S}_{CGE}(x_i|z_{1};\theta))$ is constant for all $x_i>x^*$, while $\log(\hat{S}_{CGE}(x_i|z_{2};\theta))$ continues to decrease with $x_i$. In this subset of $x_i$, $\hat{b}_{i}$ is downward biased and the expected value of $\hat{b}_{i}$ will be smaller than the actual value $b_i$. To fix this issue, we recommend a preliminary checking for the estimated $\hat{S}_{CGE}(x_i|z;\theta)$ and only use the observations with $x_i\leq x^*$.

The second remark is that  $ \hat{b}_{i}$ in (\ref{ebeta}) can be undefined for some $i$ with $x_i<x^{**}$ due to a numerical issue. This issue occurs for $x_i<x^{**}$ when there are observed exits to the event of interest for the observations with $z=z_{1}$, while none of the observations with $z=z_{2}$ has failed before $x^{**}$. In this case $\hat{S}_{CGE}(x_i|z_{1};\theta) <1$ while $\hat{S}_{CGE}(x_i|z_{2};\theta) = 1$. $\hat{b}_{i}$ is then a logarithm of zero as $\log(\hat{S}_{CGE}(x_i|z_{2});\theta)=0$ and $\log(\hat{S}_{CGE}(x_i|z_{1};\theta))<0$ and there is no meaningful estimate for $\hat{b}_{i}$ for all $i$ with $x_i <x^{**}$. The remedy for this issue is to drop the affected set of observations for solving (\ref{thebeta}). The value of $x^{**}$ is usually very small when the difference between $z_{1}$ and $z_{2}$ is small. The affected number of observations is therefore small in typical applications.

\newpage
\section*{S.V: Simulation results}
In this supplement we investigate the finite sample performance of the suggested parametric and semiparametric estimation procedures. We use nonparametric estimators for $\pi(t|z)$ and $f_t(t|z)$ in the first stage.

\subsection*{S.V.1 Finite sample performance}
We simulate latent marginals $S(t|z;\alpha_1,\beta_1, \sigma_1)$ and $R(c|z;\alpha_2,\beta_2, \sigma_2)$ from AFT models that are exemplified in Subsection \ref{eAFT}, in particular the (i) exponential, (ii) Weibull, and (iii) log-logistic model. $z$ is a binary variable with $\Pr(z=1)=0.3$ and $(\alpha_j,\beta_j, \sigma_j) =  \chi_j =(1,1,1.5)$ for $j \in\{s,r\}$. $\Kscr_{\theta}$ is a Clayton copula. We do not simulate models with other copulas to avoid excessive length. Nevertheless, previous simulation studies for copula duration models have found that the choice of an Archimedean copula only plays a limited role for the results (see e.g., Lo and Wilke, 2014). To evaluate the bias and MSE of our estimators, we simulate $500$ samples of different sizes $n$. We report the estimated MSE for $\hat{\tau}$ and $\hat{\beta}_1$ as these are the two parameters of interest. Results for $\hat{\alpha}_1$ and $\hat{\sigma}_1$ are given in supplementary material  S.VI.

\subsubsection*{S.V.1.1 AFT model}
To analyse the finite sample behaviour of the parametric 3SE, we conduct three different sets of simulations that serve different purposes:
\begin{enumerate}
\item $S(t|z)$ and $R(c|z)$ belong to the same AFT model and the estimated model for $S(t|z)$ is correctly specified. We simulate the data using different values of $\tau$, namely 0.8, 0.3, -0.3, -0.8, to analyse whether the direction or the degree of dependence between the two risks matters. This part forms the benchmark scenario for the further simulations. (Tables \ref{para1weib} and \ref{para1ellog})

\item $S(t|z)$ and $R(c|z)$ are different AFT models. The estimated model for $S(t|z)$ is correctly specified. This scenario is interesting as it is very likely in applications that $T$ and $C$ have different distributions. (Table \ref{para1diffSR})

\item The model for $S(t|z)$ is misspecified. This is likely the case in applications with limited prior information about the functional form. It informs us about the extent of potential bias and gives researchers some grip about the informativeness of estimation results. (Table \ref{para1miss})

\end{enumerate}

The results are as follows:
\begin{enumerate}
\item Table \ref{para1weib} illustrates the benchmark case when both latent variables follow a Weibull model and the model is correctly specified.
\begin{table}
\centering
\caption{Weibull model - correct specification, different values of $\tau$.}
\hspace{12 pt}
	\begin{tabular}{cccccccccc}
		\hline\hline
		\multicolumn{10}{c}{DGP: $S(t|z), R(c|z) \sim$ weib; Estimated model $\hat{S}(t|z) \sim$ weib } \\
		\hline
		& \multicolumn{4}{c}{$\hat{\tau}$} & & \multicolumn{4}{c}{$\hat{\beta}$} \\ \hline
$n=500$		& $\tau= -0.8$ & $\tau= -0.3$ & $\tau= 0.3$ & $\tau= 0.8$ &  & $\tau= -0.8$ & $\tau= -0.3$ & $\tau= 0.3$ & $\tau= 0.8$ \\ \hline
	$\text{Bias}^2$ & 0.0009 &  0.0092 &  0.0256 & 0.0003 & & 0.0007 & 0.0001 & 0.0007 & 0.0000 \\ \hline
	MSE             & 0.0181 &  0.0378 &  0.0690 & 0.0155 & & 0.0447 & 0.0416 & 0.0414 & 0.0258 \\ \hline
$n=1,000$		& $\tau= -0.8$ & $\tau= -0.3$ & $\tau= 0.3$ & $\tau= 0.8$ &  & $\tau= -0.8$ & $\tau= -0.3$ & $\tau= 0.3$ & $\tau= 0.8$ \\ \hline
	$\text{Bias}^2$ & 0.0007 &  0.0027 &  0.0061 & 0.0004 & & 0.0000 & 0.0001 & 0.0004 & 0.0000 \\ \hline
	MSE             & 0.0127 &  0.0109 &  0.0420 & 0.0095 & & 0.0193 & 0.0197 & 0.0191 & 0.0117 \\ \hline
$n=2,000$		& $\tau= -0.8$ & $\tau= -0.3$ & $\tau= 0.3$ & $\tau= 0.8$ &  & $\tau= -0.8$ & $\tau= -0.3$ & $\tau= 0.3$ & $\tau= 0.8$ \\ \hline
	$\text{Bias}^2$ & 0.0006 &  0.0007 &  0.0006 & 0.0001 & & 0.0001& 0.0000 & 0.0001 & 0.0000 \\ \hline
	MSE             & 0.0115 &  0.0064 &  0.0209 & 0.0047 & & 0.0106 & 0.0110 & 0.0100 & 0.0067 \\ \hline
 \hline
\end{tabular}
\label{para1weib}
\end{table}

\begin{enumerate}
\item The finite sample properties of $\hat{\tau}$ are rather sensitive to $\tau$. Interestingly, for $\tau$ far away from zero (e.g. $\tau = -0.8, +0.8$), the bias is close to zero. Specifically, the bias of $\hat{\tau}$ when $\tau =+0.8$ is only 0.017 $(=\sqrt{0.0003})$ when $n = 500$, and it is even smaller $(0.01=\sqrt{0.0001})$ when $n=2,000$. These biases are negligible relative to the actual sizes of $\tau$. However, for $\tau$ closer to zero (e.g. $\tau= 0.3)$, the bias is much larger. It is 0.16 $(=\sqrt{0.0256})$ for the smallest sample ($n=500$), although it drops quickly down to 0.02 $(=\sqrt{0.0001})$ for $n=2,000$. The MSE for $\tau$ has similar patterns as the bias.

To understand these patterns better, Figure \ref{fig:obj} shows the objective functions (\ref{estaft}) in $\tau$. In the right panel ($\tau= 0.8$) the objective function has a sharp global minimum at the correct value. In the left panel ($\tau= 0.3$), however, the objective function is rather flat at the true value. In consequence, the estimated $\hat{\tau}$ has a larger bias and variance, although the MSE converges to zero with a rate faster than $\sqrt{n}$. Specifically, when the sample size increases 4 times from 500 to 2000, the MSE drops $1/4$ $(\approx 0.0209/0.0690)$ when $\tau = 0.3$. For $\tau= 0.8$, the MSE drops $1/3.3$ $(\approx 0.0047/0.0155)$, which is a bit slower than the previous case, but it is still faster than $1/\sqrt{4}$ when the sample size increases 4 times from 500 to 2000. Note also that the objective function has another local minimum at around 0.1 for $\tau=0.8$. The objective functions for the other parametric models, including log-logistic and exponential models, possess similar patterns and are given in Figures S1 and S2 in the supplementary material.

\begin{figure}
     \centering
     \begin{subfigure}[b]{0.45\textwidth}
         \centering
         \caption{$\tau = 0.3$}
         \includegraphics[width=\textwidth]{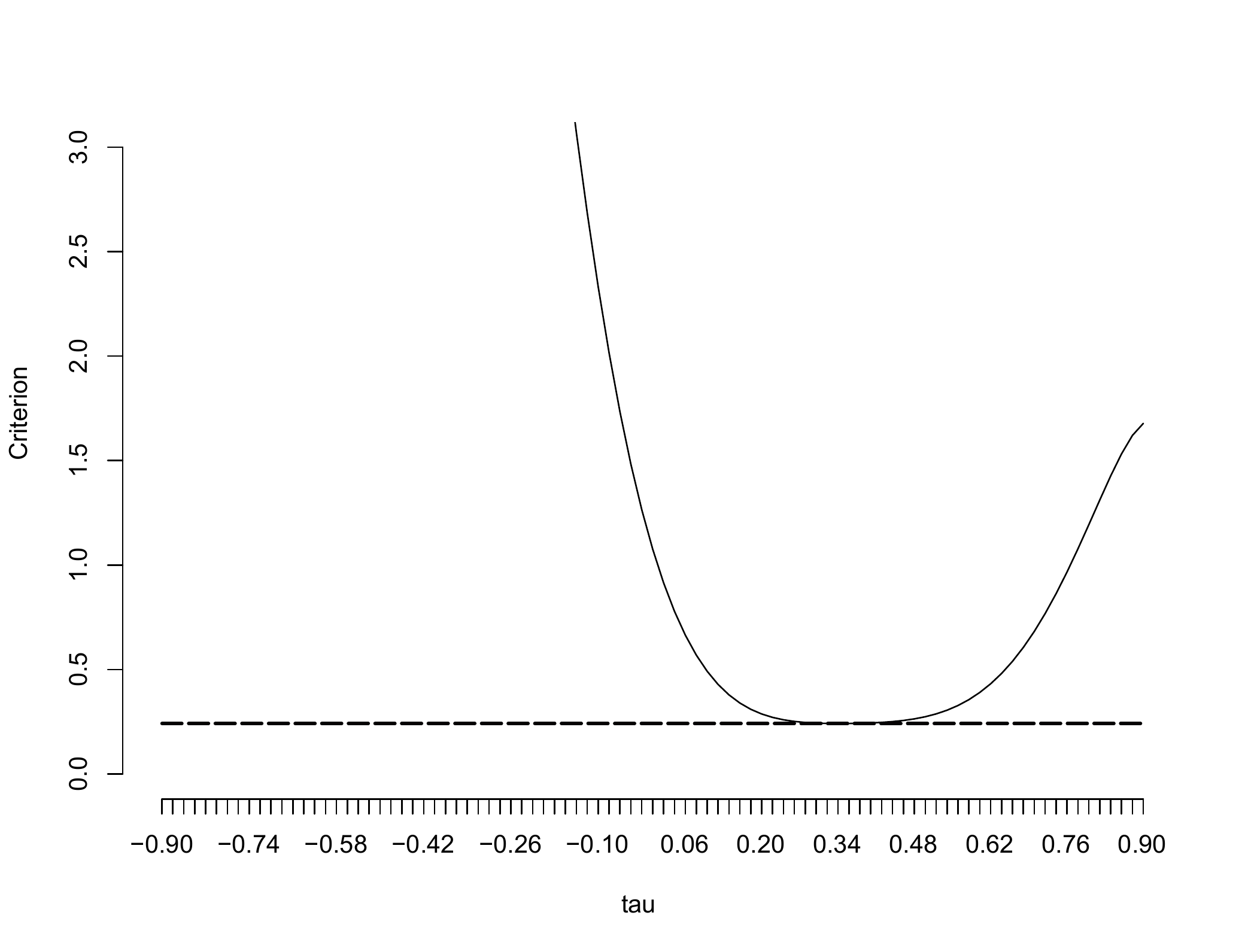}
         \label{fig:obj03}
     \end{subfigure}
     \begin{subfigure}[b]{0.45\textwidth}
         \centering
         \caption{$\tau = 0.8$}
         \includegraphics[width=\textwidth]{weib08.pdf}
         \label{fig:obj08}
     \end{subfigure}
        \caption{Objective function (\ref{estaft}) evaluated at different $\tau$}
        \label{fig:obj}
\end{figure}

\item Table \ref{para1ellog} compares the benchmark Weibull model with two other correctly specified AFT models - exponential and log-logistic for $\tau = 0.3$. The bias and the variance of $\hat{\tau}$ are obviously smaller in the simplest exponential model, where there is only one unknown parameter in the baseline survival function. A greater flexibility of the model therefore comes at the cost of larger finite sample bias and variances.

\begin{table}
\centering
\caption{Various correctly specified AFT models, $\tau = 0.3$.}

\hspace{12 pt}

\begin{tabular}{cccccccc}
		\hline\hline
DGP: $S(t|z), R(c|z)\sim$ &	expo & weib & llog & \hspace{2 pt}	& expo & weib & llog \\
Estimated $S(t|z)\sim $ 	&	expo & weib & llog & \hspace{2 pt}	& expo & weib & llog \\
\hline
$n=500$		& \multicolumn{3}{c}{$\hat{\tau}$}  & & \multicolumn{3}{c}{$\hat{\beta}_s$} \\ \hline
	$\text{Bias}^2$ & 0.0025 & 0.0256  & 0.0244 & & 0.0004 & 0.0000 & 0.0002\\ \hline
	MSE             & 0.0432 & 0.0690  & 0.0570 & & 0.0299 & 0.0258 & 0.0227 \\ \hline
$n=1,000$	& \multicolumn{3}{c}{$\hat{\tau}$} 	& & \multicolumn{3}{c}{$\hat{\beta}_s$} \\ \hline
	$\text{Bias}^2$ & 0.0009 & 0.0061  & 0.0136 & & 0.0001 & 0.0000 & 0.0001 \\ \hline
	MSE             & 0.0241 & 0.0420  & 0.0513 & & 0.0241 & 0.0117 & 0.0128 \\ \hline
$n=2,000$	& \multicolumn{3}{c}{$\hat{\tau}$} 	& & \multicolumn{3}{c}{$\hat{\beta}_s$} \\ \hline	
$\text{Bias}^2$ 	& 0.0001 & 0.0006  & 0.0093 & & 0.0001 & 0.0000 & 0.0000 \\ \hline
	MSE           	& 0.0079 & 0.0209  & 0.0540 & & 0.0069 & 0.0067 & 0.0068 \\ \hline
 \hline
\end{tabular}
\label{para1ellog}
\end{table}

\item While the finite sample properties of $\hat{\tau}$ depend on the actual $\tau$ in Table \ref{para1weib}, they are rather invariant for $\hat{\beta}$.  Relative to its true value ($\beta=1$), both the bias (ranging from 0 to 0.026 = $\sqrt{0.0007}$) and MSE (ranging from 0.0258 to 0.0447) are negligible even when the sample size is as small as 500. Similarly in Table \ref{para1ellog}, the performance of the estimated $\hat{\beta}$ does not vary much across the different models.
\end{enumerate}

\begin{table}
\centering
\caption{Different AFT models for $S(t|z), R(c|z)$ - correct specification, $\tau = 0.3$.}

\hspace{12 pt}

	\begin{tabular}{cccccccccc}
		\hline\hline
															& (1) & (2) & (3) &	&	(4) & (5) & (6)  \\
DGP: $S(t|z)\sim$ &	weib & weib & weib & \hspace{2 pt}& weib & weib & weib  \\
	DGP: $R(c|z)\sim$ &	weib & expo & llog & \hspace{2 pt}& weib & expo & llog  \\
Estimated $S(t|z)\sim$: 	&	weib & weib 					& weib 			& \hspace{2 pt} & weib	& weib & weib  \\
\hline
$n=500$		& \multicolumn{2}{c}{$\hat{\tau}$} &  & \multicolumn{2}{c}{$\hat{\beta}$} \\ \hline
	$\text{Bias}^2$ & 0.0256 & 0.0832 & 0.1273 & & 0.0007 & 0.0178 & 0.0197 &\\ \hline
	MSE             & 0.0690 & 0.1287 & 0.1777 & & 0.0414 & 0.1141 & 0.0737 &\\ \hline
$n=1,000$		& \multicolumn{2}{c}{$\hat{\tau}$}   & \multicolumn{2}{c}{$\hat{\beta}$} \\ \hline
	$\text{Bias}^2$ & 0.0061 & 0.0377 & 0.0722 & & 0.0004 & 0.0081 & 0.0081 &\\ \hline
	MSE             & 0.0420 & 0.0838 & 0.1130 & & 0.0191 & 0.0720 & 0.0424 & \\ \hline
$n=2,000$		& \multicolumn{2}{c}{$\hat{\tau}$}   & \multicolumn{2}{c}{$\hat{\beta}$} \\ \hline
	$\text{Bias}^2$ & 0.0006 & 0.0069 & 0.0289 & & 0.0001 & 0.0012 & 0.0046 &\\ \hline
	MSE             & 0.0209 & 0.0388 & 0.0715 & & 0.0100 & 0.0384 & 0.0278 & \\ \hline
\end{tabular}
\label{para1diffSR}
\end{table}

\item Table \ref{para1diffSR} shows the situation when $S(t)$ and $R(c)$ belong to different models. Compared to the benchmark case in column (1), where both $S(t)$ and $R(c)$ are Weibull, $\hat{\tau}$ in columns (2) and (3) has a larger bias (0.0832 and 0.1273 compared to 0.0256) and MSE (0.1287 and 0.1777 compared to 0.0690) when $n= 500$. For larger samples, the MSE drops quickly at a similar rate as in the benchmark case. For instance, when $S(t)$ and $R(c)$ are Weibull and log-logistic respectively in column (3), the MSE drops by $40\% (\approx 0.0715/0.1777)$ when $n=2,000$. This suggests that the finite sample bias vanishes quickly, but the estimates are less precise when the two latent survival functions come from different families. The results for the estimated $\hat{\beta}$ in columns (4) - (6) show similar patterns. To conclude, although our approach does not model $R(c)$, its functional form affects the precision of the estimates.

\item Table \ref{para1miss} shows the case when the estimated model is misspecified. The results in column (2) are for a Weibull model for $S(t|z)$ that is estimated by an exponential model. Compared to the benchmark case in column (1), the increase in bias and MSE are apparent, although the increase in the latter comes from the increase in the bias, as the Bias$^2$ is almost equal to the MSE. More importantly, this bias does not decrease with sample size, providing evidence of inconsistency. Even for $n=2,000$ the bias of $\hat{\tau}$ is around 0.3 ($=\sqrt{0.0906}$), which is 100\% of the actual $\tau$. When a log-logistic model is fitted to data from a Weibull model (column (3)), the bias of $\hat{\tau}$ is even 0.50 ($=\sqrt{0.2508}$) at $n=2,000$. Similar patterns exist for $\hat{\beta}$ in column (4) - (6). Our model is flexible as it is compatible with various parametric models for $S(t|z)$. However, as usual with parametric duration models, sizable biases in the estimates can be expected when the wrong functional form has been chosen.

\begin{table}
\centering
\caption{Incorrect specifications, $\tau = 0.3$.}

\hspace{12 pt}

	\begin{tabular}{cccccccc}
		\hline\hline
															& (1) & (2) & (3) &	&	(4) & (5) & (6)  \\
DGP: $S(t|z)\sim$ &	weib & weib & weib & \hspace{2 pt}& weib & weib & weib  \\
DGP: $R(c|z)\sim$ &	weib & weib & weib & \hspace{2 pt}& weib & weib & weib  \\
Estimated $S(t|z)\sim$: &	weib	&	expo 	& llog &			&	weib & expo	& llog  \\
\hline
$n=500$		& \multicolumn{3}{c}{$\hat{\tau}$}  & & \multicolumn{3}{c}{$\hat{\beta}_s$} \\ \hline
	$\text{Bias}^2$ & 0.0256 & 0.0909 & 0.2885 & & 0.0007  & 0.9980 & 0.1118 \\ \hline
	MSE             & 0.0690 & 0.0919 & 0.2968 & & 0.0414 & 0.9985 & 0.1313 \\ \hline
$n=1,000$		& \multicolumn{3}{c}{$\hat{\tau}$}  & & \multicolumn{3}{c}{$\hat{\beta}_s$} \\ \hline
	$\text{Bias}^2$ & 0.0061 & 0.0906 & 0.2688 & & 0.0004 & 0.9966 & 0.1123 \\ \hline
	MSE             & 0.0420 & 0.0911 & 0.2726 & & 0.0191 & 0.9981 & 0.1218 \\ \hline

$n=2,000$		& \multicolumn{3}{c}{$\hat{\tau}$}  & & \multicolumn{3}{c}{$\hat{\beta}_s$} \\ \hline
	$\text{Bias}^2$ & 0.0006 & 0.0906 & 0.2508 & & 0.0001 & 0.9970 & 0.1088 \\ \hline
	MSE             & 0.0209 & 0.0911 & 0.2526 & & 0.0100 & 0.9981 & 0.1142 \\ \hline

\end{tabular}
\label{para1miss}
\end{table}
\end{enumerate}

\subsubsection*{S.V.1.2 Semiparametric model}
To investigate the finite sample performance of the semiparametric 2SE of Section \ref{esemi}, we simulate 500 samples from the Weibull model for $S(t|z)$ and $R(c|z)$ with 500, 1,000 and 2,000 observations and for different values of $\tau$. The results are given in Table \ref{semiparaweib}. It can be seen that the bias of $\hat{\tau}$ is more sizable than in the parametric model (Table \ref{para1weib}), in particular for the smallest sample size ($n$=500) but it declines with sample size. For instance, the largest absolute bias is  0.30 (=$\sqrt{0.0899}$) for $\tau =-0.8$ but it reduces to 0.23 (=$\sqrt{0.0532}$) for $n=2,000$. The bias is also relatively large when $\tau = 0.3$, but it drops from $0.31 = \sqrt{0.0989}$ to 0.11 ($=\sqrt{0.0124}$) when the sample size increases from $n=500$ to $2,000$. For the other two values of $\tau$, the bias is negligible when sample size is 2,000. The MSE of the semiparametric 2SE is also greater than in the parametric case (Table \ref{para1weib}). While it decreases with sample size, it is not dropping as fast as in the parametric case. For instance, when $\tau = -0.8$, the MSE for $n=2,000$ reduces to about 2/3 (=0.2394/0.3550) of that for $n=500$. It is apparent that the main source of MSE is the variance and not the bias.

\begin{table}
\centering
\caption{Semiparametric model under correct specification, different values of $\tau$.}
\hspace{12 pt}
	\begin{tabular}{cccccccccc}
		\hline\hline
		\multicolumn{10}{c}{DGP: $S(t|z), R(c|z) \sim$ weib; Estimated model $\hat{S}(t|z) \sim$ semiparametric } \\
		\hline
		& \multicolumn{4}{c}{$\hat{\tau}$} & & \multicolumn{4}{c}{$\hat{\beta}$} \\ \hline
$n=500$		& $\tau= -0.8$ & $\tau= -0.3$ & $\tau= 0.3$ & $\tau= 0.8$ &  & $\tau= -0.8$ & $\tau= -0.3$ & $\tau= 0.3$ & $\tau= 0.8$ \\
	$\text{Bias}^2$ & 0.0899 & 0.0122 & 0.0989 & 0.0225 & & 0.0033 & 0.0002 & 0.0064 & 0.0001 \\
	MSE             & 0.3550 & 0.2602 & 0.3158 & 0.0907 & & 0.0931 & 0.0738 & 0.0718 & 0.0320 \\[0.2cm]
$n=1,000$		& $\tau= -0.8$ & $\tau= -0.3$ & $\tau= 0.3$ & $\tau= 0.8$ &  & $\tau= -0.8$ & $\tau= -0.3$ & $\tau= 0.3$ & $\tau= 0.8$ \\
	$\text{Bias}^2$ & 0.0595 & 0.0008 & 0.0409 & 0.0128 & & 0.0018 & 0.0000 & 0.0039 & 0.0000 \\
	MSE             & 0.2781 & 0.2427 & 0.1870 & 0.0650 & & 0.0362 & 0.0493 & 0.0418 & 0.0167 \\[0.2cm]
$n=2,000$		&  \\
	$\text{Bias}^2$ & 0.0532 & 0.0000 & 0.0124 & 0.0028 & & 0.0011 & 0.0002 & 0.0017 & 0.0000 \\
	MSE             & 0.2394 & 0.1757 & 0.0908 & 0.0278 & & 0.0302 & 0.0299 & 0.0227 & 0.0088 \\
 \hline
 \hline
\end{tabular}
\label{semiparaweib}
\end{table}

Similar to the parametric case, the finite sample properties depend on $\tau$. Figure \ref{F:semi} shows the criterion function in (\ref{thebeta}) for different $\tau$. The criterion function is sometimes flat around the true value, e.g. for $\tau=-0.8$. The criterion function is however rather steep for $\tau= 0.8$. This explains the different variances for different values of $\tau$. Overall, a sample size of $2,000$ is still rather small for a semiparametric competing risks model. In this regard, the presented results show an encouraging performance even when sample size is small.

\begin{figure}[!htbp]
	\centering
(a) $\tau=-0.8$ \hspace{6cm} (b) $\tau=-0.3$\\
	\includegraphics[scale=0.45]{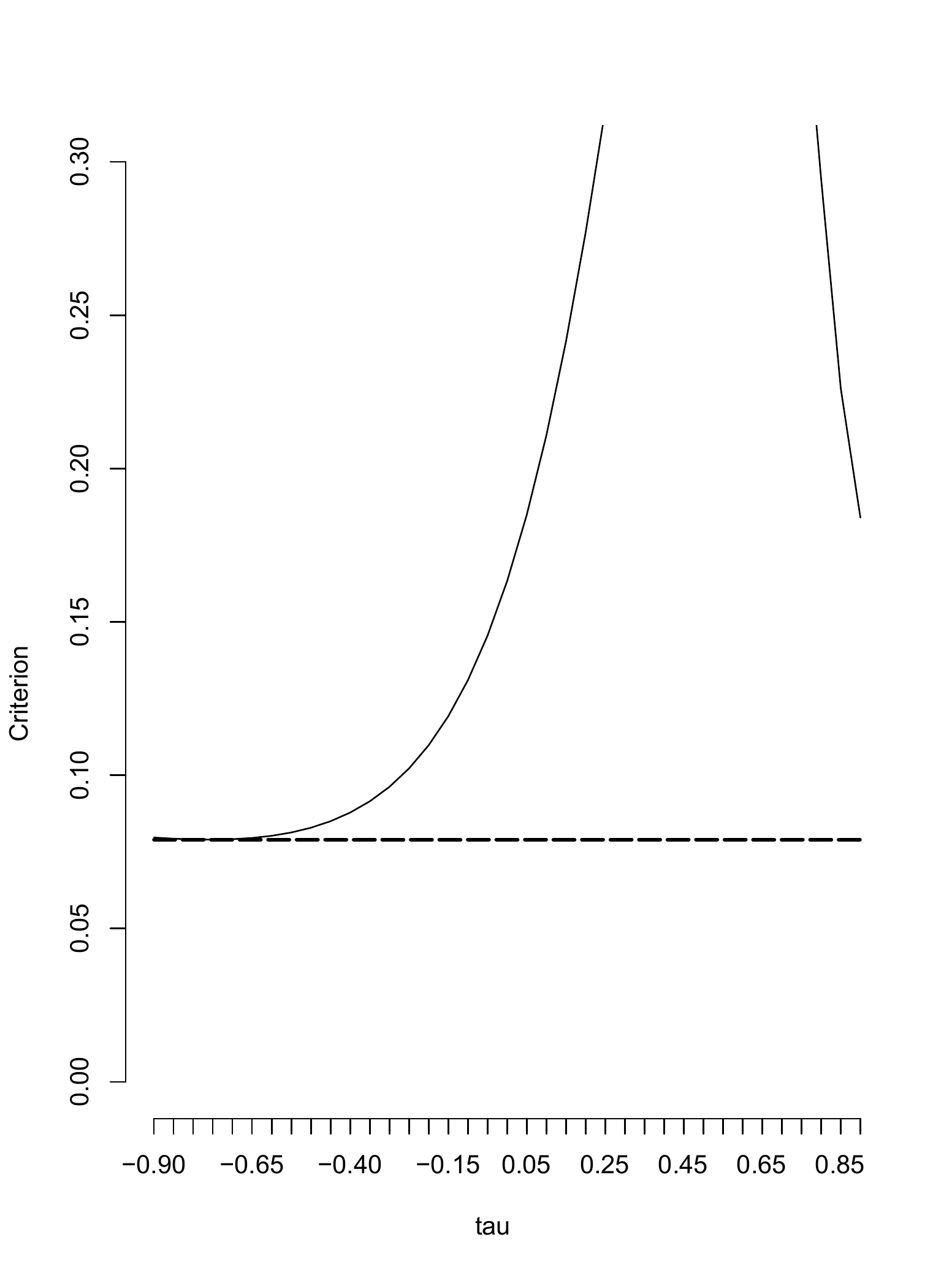}\hspace{1cm}
\includegraphics[scale=0.45]{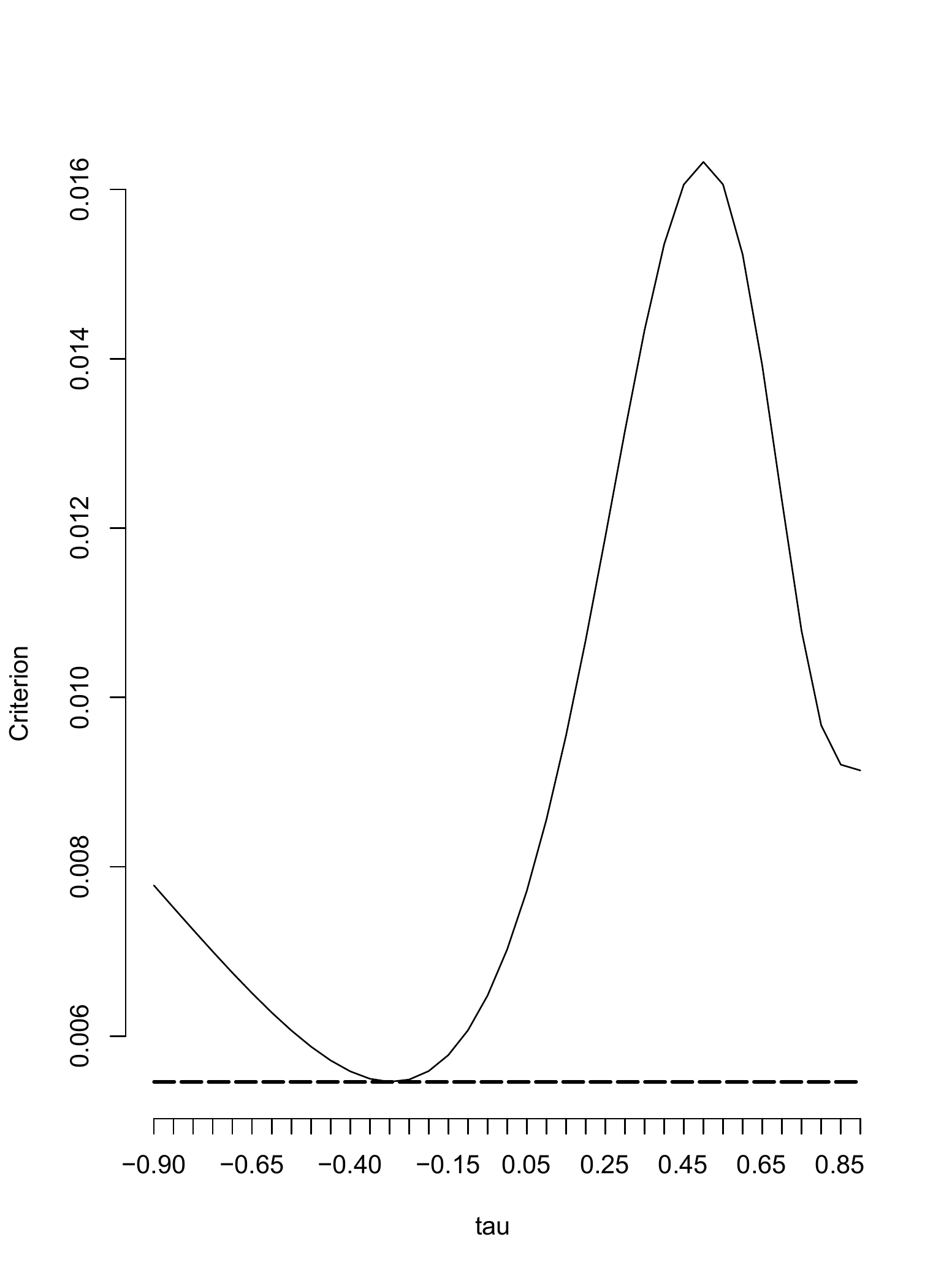}\hspace{1cm}\\
(c) $\tau=0.3$ \hspace{6cm} (d) $\tau=0.8$\\
	\includegraphics[scale=0.45]{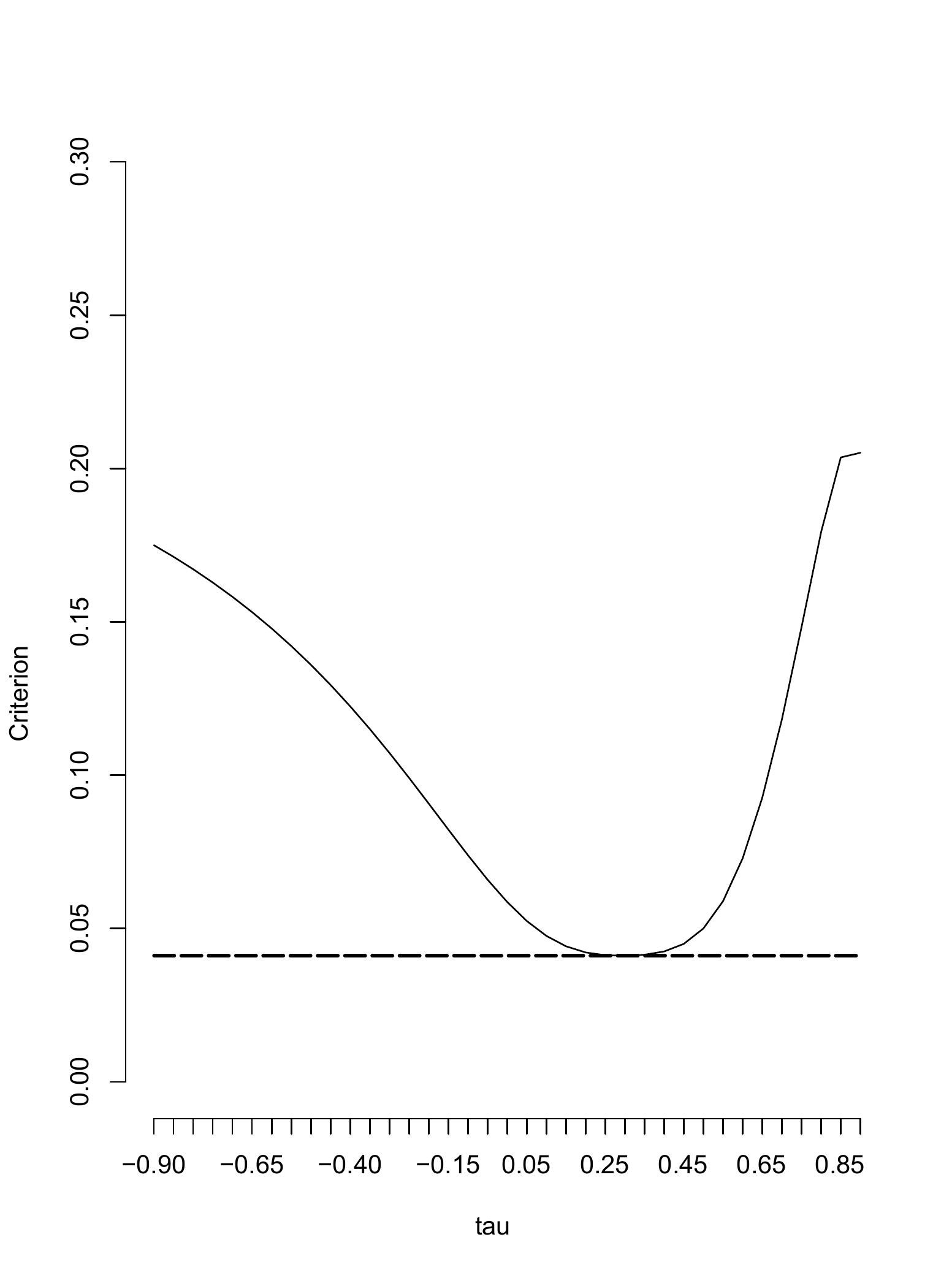}\hspace{1cm}
	\includegraphics[scale=0.45]{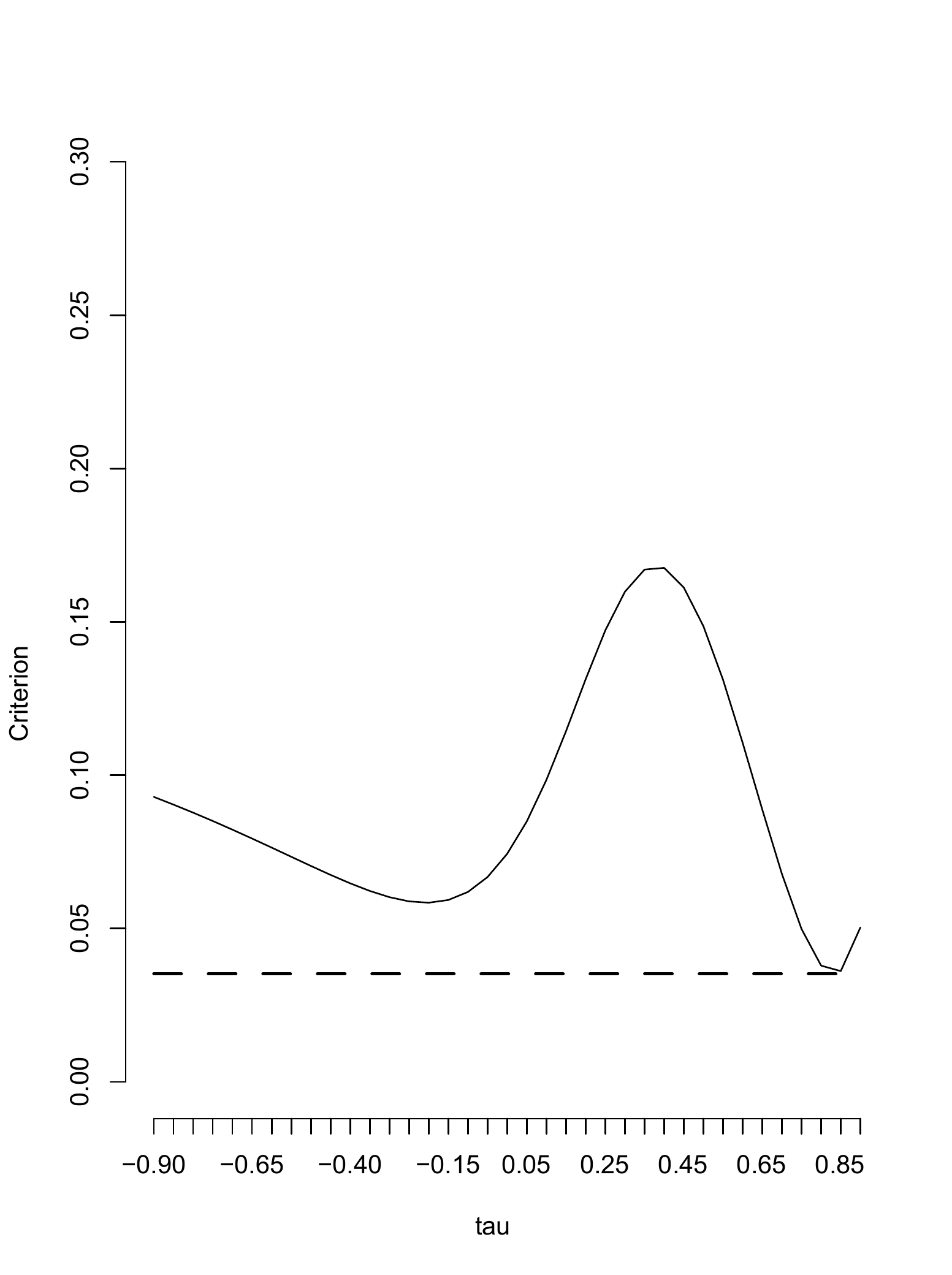}
	\caption{Criterion (\ref{thebeta}) for different values of $\tau$.}
	\label{F:semi}
\end{figure}

\subsection*{S.V.2 Model comparisons}
In this section, we compare the parametric 3SE and semiparametric 2SE with other existing methods, including: (1) full MLE, (2) MMPHM, (3) semiparametric Cox PH model with independent risks (Cox), (4) PWC PH model with independent risks, and (5) PM with independent risks. Under the assumption of independence between $S(t|z)$ and $R(c|z)$ in (3)-(5) it is only necessary to specify $S(t|z)$, while $R(c|z)$ can be ignored as in our approach.

\begin{enumerate}
\item \textbf{Full MLE}. This approach requires known parametric functional forms of $S(t|z,\chi_1)$, $R(c|z,\chi_1)$ and $\mathcal{K}_{\theta}$ with unknown parameters $\chi_1$, $\chi_2$ and $\theta$. Estimation is done by ML and the corresponding log-likelihood is
\begin{eqnarray}
l( \theta, \chi_1, \chi_2; x_i,z_i,\delta_i) &=& \sum_{i=1}^n \delta_i \log k_t + (1-\delta_i) \log k_c,
\end{eqnarray}
\nin where $k_t = \partial \mathcal{K}_{\theta}[S_{\theta}(t|z,\chi_1), R_{\theta}(c|z,\chi_2)] / \partial t |_{t=c=x}$ and $k_c$ is defined analogously. The main disadvantage of MLE compared to our approach is that it requires correct specification of $R(c|z)$. We do the comparison for different scenarios. In the first scenario (a), we apply the MLE when the AFT models for $S(t|z)$ and $R(t|z)$ are both correctly specified. In this case, MLE is expected to be more efficient, as the MLE is a one step estimator and it also makes use of information about $R(c|z)$. In the second scenario (b), we mimic the situation where there is only limited knowledge about $R(c|z)$. In particular, while $S(t|z)$ being correctly specified in the two models, the MLE fits a misspecified model for $R(c|z)$. The result for these comparisons are shown in Table \ref{para1mletau}.

\begin{enumerate}
\item Columns (1) and (4) contain the results when both methods are correctly specified. The MLE ($\text{Bias}^2 = 0.0071<0.0256$ and MSE = $0.0170 < 0.0690$) is considerably more efficient than the three-step estimator for smaller samples $(n=500)$. This advantage, however, declines with sample size. For $n=4,000$, the bias of the 3SE approaches zero with the MSE being smaller than that of MLE ($\text{Bias}^2 = 0.0000<0.0052$ and MSE = $0.0120 < 0.0129$). This advantage of the 3SE becomes even more evident when the sample size increases to 8,000. We explain this by the fact that the number of parameters is larger for MLE than for our approach, as the former uses a model for both risks while the latter only for one. For example, when both $S(t|z)$ and $R(c|z)$ are Weibull, the full likelihood contains seven unknown parameters ($\theta$, and $\lambda_j, \beta_j, \sigma_j$ for both risks $j= 1, 2$), while the 3SE only four ($\theta, \lambda_1, \beta_1, \sigma_1$).

\item The advantage of the suggested 3SE over MLE is even more obvious when the model for $R(c|z)$ is misspecified in the likelihood. Columns (2) and (5) of Table \ref{para1mletau} contain the results when the MLE fits a Weibull model for $R(c|z)$ but the true model is exponential.  Although the bias and MSE for MLE are smaller than for the 3SE for $n=500$ ($\text{Bias}^2 = 0.0832>0.0280$ and MSE = $0.1287>0.0422$), this reverses for a sample size of 2,000 ($\text{Bias}^2 = 0.0069 <0.0308$ and MSE = $0.0388<0.0452$). While the bias and MSE for the 3SE goes down to zero with larger sample size, this is not the case for MLE due to misspecification. The discrepancies between the two approaches increase with sample size, which makes 3SE outperform the MLE with a larger sample size.  Similar patterns can be found in columns (3) and (6) of Table \ref{para1mletau}, when $R(c|z)$ is log-logistic but is mistakenly estimated by Weibull model. Overall, it is found that the suggested 3SE outperforms the misspecified MLE in terms of bias and MSE for larger $n$, because MLE is inconsistent. Results for the estimated $\hat{\beta}$, $\alpha$ and $\sigma$ show similar patterns and are provided in Appendix A.II.

\begin{table}
\centering
\caption{Suggested three-step estimator vs. MLE, $\tau = 0.3$.}

\hspace{12 pt}
\begin{tabular}{ccccccccc}
		\hline\hline
													& \multicolumn{3}{c}{3SE} &\hspace{2 pt} & \multicolumn{3}{c}{MLE}  \\ \hline
													& (1) & (2) & (3) &	&	(4) & (5) & (6)  \\
DGP: $S(t|z)\sim$ 				& weib & weib & weib &	&	weib & weib & weib  \\
DGP: $R(c|z)\sim$ 				& weib & expo & llog &	&	weib & expo & llog  \\
Estimated $S(t|z)\sim$: 	& weib & weib & weib & 	&	weib & weib & weib  \\
Estimated $R(c|z)\sim$: 	&	-		 & 	- 	& -		 &	&	weib & weib & weib  \\
\hline
$n=500$		& \multicolumn{7}{c}{$\hat{\tau}$}   \\ \hline
	$\text{Bias}^2$ & 0.0256 & 0.0832 & 0.1273 & & 0.0071 & 0.0280 & 0.0135 \\ \hline
	MSE             & 0.0690 & 0.1287 & 0.1777 & & 0.0170 & 0.0422 & 0.0331 \\ \hline
$n=2,000$		& \multicolumn{7}{c}{$\hat{\tau}$}   \\ \hline
	$\text{Bias}^2$ & 0.0006 & 0.0069 & 0.0289& & 0.0061 & 0.0308 & 0.0203 \\ \hline
	MSE             & 0.0209 & 0.0388 & 0.0715& & 0.0124 & 0.0452 & 0.0394  \\ \hline
$n=4,000$		& \multicolumn{7}{c}{$\hat{\tau}$}   \\ \hline
	$\text{Bias}^2$ & 0.0000 & 0.0009 & 0.0034 & & 0.0051 & 0.0329 & 0.0189 \\ \hline
	MSE             & 0.0120 & 0.0162 & 0.0299 & & 0.0129 & 0.0464 & 0.0402    \\ \hline
$n=8,000$		& \multicolumn{7}{c}{$\hat{\tau}$}   \\ \hline
	$\text{Bias}^2$ & 0.0000 & 0.0001 & 0.0002 & & 0.0052 & 0.0324 & 0.0214 \\ \hline
	MSE             & 0.0070 & 0.0074 & 0.0097 & & 0.0084 & 0.0480 & 0.0397  \\  \hline
 \hline	
	\end{tabular}
\label{para1mletau}
\end{table}
\end{enumerate}

\item \textbf{MMPHM}. In this model the dependence between the two latent variables $T$ and $C$ is triggered by a frailty distribution, $F(v)$. The joint distribution for $(T,C)$ is
\begin{eqnarray}
H(t,c|z) &=& \int_v \exp[-(\tilde{\Lambda}(t|z)+\tilde{\Lambda}(c|z))v] dF(v). \non
\end{eqnarray}
\nin To make a reasonable comparison with 3SE, we use a MMPHM with parametric $\tilde{\Lambda}(t|z)$ and $\tilde{\Lambda}(c|z)$ - both have Weibull distributions. Since the frailty distribution is often unknown in applications, we use a discrete mass point distribution to approximate the unknown distribution of frailty in the MMPHM. Specifically, we assume that the frailty takes on two unknown values $v_1$ and $v_2$ with unknown probabilities $p_1$ and $1-p_1$, respectively. It allows for flexibility in the MMPHM to mitigate the problem of misspecification caused by a wrong parametric $F(v)$ (Heckman and Singer, 1984). The estimation results are reported in Table \ref{para1mletauIII}. To ease comparison, we also include the estimated $\beta$ from the semiparametric 2SE in the Table. The result for $\tau$ are not presented as it is not estimated by the MMPHM.
\begin{table}
\centering
\caption{Suggested 3SE, 2SE estimator vs. MMPHM, $\tau = 0.3$.}
\hspace{12 pt}
\begin{tabular}{cccccccccc}
		\hline\hline
												  & \multicolumn{3}{c}{3SE} &\hspace{2 pt} & \multicolumn{3}{c}{MMPHM} &\hspace{2 pt} & 2SE \\ \hline
													& (1) & (2) & (3) &	&	(4) & (5) & (6)  && (7) \\	\hline											
$n=500$		& $\hat{\beta}$ & $\hat{\alpha}$ & $\hat{\sigma}$ &	& $\hat{\beta}$ & $\hat{\alpha}$ & $\hat{\sigma}$ & & $\hat{\beta}$\\ \hline
	$\text{Bias}^2$ & 0.0007 & 0.0067 & 0.0044 & & 0.0671 & 0.0050 & 0.0017 && 0.0064 \\ \hline
	MSE             & 0.0414 & 0.0270 & 0.0228 & & 0.3083 & 0.2506 & 0.0451 && 0.0718\\ \hline
$n=1,000$	& $\hat{\beta}$ & $\hat{\alpha}$ & $\hat{\sigma}$ &	& $\hat{\beta}$ & $\hat{\alpha}$ && $\hat{\sigma}$ \\ \hline
	$\text{Bias}^2$ & 0.0004 & 0.0017 & 0.0014 & & 0.0656 & 0.0050 & 0.0007 && 0.0039\\ \hline
	MSE             & 0.0191 & 0.0166 & 0.0136 & & 0.2811 & 0.2043 & 0.0315 && 0.0418 \\ \hline
$n=2,000$		& $\hat{\beta}$ & $\hat{\alpha}$ & $\hat{\sigma}$ &	& $\hat{\beta}$ & $\hat{\alpha}$ && $\hat{\sigma}$ \\ \hline
	$\text{Bias}^2$ & 0.0001 & 0.0001 & 0.0001 & & 0.0517 & 0.0000 & 0.0000 && 0.0017 \\ \hline
	MSE             & 0.0100 & 0.0087 & 0.0069 & & 0.2480 & 0.0738 & 0.0323 && 0.0227  \\ \hline
 \hline	
	\end{tabular}
\label{para1mletauIII}
\end{table}
It is apparent from Columns (4)-(6) that MMPHM estimates for $\beta$ are biased, while this is not the case for $\hat{\alpha}$ and $\hat{\sigma}$. For sample size 2,000 the squared bias of $\hat{\beta}$ is 0.0517, while it is close to zero for $\hat{\alpha}$ and $\hat{\sigma}$. The bias of $\hat{\beta}$ of the MMPHM is found to be much greater than that for the 3SE for all sample sizes ($0.0671>0.0007$ when $n=500$, $0.0517>0.0001$ when $n=2,000$). Although, the discrepancy in the bias of $\hat{\alpha}$ and $\hat{\sigma}$ between the two models is less pronounced, the MSE of the MMPHM estimates is many times larger than the MSE of the 3SE. We explain the lower efficiency of the MMPHM by the fact that a discrete mass point distribution is being used  and similar to MLE both risks are modelled in the MPHMM. Column (7) reports the results for the semiparametric 2SE. It is also obvious that the 2SE outperforms the MMPHM in both bias and MSE.

\item \textbf{COX PH Model (COX)} Another classical survival model is the Cox model, which uses the semiparametric PH model for $S(t|z)$ given by (\ref{aft00}) and (\ref{ph00}).
The difference from our model is that it assumes the independence copula. In practice, only $S(t|z)$ is estimated and $R(t|z)$ is ignored in the modelling. $\beta$ is estimated in a first step by partial ML, which gives numerically stable and fast solutions. The model is very popular in empirical research because of its flexible specification of $S(t|z)$ that leads to a lower risk of misspecification of $S(t|z)$ and no risk of misspecification of $R(t|z)$. Our main focus is therefore on the consequences of misspecifying $\mathcal{K}_{\theta}$. The estimation results are reported in Column (2) of Table \ref{para1mletauIV}. We restrict the presentation to $\beta$ as it is the only parameter in the Cox model. The results show that the Cox model is biased and the bias does not vanish with sample size. Although it has a smaller MSE than the 2SE and 3SE for $n=500$, this reverses for larger samples sizes due to the bias. Our recommendation is therefore to work with the Cox model when $|\tau|$ and $n$ are sufficiently small. Our procedure is expected to be superior the larger $|\tau|$ and $n$. Use the Cox model, if $H_0:\tau=0$ cannot be rejected on the grounds of the 2SE results.

\begin{table}
\centering
\caption{Suggested 3SE, 2SE estimator vs. other models, $\tau = 0.3$.}
\hspace{12 pt}
\begin{tabular}{cccccc}
		\hline\hline
													& (1) & (2) & (3) &		(4)  & (5) \\
Estimator 								& 3SE & 2SE & Cox & PWC &		PM  \\
\hline
$n=500$		& \multicolumn{5}{c}{$\hat{\beta}$}   \\ \hline
	$\text{Bias}^2$ & 0.0007 & 0.0064 & 0.0163 & 0.0175 &  0.0671  \\ \hline
	MSE             & 0.0414 &  0.0718 & 0.0367 & 0.0401 &  0.3083  \\ \hline
$n=1,000$		& \multicolumn{5}{c}{$\hat{\beta}$}   \\ \hline
	$\text{Bias}^2$ & 0.0004 & 0.0039 & 0.0189 & 0.0200 &  0.0656  \\ \hline
	MSE             & 0.0191 & 0.0418 & 0.0296 & 0.0316 &  0.2811  \\ \hline
$n=2,000$		& \multicolumn{5}{c}{$\hat{\beta}$}   \\ \hline
	$\text{Bias}^2$ & 0.0000 & 0.0017  & 0.0213 & 0.0231 &  0.0517 \\ \hline
	MSE             & 0.0100 & 0.0227& 0.0256 & 0.0287 &  0.2480   \\ \hline
 \hline	
	\end{tabular}
\label{para1mletauIV}
\end{table}

\item \textbf{PWC hazard model} The piecewise constant hazard model (see, e.g., Lancaster, 1990) is a frequently applied alternative to the Cox model. It approximates the baseline hazard function by a step function. The more cutting points are included in the model, the more flexible is the model and the closer it becomes to the Cox model. Adding more interval points at the same time causes a loss of precision. In our simulation, we use 60 cut-off points. The results in column (3) of Table \ref{para1mletauIV} are basically the same as that for the Cox model.

\item \textbf{PM} Another commonly used method is to assume a parametric survival function under the assumption of an independent copula. In this case, it suffices to model $S(t|z)$ while $R(c|z)$ can be ignored similar to the Cox model. We consider the model under correct specification of $S(t|z)$ and the focus is therefore to analyse the consequences of incorrect specification of $\mathcal{K}_{\theta}$. The results in column (4) of Table \ref{para1mletauIV} are in line with the results of the Cox model. The bias and MSE are smaller than for the Cox model and piecewise constant model but the bias and MSE once again do not vanish as $n$ increases. The parametric specification therefore leads to improved efficiency but the misspecification of the copula makes the estimates inconsistent.

\end{enumerate}
We summarise our findings as follows. Estimation of the competing risks model is generally sensitive to the assumed model for the latent marginals and the assumed dependency. For this reason, it is desirable to work with milder parametric restrictions if they are not known to hold. In this regard, our three-step estimator has a clear advantage over existing methods in competing risk models that require assumptions on both risks or an independence assumption.

\newpage
\section*{S.VI: Simulation results for $\alpha$ and $\sigma$}

This supplement provides the simulation results for the parameters $\alpha$ and $\sigma$ in the AFT models.
\begin{enumerate}
\item Table \ref{para1weibII} illustrates the benchmark case when both latent variables follow a Weibull model and the model is correctly specified. Similar to the estimated $\tau$, the bias and MSE depends on the true $\tau$. They are largest at $\tau = 0.3$. However, these bias and variance go down rapidly with a larger sample size.
\begin{table}[h!]
\centering
\caption{Weibull model - correct specification, different values of $\tau$.}
\hspace{12 pt}
	\begin{tabular}{cccccccccc}
		\hline\hline
		\multicolumn{10}{c}{DGP: $S(t|z), R(c|z) \sim$ weib; Estimated model $\hat{S}(t|z) \sim$ weib } \\
		\hline
		& \multicolumn{4}{c}{$\hat{\alpha}$} & & \multicolumn{4}{c}{$\hat{\sigma}$} \\ \hline
$n=500$		& $\tau = -0.8$ & $\tau= -0.3$ & $\tau= 0.3$ & $\tau= 0.8$ &  & $\tau= -0.8$ & $\tau= -0.3$ & $\tau= 0.3$ & $\tau= 0.8$ \\ \hline
	$\text{Bias}^2$ & 0.0001 &  0.0010 &  0.0067 & 0.0003 & & 0.0001 & 0.0007 & 0.0044 & 0.0007 \\ \hline
	MSE             & 0.0089 &  0.0138 &  0.0270 & 0.0045 & & 0.0165 & 0.0199 & 0.0228 & 0.0087 \\ \hline
$n=1,000$		& $\tau= -0.8$ & $\tau= -0.3$ & $\tau= 0.3$ & $\tau= 0.8$ &  & $\tau= -0.8$ & $\tau= -0.3$ & $\tau= 0.3$ & $\tau= 0.8$ \\ \hline
	$\text{Bias}^2$ & 0.0000 &  0.0004 &  0.0017 & 0.0001 & & 0.0083 & 0.0005 & 0.0014 & 0.0004 \\ \hline
	MSE             & 0.0049 &  0.0062 &  0.0166 & 0.0024 & & 0.0084 & 0.0097 & 0.0136 & 0.0048 \\ \hline
$n=2,000$		& $\tau= -0.8$ & $\tau= -0.3$ & $\tau= 0.3$ & $\tau= 0.8$ &  & $\tau= -0.8$ & $\tau= -0.3$ & $\tau= 0.3$ & $\tau= 0.8$ \\ \hline
	$\text{Bias}^2$ & 0.0000 &  0.0001 &  0.0001 & 0.0001 & & 0.0000 & 0.0001 & 0.0001 & 0.0001 \\ \hline
	MSE             & 0.0031 &  0.0033 &  0.0087 & 0.0012 & & 0.0048 & 0.0050 & 0.0069 & 0.0025 \\ \hline
 \hline
\end{tabular}
\label{para1weibII}
\end{table}

\item Table \ref{para1ellogII} compares the benchmark Weibull model with two other correctly specified AFT models - exponential and log-logistic for $\tau = 0.3$. Like the estimated $\hat{\tau}$ and $\hat{\beta}$, the bias of $\hat{\alpha}$ are smaller in the simplest exponential model, where there is only one unknown parameter in the baseline survival function. The MSE of $\hat{\alpha}$ is not obvious, it is larger than the Weibull model when sample size is small, while it is smaller when sample size is large. Nevertheless, the log-logistic model has the largest bias and variance with all sample sizes.

\begin{table}
\centering
\caption{Various correctly specified AFT models, $\tau = 0.3$.}

\hspace{12 pt}

\begin{tabular}{cccccccc}
		\hline\hline
DGP: $S(t|z), R(c|z)\sim$ &	expo & weib & llog & \hspace{2 pt}	& expo & weib & llog \\
Estimated $S(t|z)\sim $ 	&	expo & weib & llog & \hspace{2 pt}	& expo & weib & llog \\
\hline
$n=500$		& \multicolumn{3}{c}{$\hat{\alpha}$}  & & \multicolumn{3}{c}{$\hat{\sigma}$} \\ \hline
	$\text{Bias}^2$ & 0.0037 & 0.0067  & 0.0281 & & - & 0.0044 & 0.0134\\ \hline
	MSE             & 0.0434 & 0.0270  & 0.0664 & & - & 0.0228 & 0.0348 \\ \hline
$n=1,000$	& \multicolumn{3}{c}{$\hat{\alpha}$}  & & \multicolumn{3}{c}{$\hat{\sigma}$} \\ \hline
	$\text{Bias}^2$ & 0.0012 & 0.0017  & 0.0195 & & - & 0.0014 & 0.0082 \\ \hline
	MSE             & 0.0200 & 0.0166  & 0.0580 & & - & 0.0136 & 0.0287 \\ \hline
$n=2,000$	& \multicolumn{3}{c}{$\hat{\alpha}$}  & & \multicolumn{3}{c}{$\hat{\sigma}$} \\ \hline	
$\text{Bias}^2$ 	& 0.0001 & 0.0001  & 0.0174 & & - & 0.0001 & 0.0061 \\ \hline
	MSE           	& 0.0078 & 0.0087  & 0.0589 & & - & 0.0069 & 0.0251 \\ \hline
 \hline
\end{tabular}
\label{para1ellogII}
\end{table}

\item Table \ref{para1diffSRII} shows the situation when $S(t)$ and $R(c)$ belong to different models. Compared to the benchmark case in column (1), where both $S(t)$ and $R(c)$ are Weibull, $\hat{\alpha}$ in columns (2) and (3) has a larger bias and MSE. Similar to the estimated $\tau$, the finite sample bias vanishes quickly, but the estimates are less precise when the two latent survival functions come from different families. The results for the estimated $\hat{\sigma}$ in columns (4) -(6) show similar patterns. To conclude, although our parametric estimator does not require any knowledge about the unknown $R(c)$, its functional form affects the precision of the estimates.

\begin{table}
\centering
\caption{Different AFT models for $S(t|z), R(c|z)$ - correct specification, $\tau = 0.3$.}

\hspace{12 pt}

	\begin{tabular}{cccccccccc}
		\hline\hline
															& (1) & (2) & (3) &	&	(4) & (5) & (6)  \\
DGP: $S(t|z)\sim$ &	weib & weib & weib & \hspace{2 pt}& weib & weib & weib  \\
	DGP: $R(c|z)\sim$ &	weib & expo & llog & \hspace{2 pt}& weib & expo & llog  \\
Estimated $S(t|z)\sim$: 	&	weib & weib 					& weib 			& \hspace{2 pt} & weib	& weib & weib  \\
\hline
$n=500$		& \multicolumn{2}{c}{$\hat{\alpha}$} &  & \multicolumn{2}{c}{$\hat{\sigma}$} \\ \hline
	$\text{Bias}^2$ & 0.0067 & 0.0335 & 0.0197 & & 0.0044 & 0.0119 & 0.0119 &\\ \hline
	MSE             & 0.0270 & 0.0571 & 0.0324 & & 0.0228 & 0.0339 & 0.0274 &\\ \hline
$n=1,000$		& \multicolumn{2}{c}{$\hat{\alpha}$} &  & \multicolumn{2}{c}{$\hat{\sigma}$} \\ \hline
	$\text{Bias}^2$ & 0.0017 & 0.0154 & 0.0115 & & 0.0014 & 0.0048 & 0.0056 &\\ \hline
	MSE             & 0.0166 & 0.0377 & 0.0199 & & 0.0136 & 0.0204 & 0.0158 & \\ \hline
$n=2,000$		& \multicolumn{2}{c}{$\hat{\alpha}$} &  & \multicolumn{2}{c}{$\hat{\sigma}$} \\ \hline
	$\text{Bias}^2$ & 0.0001 & 0.0026 & 0.0046 & & 0.0001 & 0.0008 & 0.0024 &\\ \hline
	MSE             & 0.0087 & 0.0180 & 0.0128 & & 0.0069 & 0.0104 & 0.0093 & \\ \hline
\end{tabular}
\label{para1diffSRII}
\end{table}

\item Table \ref{para1mletauII} compares our three-step parametric estimator with the one-step full MLE estimator for the parameter $\beta$, $\alpha$ and $\sigma$.  Columns (1) and (4) of Table \ref{para1mletauII} contain the results when both methods are correctly specified. When sample size is small ($n=500$), MLE is generally more efficient than the three-step estimator. This advantage, however, declines and becomes less obvious with larger sample size. The advantage of the three-step procedure over full MLE is best demonstrated when the full ML contains a misspecified model for $R(c|z)$. Columns (2) and (5) of Table \ref{para1mletauII} contain the results when the MLE fits a Weibull model for $R(c|z)$ but the true model is exponential. While the bias and MSE for $\beta$, $\alpha$, and $\sigma$ in column (2) drop rapidly to zero with larger sample size, they do not vanish in column (5). For sample of size $n = 8000$, the bias and MSE for all three parameters in column (2) are all smaller than that in column (5). Similar pattern can be found by comparing columns (3) and (6) of Table \ref{para1mletauII}, when $R(c|z)$ is log-logistic but is mistakenly estimated by Weibull model. Overall, the three-step parametric estimator outperforms the fully parametric MLE estimator when the distribution of the censoring is unknown, which is often the case in an application. Even if the distribution of the censoring is unknown, the efficiency of the three-step parametric estimator improves rapidly with sample size, while the bias of MLE cannot be eliminated at all.

\begin{table}
\centering
\caption{3SE vs. full MLE, $\tau = 0.3$.}

\hspace{12 pt}
\begin{tabular}{ccccccccc}
		\hline\hline
													& \multicolumn{3}{c}{3SE} &\hspace{2 pt} & \multicolumn{3}{c}{MLE}  \\ \hline
													& (1) & (2) & (3) &	&	(4) & (5) & (6)  \\
DGP: $S(t|z)\sim$ 				& weib & weib & weib &	&	weib & weib & weib  \\
DGP: $R(c|z)\sim$ 				& weib & expo & llog &	&	weib & expo & llog  \\
Estimated $S(t|z)\sim$: 	& weib & weib & weib & 	&	weib & weib & weib  \\
Estimated $R(c|z)\sim$: 	&	-		 & 	- 	& -		 &	&	weib & weib & weib  \\
\hline
$n=500$		& \multicolumn{5}{c}{$\hat{\beta}$}   \\ \hline
	$\text{Bias}^2$ & 0.0007 & 0.0178 & 0.0197 & & 0.0002 & 0.0024 & 0.0018 \\ \hline
	MSE             & 0.0414 & 0.1141 & 0.0737 & & 0.0182 & 0.0272 & 0.0182 \\ \hline
$n=2,000$		& \multicolumn{5}{c}{$\hat{\beta}$}   \\ \hline
	$\text{Bias}^2$ & 0.0001 & 0.0012 & 0.0046 & & 0.0001 & 0.0039 & 0.0009 \\ \hline
	MSE             & 0.0100 & 0.0384 & 0.0278 & & 0.0054 & 0.0175 & 0.0081  \\ \hline
$n=8,000$		& \multicolumn{5}{c}{$\hat{\beta}$}   \\ \hline
	$\text{Bias}^2$ & 0.0000 & 0.0000 & 0.0000 & & 0.0000 & 0.0046 & 0.0015 \\ \hline
	MSE             & 0.0027 & 0.0093 & 0.0051 & & 0.0019 & 0.0140 & 0.0063  \\
	\hline	\hline
$n=500$		& \multicolumn{5}{c}{$\hat{\alpha}$}   \\ \hline
	$\text{Bias}^2$ & 0.0067 & 0.0335 & 0.0197 & & 0.0015 & 0.0050 & 0.0012 \\ \hline
	MSE             & 0.0270 & 0.0571 & 0.0324 & & 0.0059 & 0.0154 & 0.0052 \\ \hline
$n=2,000$		& \multicolumn{5}{c}{$\hat{\alpha}$}   \\ \hline
	$\text{Bias}^2$ & 0.0001 & 0.0026 & 0.0046 & & 0.0009 & 0.0104 & 0.0011 \\ \hline
	MSE             & 0.0087 & 0.0180 & 0.0128 & & 0.0031 & 0.0175 & 0.0040  \\ \hline
$n=8,000$		& \multicolumn{5}{c}{$\hat{\alpha}$}   \\ \hline
	$\text{Bias}^2$ & 0.0000 & 0.0000 & 0.0000 & & 0.0000 & 0.0098 & 0.0012 \\ \hline
	MSE             & 0.0026 & 0.0035 & 0.0016 & & 0.0019 & 0.0163 & 0.0035  \\ 	\hline	\hline
$n=500$		& \multicolumn{5}{c}{$\hat{\sigma}$}   \\ \hline
	$\text{Bias}^2$ & 0.0044 & 0.0119 & 0.0119 & & 0.0071 & 0.0019 & 0.0001 \\ \hline
	MSE             & 0.0228 & 0.0339 & 0.0274 & & 0.0170 & 0.0089 & 0.0051 \\ \hline
$n=2,000$		& \multicolumn{5}{c}{$\hat{\sigma}$}   \\ \hline
	$\text{Bias}^2$ & 0.0001 & 0.0008 & 0.0024 & & 0.0000 & 0.0026 & 0.0001 \\ \hline
	MSE             & 0.0069 & 0.0104 & 0.0093 & & 0.0015 & 0.0077 & 0.0023 \\  \hline
$n=8,000$		& \multicolumn{5}{c}{$\hat{\beta}$}   \\ \hline
	$\text{Bias}^2$ & 0.0000 & 0.0000 & 0.0000 & & 0.0000 & 0.0026 & 0.0000 \\ \hline
	MSE             & 0.0018 & 0.0022 & 0.0013 & & 0.0004 & 0.0066 & 0.0017  \\ 	\hline	\hline
		
	\end{tabular}
\label{para1mletauII}
\end{table}
\end{enumerate}

\newpage
\section*{S.VII Application: Parametric Model for Employment duration}
We consider a two risks model with $T$ being the employment duration terminated by known reasons (risk 1) and $C$ being the employment duration terminated by unknown reasons (risk 2). We use data collected from the U.S. National Longitudinal Surveys from 1979 (round 1) until 2018 (round 28). These data include 12,686 young Americans who are born between 1957-1964. The participants in these cohorts have normally finished their schooling and entered the labour market for the first time. In this example, we extract the employment duration for the first job spell for each individual. Recorded termination reasons for employment include quitting the job and layoff. However, the reasons are missing for around 30\% of the spells, because the question was skipped during the interview for some unknown reasons. It is common practice in empirical research to assume independence of $T$ and $C$ as the missing information was non-systematic and not related to employment duration. It is therefore of interest to apply the suggested approach to estimate Kendall's tau between $T$ and $C$ and therefore to scrutinise common practice.

The observed duration $X=\min\{T,C\}$ has an average value of 321 days (almost a year) with minimum 1 day and maximum 7,408 days (around 20 years). To get a first impression on the distribution of $X$, we plot the estimated density function $f_t(t)$ and $f_c(c)$ in Figure \ref{ex2:f}. These correspond to the probability of terminating an employment spell for known reasons at time $t$ and the probability of terminating an employment spell with unknown reasons at time $c$. While the density for risk 1 could be compatible with Weibull or log-logistic, this is unfortunately less clear for risk 2 as it is almost discontinuous at around 100 days.

\begin{figure}[!htbp]
	\centering
(a) $\hat{f}_t(t)$ \hspace{6cm} (b) $\hat{f}_c(c)$\\
	\includegraphics[scale=0.45]{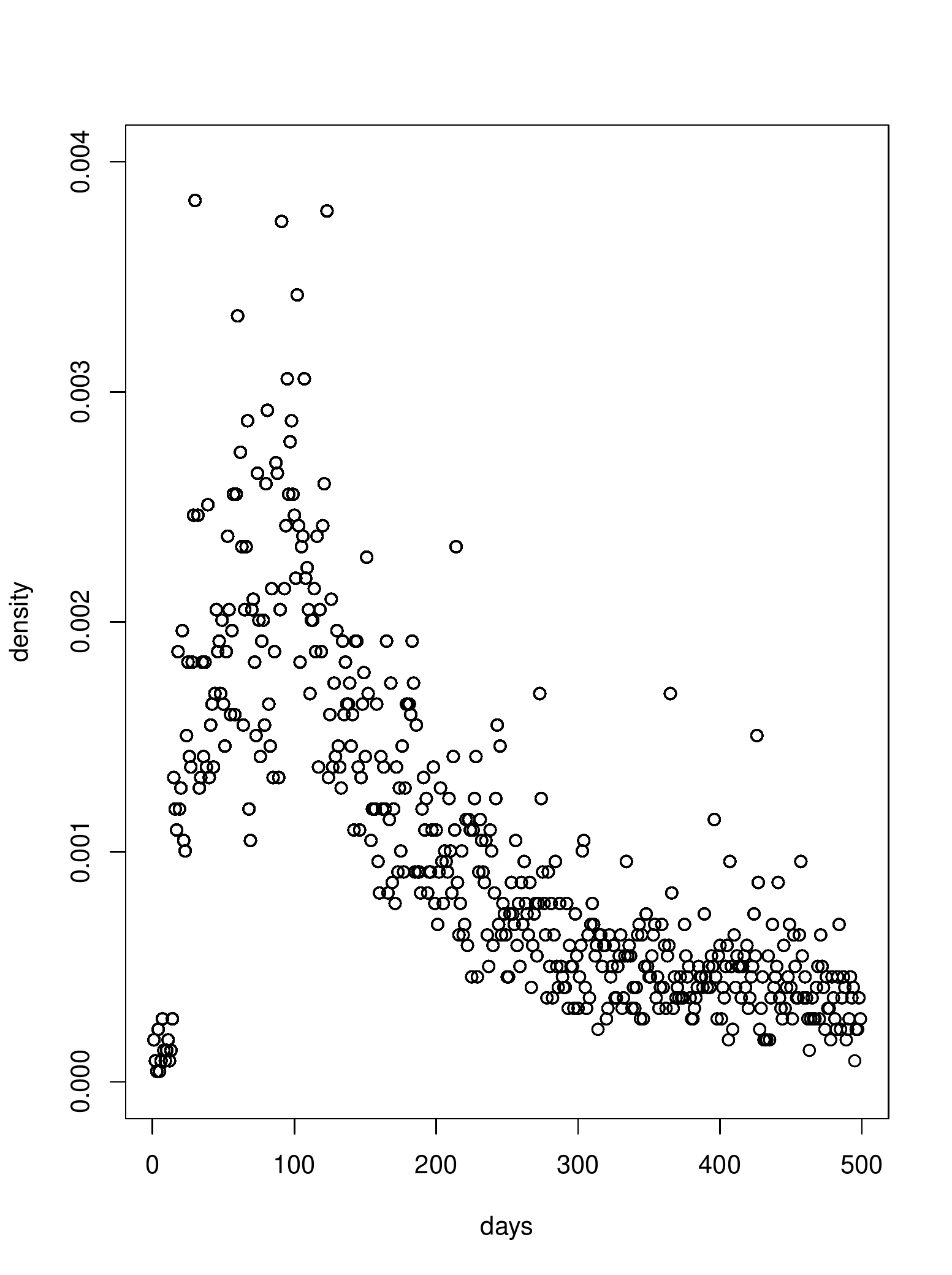}\hspace{1cm}
  \includegraphics[scale=0.45]{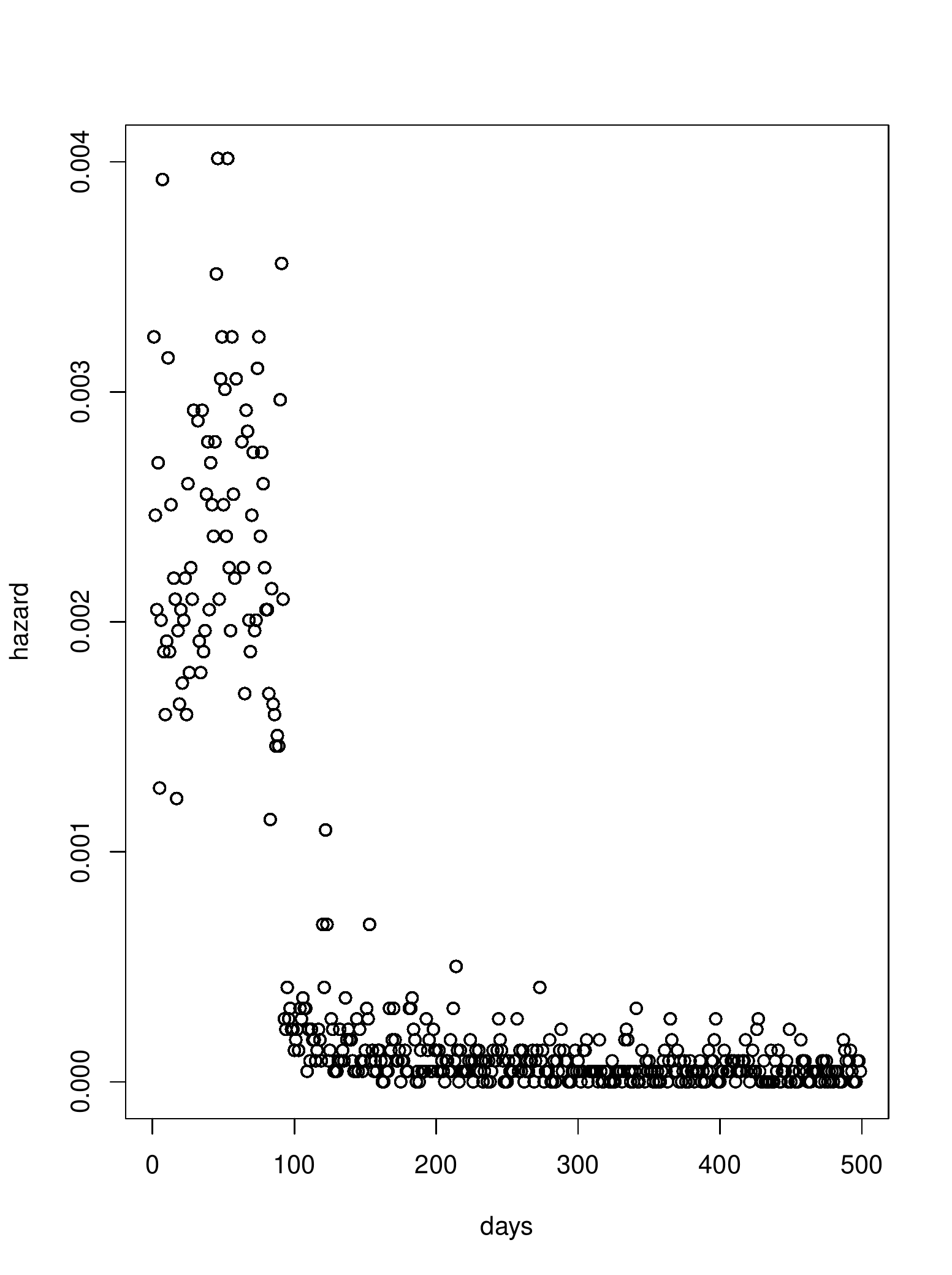}\hspace{1cm}\\
	\caption{Estimated density functions.}
	\label{ex2:f}
\end{figure}

Although $f_t(t)$ and $f_c(c)$ are not the same as the pdf for the latent duration $T$ and $C$, we expect that they have similar shape if $T$ and $C$ were independent. For this reason, we use our proposed method to model the latent marginal distribution for $T$ using the Weibull or log-logistic distribution for $T$ without the need to specify the distribution of $C$.

Figure \ref{ex2:f2} (a) illustrates the objective function for the Weibull model. There is a distinct global minimum at $\hat{\tau} =0.090$. Based on 500 bootstrap resamples, the mean estimated $\tau$ is 0.0896 with 95\% confidence interval $[0.041,0.145]$. $\tau$ is therefore estimated to be significantly greater than zero but dependence is rather weak. Figure \ref{ex2:f2} (b) illustrates the pdf for $\hat{\tau}^*$ for the 500 bootstrap samples, which resembles a normal distribution.

\begin{figure}[!htbp]
	\centering
(a)  \hspace{6cm} (b) \\
	\includegraphics[scale=0.45]{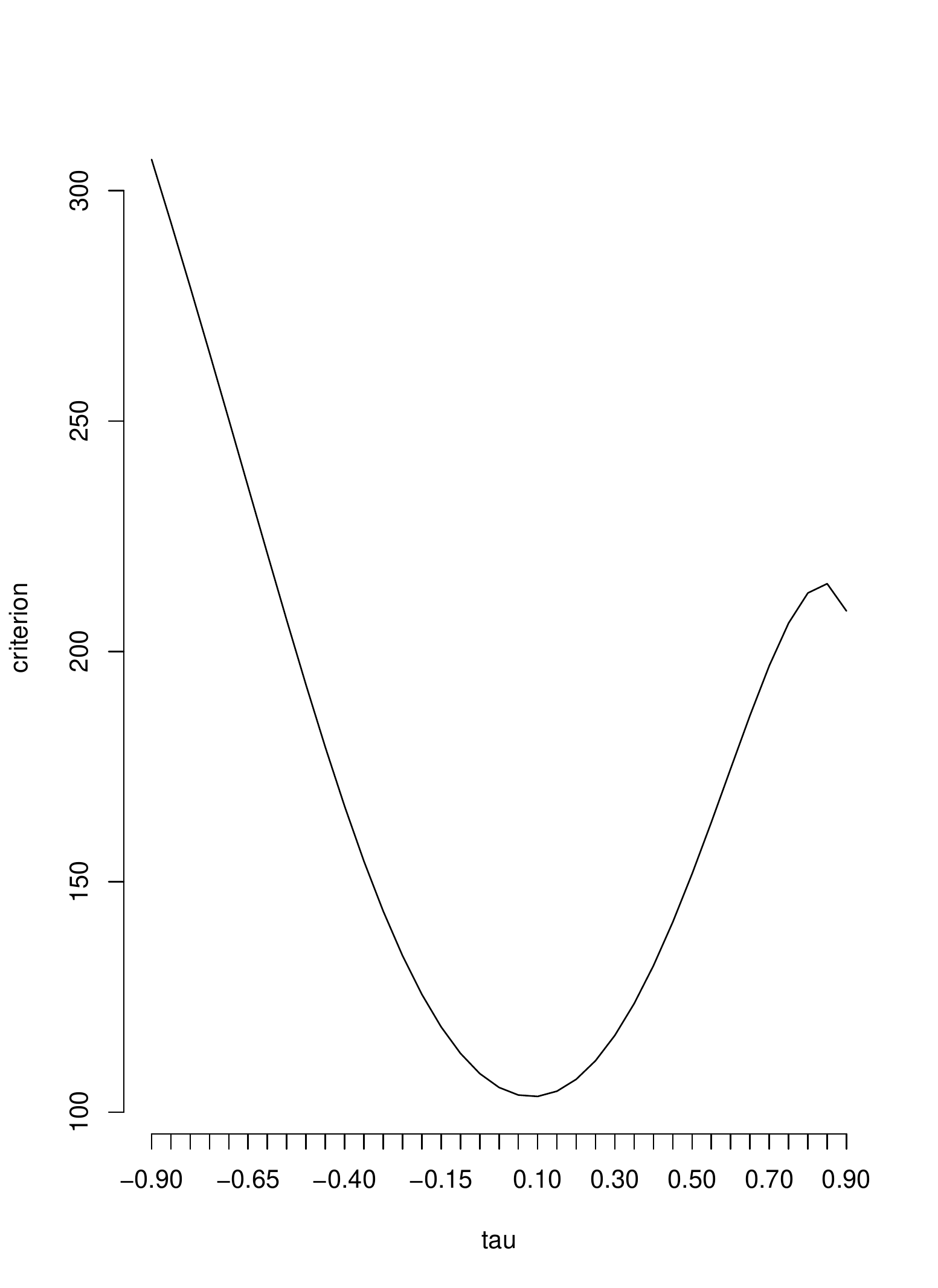}\hspace{1cm}
  \includegraphics[scale=0.45]{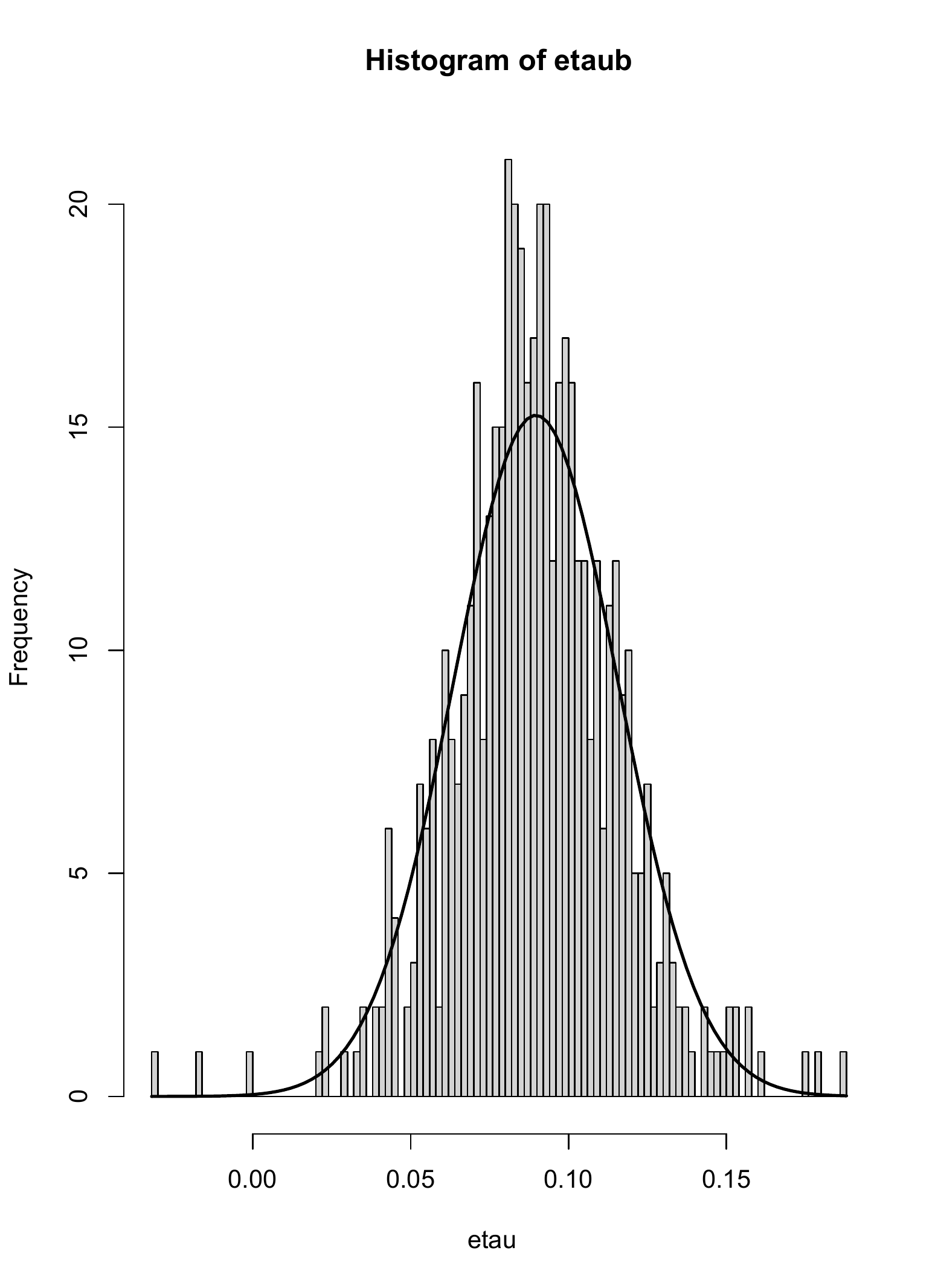}\hspace{1cm}\\
	\caption{(a) The objective function in $\tau$ and (b) the bootstrap pdf of $\hat{\tau}^*$.}
	\label{ex2:f2}
\end{figure}

As a robustness check, we fit the log-logistic model to the data. The estimated $\hat{\tau}$ is 0.136 with bootstrap 95 \% confidence interval  $[0.013,0.278]$, which suggests a slightly stronger dependency. To compare these two results, we plot the estimated $S_{AFT}$ (solid line) and $S_{CGE}$ (dashed line) in Figure \ref{ex2:f3}. The two estimates are more similar for the log-logistic model than the Weibull model, which suggests that the former fits the data better than the latter.

\begin{figure}[!htbp]
	\centering
    (a)  \hspace{6cm} (b) \\
	\includegraphics[scale=0.45]{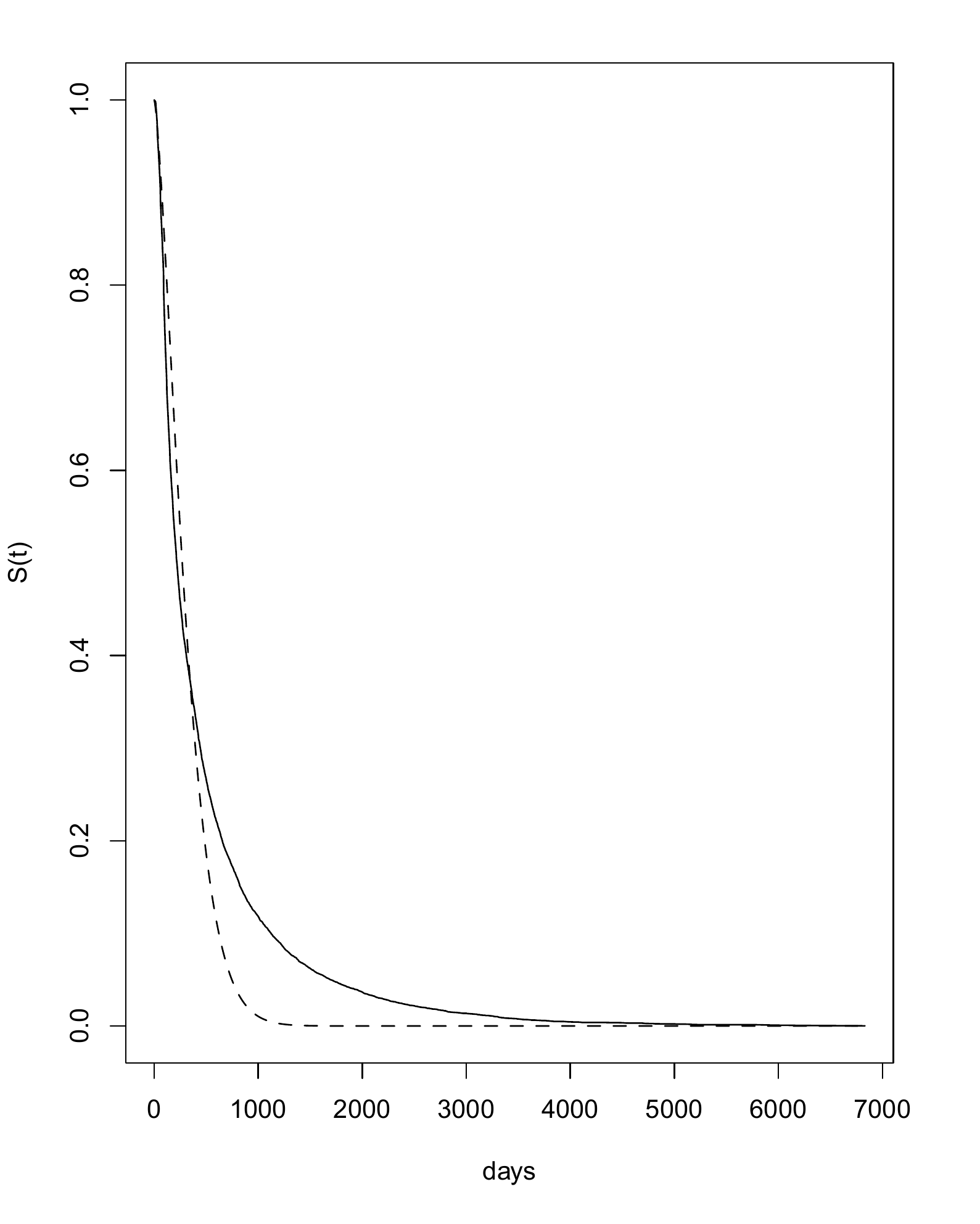}\hspace{1cm}
  \includegraphics[scale=0.45]{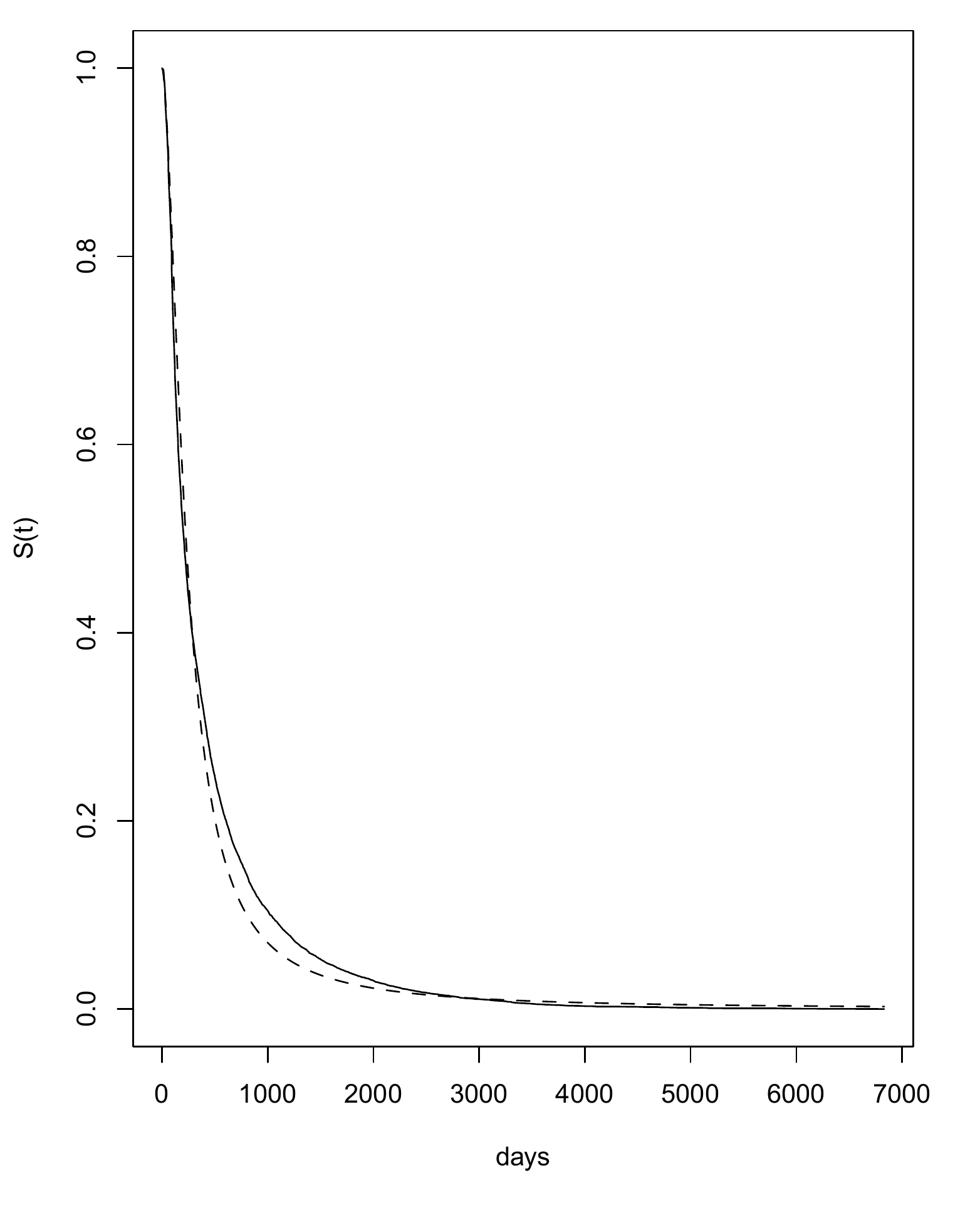}\hspace{1cm}\\
	\caption{The estimated $S_{AFT}$ (solid line) and $S_{CGE}$ (dashed line) for (a) Weibul model and (b) log-logistic model.}
	\label{ex2:f3}
\end{figure}

\begin{table}
\centering
\caption{Effect of gender on $S(t|z)$. Bootstrap sample = 500.}
\hspace{12 pt}
\begin{tabular}{ccccc}
		\hline\hline
														& (1) 			& (2) 			& (3)			&  		\\\hline
Estimator 									& 3SE 			& 3SE 			& MLE			& 		\\
$S$ 												& llog 			& weib 			& weib		&	  	\\
$R$ 												& 	- 			& 	- 			& weib		&   	\\
Dependency 									& Clayton 	& Clayton 	& Clayton	&  \\
\hline
	$\hat{\beta}$ 			&  -0.0582	&	-0.0473 	& -0.716 & \\ \hline
	s.e.($\hat{\beta}$) 	& 0.0016		& 0.0001		& 0.2239 & \\ \hline
	95\% CI for $\hat{\beta}$	& (-0.1401, 0.0153)	& (-0.1061, 0.0102) & (-1.5255, 0.3044) & \\ \hline
 \hline	
													& (4) 						& (5) 						& (6) 						&		(7)   \\\hline
Estimator 								& MPHM 						& Cox 						& PWC							&		PM  \\
$S$ 											& weib 						& nonparametric 	& step function 	&		weib  \\
$R$ 											& weib 						& - 							& - 							&		-  \\
Dependency								& two mass points & independent 		& independent 		&   independent\\
\hline
	$\hat{\beta}$			& -0.8303 				& -0.1676     	&  -0.1590 				& -0.1751 \\ \hline
	s.e.($\hat{\beta}$) 	& 0.2198        	& 0.0001     		&  0.0001 				&  0.0001 \\ \hline
	95\% CI for $\hat{\beta}$	& (-1.347, 0.401) & (-0.2129, -0.1216) & (-0.2069, -0.1093) & (-0.1265, -0.2219) \\ \hline
 \hline	
	\end{tabular}
\label{para1mletauV}
\end{table}

Next, we consider the role of gender for job tenure by including a female dummy as covariate. We apply 3SE and several classical models for comparison.
The results for the coefficient on female are reported in Table \ref{para1mletauV}. A negative $\beta$ corresponds to longer job duration for females than for males. In column (1) of Table \ref{para1mletauV}, the 3SE estimate using the log-logistic model is -0.0582 with 95\% bootstrap confidence interval $[-0.1401,0.0153]$. It suggests that females' hazard rate of job termination for known reasons is $5.7\%$ $(= 1-\exp(-0.0582))$ lower than for males. As the 95 \% confidence interval covers zero, it is not significant. We estimate the 3SE with a Weibull model as a robustness check and the results in column (2) are very similar.


The results for parametric full MLE and the MMPHM with Weibull models for both risks are reported in columns (3) and (4), respectively. These models also allow for risk dependency. The estimated effect of female is more negative than for the 3SE. Females are estimated to have 51\% $(=1 - \exp(-0.716))$ (MLE) and 56\% $(=1 - \exp(-0.8303))$ (MMPHM) lower hazard of terminating a job than males. These estimates are, however, much less precise due to much larger variances. As discussed above, two issues may affect MLE and MMPHM results. First, they are subject to greater risk of misspecification as it requires $R(c)$ to be known. In fact, Figure \ref{ex2:f} suggests that the distribution of $C$ is not Weibull, because of the flat right tail. This misspecification likely causes the MLE and MMPHM results to deviate from 3SE results. Second, the S.E. are larger, because they contain more parameters than the 3SE. In the case of the MMPHM, the imprecision can also be related to the specification of a two mass point approximation of the frailty distribution.

The results in columns (5) to (7) are for models that assume independent risks. As the 3SE results suggest that the two risks are significantly correlated with $\hat{\tau}\approx 0.1$, the results for the latter three models are expected to be somewhat different from 3SE. $\hat{\beta}$ for these models ranges from -0.1590  to -0.1751. Female is therefore estimated to lower the hazard by 14.7\% (=$1-\exp(-0.1590)$)  to 16.0\% (=$1-\exp(-0.1751)$), which is a bit larger in size than for the 3SE. The coefficients of these models are similarly precisely estimated as for the 3SE due to their partial nature. It is worth mentioning that the Cox and PWC model, despite their flexible functional form of $S(t)$, give statistically different results than the 3SE, despite that estimated dependency is rather low. This points to the relevance of estimating the dependence structure to avoid misspecification bias.

We repeat the analysis by restricting the sample to non-degree holders. $\hat{\tau}$ of the 3SE for non-degree holders is 0.0452 with 95\% CI $[-0.1007, 0.1902]$. Since $\hat{\tau}$ is not significantly different from zero, as expected, the estimated gender effects from the 3SE (-0.1377) are almost identical to the estimates using the COX (-0.1524), PWC (-0.1337) and PM (-0.1667) models, which assume independent risks (see Table \ref{tab:sub2}). Moreover, the bootstrap S.E. for the 3SE (0.0017) are very close to those of the other models (0.0014 to 0.0018), providing no evidence of a practically relevant loss of efficiency when using 3SE. For the degree holders, $\hat{\tau}$ is 0.1247 and significant. As expected, $\hat{\beta}$ using the 3SE (0.1324 and significant) is rather different from that of the COX, PWC, and PM models (negative and insignificant). To conclude, the results of 3SE are similar to classical models when the risks are (nearly) independent, but they deviate more the stronger the dependence.

\begin{table}
\centering
\caption{Gender effect on $S(t|z)$ for non-degree holders. Bootstrap sample = 500.}
\hspace{12 pt}
\begin{tabular}{cccccc}
		\hline\hline
														& (1) 								& (2) 						& (3) 						&		(4)   \\\hline
Estimator 									& 3SE 								& Cox							&  PWC						& PM	\\
\hline
	$\hat{\beta}$ 						& -0.1377							& -0.1524	 						& -0.1337		& -0.1667	\\ \hline
	s.e.($\hat{\beta}$) 	& 0.0017							& 0.0014 							& 0.0018		&	0.0015 	\\ \hline
	95\% CI for $\hat{\beta}$	& (-0.2187, -0.0612)	& (-0.2261, -0.0797) 	&	(-0.2264, -0.0465)	& (-0.2405, -0.0926)\\ \hline

 \hline	
	\end{tabular}
\label{tab:sub2}
\end{table}

}



\end{document}